\documentclass{elsart1}

\usepackage[T1]{fontenc}
\usepackage{exscale}
\usepackage{graphicx}
\usepackage{amsmath}
\usepackage{amssymb}
\usepackage{bm}

\newcommand{\vect}[1]{\mathbf{#1}}
\newcommand{\ten}[1]{\mbox{\textbf{\textit{\textsf{#1}}}}}
\newcommand{\cten}[1]{\mbox{\textit{\textsf{#1}}}}
\newcommand{\sprod}{\!\cdot\!}
\newcommand{\tprod}{}
\newcommand{\vprod}{\!\times\!}
\newcommand{\trace}{\mathrm{Tr}}
\newcommand{\trans}{\mathsf{T}}
\newcommand{\dif}{\mathrm{d}}
\newcommand{\mi}{\mathrm{i}}
\newcommand{\me}{\mathrm{e}}

\begin{document}

\begin{frontmatter}

\title{Dispersion forces in macroscopic quantum
electrodynamics}

\author{Stefan Yoshi Buhmann\corauthref{cor}\thanksref{thanks}},
\ead{s.buhmann@tpi.uni-jena.de}
\author{Dirk-Gunnar Welsch\thanksref{thanks}}
\address{Theoretisch-Physikalisches Institut,
Friedrich-Schiller-Universit\"{a}t Jena, Max-Wien-Platz 1,
07743 Jena, Germany}
\ead[url]{www.tpi.uni-jena.de/tpi/qophysics/qo.html}
\corauth[cor]{Corresponding author.}
\thanks[thanks]{This work was supported by the Deutsche
Forschungsgemeinschaft.}

\begin{abstract}
The description of dispersion forces within the framework of
macroscopic quantum electrodynamics in linear, dispersing and
absorbing media combines the benefits of approaches based on
normal-mode techniques of standard quantum electrodynamics and
methods based on linear-response theory in a natural way. It renders
generally valid expressions for both the forces between bodies and
the forces on atoms in the presence of bodies while showing very
clearly the intimate relation between the different types of
dispersion forces. By considering examples, the influence of various
factors like form, size, electric and magnetic properties, or
intervening media on the forces is addressed. Since the approach
based on macroscopic quantum electrodynamics does not only apply to
equilibrium systems, it can be used to investigate dynamical effects
such as the temporal evolution of forces on arbitrarily excited
atoms.
\end{abstract}


\end{frontmatter}

%%%%%%%%%%%%%%%%%%%%%%%%%%%%%%%%%%%%%%%%%%%%%%%%%%%%%%%%%%%%%%%%%%%%%%

\tableofcontents

%%%%%%%%%%%%%%%%%%%%%%%%%%%%%%%%%%%%%%%%%%%%%%%%%%%%%%%%%%%%%%%%%%%%%%

\section{Introduction}
\label{sec1}

Dispersion forces originate from the electromagnetic interaction
between electrically neutral objects which do not carry permanent
electric and magnetic moments. They have been of increasing interest
because of their important impact on many areas of science. In
particular, the extremely miniaturized components in nanotechnology
can be strongly affected by dispersion forces. Recent progress in
experimental techniques has led to accurate measurements of dispersion
forces which have confirmed some of the theoretical predictions while
posing new questions at the same time.

%%%%%%%%%%%%%%%%%%%%%%%%%%%%%%%%%%%%%%%%%%%%%%%%%%%%%%%%%%%%%%%%%%%%%%

\subsection{Dispersion forces}
\label{sec1.1}

The prediction of dispersion forces is one of the most prominent
achievements of quantum electrodynamics (QED) where they can be
regarded as being a consequence of quantum ground-state
fluctuations. In order to understand how quantum fluctuations lead
to dispersion forces, it may be helpful to first recall the
corresponding classical situation. According to classical
electrodynamics, electrically neutral, unpolarized material objects
will not interact with each other, even if they are polarizable. An
interaction can only occur if (i) at least one of the objects is
polarized or (ii) an electromagnetic field is applied to at least one
of the objects. In the former case the object's polarization will give
rise to an electromagnetic field which can induce a polarization of
the other polarizable object(s); in the latter case the applied field
induces a polarization of the object which in turn gives rise to an
electromagnetic field acting on the other object(s). Both cases result
in polarized objects interacting with each other via an
electromagnetic field, the interaction and the resulting attractive
forces between them being a consequence of the departure from the
classical ground state---unpolarized objects and vanishing
electromagnetic field.

In QED, the state that most closely corresponds to the classical
ground state is given by the material objects being in their
(unpolarized) quantum ground states and the electromagnetic field
being in its vacuum state, such that both the electromagnetic field
and the polarization of all objects vanish on the quantum average. At
first glance, one could hence expect the absence of any interaction
between the objects. However, the Heisenberg uncertainty principle
necessarily implies the existence of ground state fluctuations, i.e.,
both (i) fluctuating polarizations of the objects and (ii) a
fluctuating electromagnetic field will always be present. These
fluctuations give rise to an interaction between the objects---a
purely quantum effect which is manifested in the dispersion forces
acting on them. At finite temperatures, additional thermal
fluctuations come into play.

Thus, dispersion forces---also known as Casimir or van der Waals
forces\footnote{Often, the term Casimir force is used to denote
dispersion forces on a macroscopic level whereas dispersion forces
on a microscopic level are referred to as van der Waals
forces.}---are ever-present long-range forces between atoms and/or
macroscopic bodies, i.e., they exist even if the interacting objects
are electrically neutral and do not carry electric or magnetic
moments. Naturally, dispersion forces have many important
consequences. On a microscopic level, they influence, e.g., the
properties of weakly bound molecules \cite{0529,0531}. A prominent
macroscopic signature of dispersion forces is the well-known
correction to the equation of state of an ideal gas, leading to the
more general van der Waals equation.\footnote{In fact, it was in this
context that the existence of dispersion forces was first predicted,
for a historical review, see Ref.~\cite{0487}.} But dispersion forces
also influence the macroscopic properties of liquids and solids such
as the anomalies of water \cite{0532}, the magnetic, thermal and
optical properties of solid oxygen \cite{0544} or the melting
behavior of weakly bound crystals \cite{0530}.

The influence of dispersion forces becomes even more pronounced in
the presence of interfaces between different phases and/or media.
Atom--surface dispersion interactions drive the adsorption of inert
gas atoms to solid surfaces \cite{0117,0225,0505}, influence the
wetting properties of liquids on such surfaces
\cite{0225,0534,0533} and lead to the phenomenon of capillarity
\cite{0539}. The mutual dispersion attractions of colloidal
particles suspended in a liquid \cite{0509} influence the stability
of such suspensions \cite{0540,0541}; unless sufficiently balanced
by repulsive forces, they lead to a clustering of the particles,
commonly known as flocculation \cite{0543,0542}.\vspace*{-1ex}

The above mentioned relevance of dispersion forces to material
sciences and physical chemistry being rather obvious, it is perhaps
more surprising to note that they also play a role in astrophysics
and biology. Thus, dispersion forces initiate the preplanetary dust
aggregation leading to the formation of planets around a star
\cite{0508}. Furthermore, they are needed for an understanding of
the interaction of molecules with cell membranes \cite{0359,0188} and
of cell adhesion driven by mutual cell-membrane interactions
\cite{0359,0495}. Recently, dispersion forces have been found to be
responsible for the remarkable abilities of some gecko \cite{0112} and
spider species \cite{0111} to climb smooth, dry
surfaces.\vspace*{-1ex}

%%%%%%%%%%%%%%%%%%%%%%%%%%%%%%%%%%%%%%%%%%%%%%%%%%%%%%%%%%%%%%%%%%%%%%

\subsection{Experimental observations}\vspace*{-1ex}
\label{sec1.2}

A force between two macroscopic bodies can most easily be measured
in a quasi-static way where the bodies are brought close together
and the force to be measured is compensated by a force of known
magnitude. In the first use of this idea for measuring dispersion
forces, the compensating force was provided by a Hookean spring and
the distance of the bodies was measured by means of optical
interferometry \cite{0568}. In this way, forces between two dielectric
\cite{0568,0569,0520,0567,0565} and metal plates
\cite{0566,0570,0571,0572} were investigated. Difficulties in
aligning the plates were overcome by using alternative setups of a
plate interacting with a spherical lens \cite{0565,0581,0576,0560},
two interacting spheres \cite{0559}, and crossed cylinders
\cite{0559,0561}. Magnetic forces generated by electric
currents were also used to compensate dispersion forces where via
feedback, the force values could be inferred from current measurements
\cite{0517,0558,0585,0573}. The typical outcome of all these
experiments was the observation of attractive forces which follow
$1/z^4$ and $1/z^3$ power laws for the plate--plate and
plate--sphere\footnote{Note that the plate--sphere separation was
much smaller than the sphere radius.} geometries, respectively, with
$z$ denoting the object--object separation (see Ref.~\cite{0556} for
a review). However, owing to the rather low accuracy---the main
limitations being the presence of electrostatic forces due to
residual charges, the roughness of the samples, the lack of accurate
position control and the low resolution of the actual force
measurements---the early results remained controversial. This is
best illustrated by the fact that differing power laws
\cite{0568,0569} and even signs \cite{0566} were found.

Substantial progress was made by using a torsion-balance scheme
\cite{0575} where very smooth bodies, piezoelectric devices for
position control, and capacitive force detection, were used to
measure the force between a metal sphere and plate with high accuracy,
thereby confirming an attractive force proportional to $1/z^3$. This
experiment has been followed by a number of experiments which have
profited by recent developments in nanotechnology (for an overview,
see Refs.~\cite{0378,0135}). By attaching a microsphere to the
cantilever of an atomic force microscope (playing the role of the
spring) and monitoring the dispersion-force induced bending of the
cantilever via deflection of an optical beam, the force between the
sphere and a nearby surface was measured to obtain very accurate
results for various metals \cite{0574,0553,0555,0551,0547} and/or
dielectrics \cite{0588,0512}, including the influence of finite
conductivity \cite{0574,0553,0551} as well as surface roughness
\cite{0555,0547}. It was further demonstrated that one-dimensional
periodic surface corrugations can lead to a sinusoidally varying
tangential force in addition to the attractive normal Casimir force
\cite{0547,0591,0590}. A similar experimental setup was used to
measure the force between two dielectric cylinders \cite{0548} and
spheres \cite{0546}. Dispersion forces have also been measured by
means of a micromachined torsion-balance scheme where a small plate
suspended by two thin rods rotates in response to the Casimir force
exerted by a nearby microsphere and this tilting is monitored by
capacitive measurements. The scheme was used to study the force
between dissimilar metals \cite{0545,0516} and to demonstrate its
dependence on the thickness of the interacting objects
\cite{0587,0586}. Changing the objects' reflectivity in the visible
region by means of hydrogen deposition was found to have no observable
influence on the force \cite{0586,0580}, indicating that it should
depend on the frequency-dependent body properties in some integral
form.

Dispersion forces can also be measured in a dynamical setup, based
on the idea that any interaction will affect the relative motion of
two objects. In the first experimental realization of this idea, a
spherical lens was mounted on a loud speaker and periodically driven,
thereby inducing---by means of the Casimir force---a similar motion of
a nearby plate mounted on a microphone \cite{0564,0583}. Detection of
the amplitude of the induced oscillations led to accurate force
measurements. This idea has been applied in modern experiments to
infer the Casimir force from the periodic motion of an object such as
a microsphere oscillating at the tip of an atomic force microscope
cantilever while interacting with a surface \cite{0557,0579} or an
oscillating micromachined torsion balance interacting with a rigidly
mounted sphere \cite{0545,0516,0550,0401}. Dynamical measurements of
this kind can also provide high-precision results for atom--body
dispersion interactions, which was demonstrated by observing the
change of the oscillatory motion of a single excited ion trapped in
a standing electromagnetic wave \cite{0148} and of cold gases of
(ground-state) atoms confined in a magnetic trap \cite{0404} or an
optical lattice \cite{0403}, induced by their dispersion interaction
with a nearby surface. In the latter experiment, a temperature
dependence of dispersion forces was observed.

Controlled dynamical measurements of dispersion forces on atoms have
only become feasible recently due to the availability of efficient
techniques to cool and trap single atoms. In the early experiments,
scattering techniques were employed which are of course much simpler
to implement. Atom--atom interactions have been observed by scattering
a beam of ground-state atoms with known narrow velocity profile off a
second beam of atoms with thermal velocity distribution
\cite{0592,0584,0589} or a stationary target gas
\cite{0582,0594,0595,0598} where an attractive $1/z^7$ force, was
found. Scattering techniques have also been employed to probe the
interactions of atoms with anisotropic molecules \cite{0597,0599,0600}
and even the interaction of excited atoms with ground-state atoms
\cite{0596} where in the latter case a strong enhancement of the
force, was observed. Evidence of atom--body forces was first found by
observing the deflection of a beam of ground-state atoms or molecules
passing near the surface of a metal or dielectric cylinder, the
results suggesting an attractive $1/z^4$ force
\cite{0137,0138,0139,0140}.\footnote{In the experiments, the minimum
atom--surface separation was so small that to a good approximation,
the cylinder surfaces can be regarded as planar.} In a similar scheme,
the deflection of atoms passing between two metal plates was monitored
by observing the atom flux losses due to the sticking of atoms to the
plates. In doing so, a strong enhancement of the force on excited
atoms was observed \cite{0141}. It was further found \cite{0143,0142}
that the distance dependence of the ground-state force changes from a
$1/z^4$ power law for small atom--surface separations (non-retarded
regime) to the more rapidly decreasing $1/z^5$ power law as soon as
the separations exceed the relevant transition wavelengths of the
atoms and the bodies (retarded regime\footnote{Note that the
above mentioned measurements of Casimir forces between  macroscopic
bodies typically operate in the retarded regime.}).

It has turned out that introducing a controllable compensating force
is also useful in the context of atom-scattering experiments; this
is the central idea of the evanescent-wave mirror: A laser beam is
incident on the surface of a dielectric from the inside at a
sufficiently shallow angle, such that total reflection leads to an
exponentially decaying electric field at its exterior. An atom
placed in the vicinity of the body will interact with this
evanescent field, leading to an optical potential. If the laser
frequency is larger than the relevant atomic transition frequency
(blue detuning), then this potential is repulsive; thus creating the
required compensating force which can be controlled by varying the
laser frequency and intensity. In this way, dispersion forces on
ground-state atoms can be measured by monitoring the reflection of
the atoms incident on evanescent-wave mirrors \cite{0172,0164,0294}.
Alternatively, compensating forces can be provided by the magnetic
fields created by magnetic films, the strength being controlled by
varying the film thickness \cite{0354}.

Effects due to the wave nature of the atomic motion become relevant
for small values of the (center-of-mass) momentum such that the
atomic de Broglie wavelength becomes sufficiently large. In this
case quantum reflection of an atom from the potential associated
with the atom--body force may occur \cite{0076}. Quantum reflection
of ground-state \cite{0444,0167,0402,0449} and excited atoms
\cite{0177,0450} incident on the surface of dielectric bodies was
observed in various experiments where a detailed measurement of the
atom--surface dispersion potential, was achieved by recording the
reflectivities at different normal velocities. Another prominent wave
phenomenon that can be exploited for the measurement of atom--body
potentials is the diffraction of an atomic wave incident on a
transmission grating forming a periodic array of parallel slits. When
passing the slits (which may be regarded as small planar cavities),
the atomic matter wave acquires a phase shift due to the dispersion
potential which affects the interference pattern forming behind the
slits. By comparing the experimental observations with theoretical
simulations, the interaction of ground-state
\cite{0160,0452,0453,0454} as well as excited atoms
\cite{0162,0161,0455} with dielectrics in the non-retarded regime has
been measured.\vspace*{-1ex}

Spectroscopic measurements provide a powerful indirect method for
studying atom--body dispersion interactions. Here, the fact is
exploited that the dispersion potential of an atom can be identified
with the position-dependent shift of the respective atomic energy
level \cite{0030}. The resulting shifts of the atomic transition
frequencies can be observed by spectroscopic means. As the shifts are
usually much more noticeable for excited levels, this approach yields
good estimates of the dispersion potentials of atoms in excited
energy eigenstates. This was demonstrated in experiments measuring
dispersion potentials of atoms inside planar \cite{0144,0456,0145}
and spherical metal cavities \cite{0146,0340}, near a dielectric half
space \cite{0457,0458}, and of an ion near a metal plate \cite{0147}.
In this context, selective reflection spectroscopy of atomic gases has
proven to be a particularly powerful method \cite{0153}. It is based
on the fact that the reflection of a laser beam incident on a gas cell
is modified due to the laser-induced polarization of the gas atoms
which in turn is strongly influenced by the dispersion interaction of
the atoms with the walls of the cell. By comparing measured
reflectivity spectra with theoretically computed ones, very accurate
information on the non-retarded dispersion interaction of atoms with
dielectric plates \cite{0151,0152,0154,0180,0459,0156} was obtained,
including the potentials of atoms in very short-lived excited energy
eigenstates which are difficult to study by scattering methods. Note
that the dispersion interaction with metal plates is much more
difficult to observe via selective reflection spectroscopy
\cite{0460}. As a major achievement, the method has shown that the
dispersion forces on excited atoms can be repulsive
\cite{0157,0159}.\vspace*{-1ex}

%%%%%%%%%%%%%%%%%%%%%%%%%%%%%%%%%%%%%%%%%%%%%%%%%%%%%%%%%%%%%%%%%%%%%%

\subsection{Applications}\vspace*{-1ex}
\label{sec1.3}

Taking advantage of the substantially improved sensitivity of
dispersion-force measurements, comparison of the experimental
results with theoretical predictions can nowadays even be used to
place constraints on other short-scale interactions of fundamental
interest, such as non-standard gravitational forces
\cite{0378,0516,0404,0554,0552,0549,0538}. In addition, dispersion
forces have become of increasing importance in applied science such as
nanotechnology and related fields. While providing a powerful tool
for surface control, e.g., in near-field scanning microscopy
\cite{0114,0115}, they can also be a disturbing factor whose
influence will become more and more pronounced with proceeding
miniaturization. In particular, they can lead to an undesired and
permanent sticking of (small) objects to surfaces
\cite{0578,0593,0577}. A similarly disturbing effect is observed when
atom traps are operated near surfaces where dispersion forces can
diminish the depth of magneto-optical traps, thereby imposing
limits upon the near-surface operation of such traps
\cite{0116,0194}. Traps based on evanescent waves
\cite{0173,0461,0176,0466,0386} necessarily operate in the
near-surface regime so that dispersion forces automatically come
into play. The influence of dispersion forces also needs to be taken
into account when constructing evanescent-wave based elements for
atom guiding \cite{0175,0465,0387,0195,0463}.

Dispersion forces are indispensable in atom optics \cite{0118}
where mirrors and beam splitters for atomic matter waves, have been
constructed based on the dispersion interactions of atoms with flat
surfaces and transmission gratings, respectively. Transmission
gratings can be used to realize Mach--Zehnder-type interferometers
for atoms \cite{0453}. Flat quantum reflective mirrors provide a
focussing mechanism when dispersion and gravitational forces are
combined with in an appropriate way \cite{0467}. In addition, by
locally enhancing the reflectivity of the mirrors via a Fresnel
reflection structure \cite{0468}, reflection holograms for atomic
matter waves can be realized \cite{0178}. The efficiency of atomic
mirrors can also be enhanced by using evanescent-wave mirrors which
can even operate quantum-state selectively \cite{0174}. As recently
reported, the quantum reflection of ultracold gases at dielectric
surfaces gives rise to interesting phenomena, such as the excitation
of solitons and vortex structures \cite{0384}.

Further impact on the application of dispersion forces has been
made by the recent proposal \cite{0477} and subsequent fabrication
\cite{0479} of materials with tailored magneto-electric properties,
also known as metamaterials.\footnote{For the current state-of-art
of metamaterial fabrication, see, e.g.,
Refs.~\cite{0482,0484,0486,0485,0483}.} Metamaterials displaying
simultaneous negative permittivity and permeability in some
frequency range, allow for the existence of traveling
electromagnetic waves whose electric-field, magnetic-field and wave
vector, form a left-handed triad \cite{0476}\footnote{For this reason
materials with these properties are commonly referred to as
left-handed materials.}, leading to a number of unusual effects. It
is yet an open question whether left-handed properties can lead to
interesting phenomena in the context of dispersion forces and to what
extent metamaterials can be exploited to tailor the shape and sign
of these forces. An interesting behavior of dispersion forces may
also occur in conjunction with soft-magnetic alloys, such as permalloy
or Mumetal \cite{0694,0635}. After heating and rapid cooling (a
process called annealing), these materials are in a state of extremely
high permeability; values of more than $5\times 10^4$ have been
reported for Mumetal \cite{0609}.

%%%%%%%%%%%%%%%%%%%%%%%%%%%%%%%%%%%%%%%%%%%%%%%%%%%%%%%%%%%%%%%%%%%%%%

\subsection{Theoretical approaches}
\label{sec1.4}

As already mentioned, dispersion forces arise from quantum
zero-point fluctuations, namely the fluctuating charge and current
distributions of the interacting objects and the vacuum fluctuations
of the (transverse) electromagnetic field. If the separation of the
objects is smaller than the wavelengths of the relevant field
fluctuations, then the latter can be disregarded, allowing for a
simplified treatment of dispersion forces. In this non-retarded
regime, dispersion forces are dominated by the Coulomb interaction
of fluctuating charge distributions. In particular, the Coulomb
interaction between two atoms may within a leading-order multipole
expansion be regarded as the interaction of two electric dipoles
$\hat{\vect{d}}$ and $\hat{\vect{d}}'$,
\begin{equation}
\label{1.1}
\hat{V}=\frac{\hat{\vect{d}}\sprod\hat{\vect{d}}'
 -3\hat{d}_z\hat{d}'_z}{4\pi\varepsilon_0z^3}\,.
\end{equation}
This approach was first used by London in conjunction with
leading-order perturbation theory to derive the potential energy of
two isotropic ground-state atoms to be
\begin{equation}
\label{1.2}
U(z)=-\frac{C}{z^6}\,,\qquad
C=\frac{1}{24\pi^2\varepsilon_0^2}
 \sum_{kk'}\frac{\left|\langle 0|\hat{\vect{d}}|k\rangle\right|^2
 \left|\langle 0'|\hat{\vect{d}}'|k'\rangle\right|^2}
 {E_k+E_{k'}-(E_0+E_{0'})}
\end{equation}
with $|k^{(\prime)}\rangle$ and $E_{k^{(\prime)}}$ denoting the
eigenstates and -energies of the unperturbed atoms \cite{0374}. The
London potential implies an attractive force proportional to $1/z^7$.
The idea of deriving dispersion forces from dipole--dipole
interactions by means of perturbation theory was later applied to
three- \cite{0084,0085,0086,0515,0510,0507,0506}, four- \cite{0511}
and $N$-atom interaction potentials \cite{0494,0119,0120,0501}.
Lennard-Jones showed \cite{0022} that the interaction of an atom with
a perfectly conducting plate can be treated on an equal footing, by
using the image-charge method. Considering the dipole--dipole
interaction of the atom with its own image in the plate (instead of a
second atom) within first-order perturbation theory, he found a
potential
\begin{equation}
\label{1.3}
U(z)=-\frac{\langle 0|\hat{\vect{d}}^2|0\rangle}
 {48\pi\varepsilon_0z^3}\,,
\end{equation}
implying an attractive $1/z^4$ force. The influence of fluctuating
magnetic dipoles \cite{0499,0496,0121}, electric quadrupoles
\cite{0514} and higher multipoles \cite{0513,0237} as well as
permanent electric \cite{0374,0515,0497,0502} and magnetic
dipoles \cite{0497} on the atom--atom interaction has also been
discussed. In this way, it was found that the force between a
magnetizable and a polarizable atom is repulsive and proportional to
$1/z^5$, in contrast to the attractive $1/z^7$ force between two
polarizable atoms. Furthermore, studying the interaction of atoms
prepared in excited energy eigenstates showed that the contributions
to the force which arise from real, resonant transitions can be
attractive or repulsive \cite{0515,0496,0497} (for further reading
regarding the atom--atom interaction, refer to
Refs.~\cite{0693,0261,0608}). Similar extensions have been
accomplished regarding the interaction of atoms with (planar) bodies.
Quadrupole \cite{0288} and higher-order multipole atomic moments
\cite{0282,0289,0412} were included in the interaction of ground-state
atoms with perfectly conducting plates and extensions of the
image-charge method to the interaction of ground-state
\cite{0029,0277} and atoms in excited energy eigenstates \cite{0347}
with planar dielectric bodies were given.

The method can be further improved by describing the atom and the
body on an equal footing in terms of their charge densities and
expressing the resulting interaction potential in terms of
electrostatic linear response functions\footnote{In contrast to the
quantities appearing in earlier attempts to treat conductors in a more
realistic way \cite{0025,0023,0026}, the two response functions are
directly accessible to measurements.} of the two systems. This was
first demonstrated for a ground-state atom interacting with a
realistic electric\footnote{The term electric is used where no
explicit distinction is made between metals (conductors) and
dielectrics (insulators). Likewise, the notion magneto-electric is
used to refer to metals or dielectrics possessing non-trivial magnetic
properties.} half space exhibiting non-local properties \cite{0027}.
The approach was demonstrated to lead to a finite value of the
interaction potential in the limit $z\to 0$
\cite{0282,0375,0269,0281,0273}.\footnote{For the dispersion
interaction between two dielectric half spaces, this is shown in
Refs.~\cite{0650,0649}.} For sufficiently large values of $z$, the
potential for an atom in front of a half space can be given by an
asymptotic power series in $1/z$ \cite{0027,0286,0276,0341,0348}
\begin{equation}
\label{1.4}
U(z)=-\frac{\hbar}{16\pi^2\varepsilon_0z^3}
 \int_0^\infty\dif\xi\,\alpha(\mi\xi)\,
 \frac{\varepsilon(\mi\xi)-1}{\varepsilon(\mi\xi)+1}
 +O(1/z^4)
\end{equation}
[$\alpha(\omega)$, dipole polarizability of the atom;
$\varepsilon(\omega)$, local permittivity of the half space] where
the leading-order term corresponds to an attractive $1/z^4$ force and
coincides with the perfect conductor result (\ref{1.3}) in the limit
of infinite permittivity. Next-order corrections are due to the atomic
quadrupole polarizability on the one hand \cite{0273} and the
leading-order non-local dielectric response on the other hand
\cite{0027}. The response-function approach has been used to study the
interaction of various ground-state objects with half spaces of
different kinds, such as the forces on an ion \cite{0298} and a
permanently polarized atom \cite{0272,0283} in front of a metal half
space, an anisotropic molecule in front of an electric half space
\cite{0290} as well as the interaction of two atoms in front of a
metal \cite{0361,0518} and an electric half space \cite{0036}.
Extensions include the interaction of an atom in an excited energy
eigenstate with an electric \cite{0285} and a birefringent dielectric
half space \cite{0028}; non-perturbative effects \cite{0287}; effects
due to a constant external magnetic field \cite{0295}; and the
interaction of single ground-state atoms/molecules with bodies of
various shapes where perfectly conducting \cite{0110},
non-local metallic \cite{0270,0060} and electric spheres \cite{0110},
non-local metallic \cite{0040,0397} and electric cylinders
\cite{0284}, perfectly conducting planar \cite{0070} and
non-local metallic spherical cavities \cite{0393,0396}, have been
considered.\vspace*{-1ex}

The interaction of two macroscopic bodies $B$ and $B'$ was first
treated by pairwise summation over the microscopic London
potentials~(\ref{1.2}) between the  atoms constituting the bodies
\cite{0642,0641},\vspace*{-1ex}
\begin{equation}
\label{1.4b}
U(z)=-\sum_{\vect{r}\in B}\sum_{\vect{r}'\in B'}
 \frac{C}{|\vect{r}-\vect{r}'|^6}\,,\vspace*{-1ex}
\end{equation}
yielding an attractive $1/z^3$ force between two dielectric half
spaces \cite{0642,0656}. Though applicable to bodies of various shapes
(cf. also Refs.~\cite{0375,0322}), the method could only yield
approximate results due to the restriction to two-atom interactions.
By modeling the body atoms by harmonic oscillators, the interaction
energy of the bodies could be shown to be a sum of all possible
many-atom interaction potentials \cite{0642,0656,0342}. Applications
to the interaction of two half spaces \cite{0120} and two spheres
\cite{0343} were studied. Microscopic calculations of the dispersion
interaction between bodies were soon realized to be very cumbersome,
in particular for more involved geometries. In an alternative approach
based on macroscopic electrostatics, the interaction energy can be
derived from the eigenmodes of the electrostatic Coulomb potential
which are subject to the boundary conditions imposed by the surfaces
of discontinuity \cite{0640,0654} (for an overview, cf.
Ref.~\cite{0135}). The method was used to calculate Casimir forces
between electric spheres \cite{0654}; electric spherical cavities
\cite{0504}; metal half spaces exhibiting non-local properties
\cite{0341}; rough electric half spaces \cite{0630,0631}; and
electrolytic half spaces separated by a dielectric
\cite{0651}.\vspace*{-1ex}

Even though electrostatic methods have been developed into a
sophisticated theory covering various aspects of dispersion forces,
they can only render approximate results valid in the non-retarded
limit where the object separations are sufficiently small so that
the influence of the transverse electromagnetic field can be
disregarded. This was first demonstrated by Casimir and Polder
\cite{0030,0373}. Using a normal-mode expansion of the quantized
electromagnetic field inside a planar cavity bounded by perfectly
conducting plates, they showed that the force between the plates can
be derived from the total zero-point energy of the modes\vspace*{-1ex}
\begin{equation}
\label{1.5}
E=\sum_k{\textstyle\frac{1}{2}}\hbar\omega_k.
\end{equation}
The difficulty that this energy is divergent was overcome by
subtracting the respective diverging energy corresponding to infinite
plate separation, the finite result implying a force per unit area
\begin{equation}
\label{1.6}
\bar{F}=\frac{\pi^2\hbar c}{240}\,\frac{1}{z^4}\,.
\end{equation}
In a similar way, they obtained the force on an atom near one of such
plates and the force between two atoms in free space from the
ground-state energy of the respective system in leading-order
perturbation theory. They recovered the results of the non-retarded
limit, Eqs.~(\ref{1.2}) and (\ref{1.3}), and found that in the
retarded limit the atom--atom and atom--plate potentials are given by
\begin{equation}
\label{1.7}
U(z)=-\frac{23\hbar c\alpha(0)\alpha'(0)}
 {64\pi^3\varepsilon_0^2z^7}
\end{equation}
and
\begin{equation}
\label{1.8}
U(z)=-\frac{3\hbar c\alpha(0)}{32\pi^2\varepsilon_0z^4}\,,
\end{equation}
respectively, which correspond to attractive $1/z^8$ and $1/z^5$
forces that decrease more rapidly than the ones in the non-retarded
limit. Casimir and Polder had thus developed a unified theory to
describe dispersion interactions over a large range of distances.

Normal-mode techniques have since been widely used to study
dispersion interactions. The two-atom interaction has been confirmed
in various ways
\cite{0011,0521,0325,0498,0047,0055,0054,0056,0320,0051,0323,0007,%
0067,0201,0200}, inter alia by basing the calculations on the
multipolar-coupling scheme \cite{0011,0325,0007} in place of the
minimal-coupling scheme originally used by Casimir and Polder, and
relativistic corrections have been considered \cite{0189}. Extensions
include the interaction of three \cite{0047,0067,0088,0106,0202,0109}
or more atoms \cite{0090,0091}, the influence of higher-order
multipole moments \cite{0490} and permanent dipole moments \cite{0492}
on the two-atom force, the interaction between anisotropically
polarizable atoms \cite{0107} and that between a polarizable and a
magnetizable atom \cite{0089,0095,0094,0097,0537,0104,0096} (for an
overview, cf.~Ref.~\cite{0620}). In particular, it was found that in
the retarded limit the force between a polarizable and a magnetizable
atom is repulsive as in the non-retarded limit, but follows the same
$1/z^8$ power law as that between two polarizable atoms. Furthermore,
the interaction of atoms in excited energy eigenstates
\cite{0527,0526,0493,0099,0098,0333,0522} and the influence of
external conditions such as finite temperature
\cite{0104,0103,0376,0105,0101}, applied electromagnetic fields
\cite{0105}, or additional bodies \cite{0107,0367,0309,0679} on the
atom--atom interaction have been studied. In particular, when the
interatomic separation exceeds the thermal wavelength, the force
decreases more slowly ($\sim 1/z^7$) than in the zero-temperature
limit. Similarly, the Casimir--Polder result for the atom--plate
interaction has been confirmed
\cite{0047,0055,0054,0056,0051,0095,0321}, atoms that carry permanent
electric dipole moments \cite{0327} or are magnetizable \cite{0293}
have been considered and the influence of finite temperature
\cite{0376,0034} as well as force fluctuations \cite{0072} has been
studied. In close analogy to the atom--atom interaction, it was found
that the interaction between a magnetizable atom and a perfectly
conducting plate is repulsive and that the force decreases more
slowly ($\sim 1/z^4$) than in the zero-temperature limit as soon as
the atom--plate separation exceeds the thermal wavelength. In contrast
to the atom--atom interaction, the atom--plate potential for an atom
in an excited energy eigenstate was found to show an oscillatory
behavior in the retarded limit \cite{0320,0367,0061,0062}, thereby
making the effect of the transverse electromagnetic field more
explicit. In addition, atoms interacting with bodies of different
shapes and materials have been considered, such as: Perfectly
conducting planar \cite{0327,0292,0278,0301,0063,0315} and
parabolic cavities \cite{0390,0389}; metal \cite{0296}, electric
\cite{0053,0031,0308,0331} and magneto-electric half spaces
\cite{0330}; electric planar \cite{0032,0033} and spherical cavities
\cite{0441}.\vspace*{-1.3ex}

Needless to say that the pioneering work of Casimir and Polder on
dispersion forces has also stimulated further studies of the problem
of body--body interactions (for reviews see Refs.~\cite{0378,0136}).
Apart from confirming and interpreting the original results
\cite{0068,0131,0746}, normal-mode techniques have been employed to
include effects that arise from finite temperatures \cite{0632},
surface roughness \cite{0747,0748,0749}, the  presence of a dielectric
medium between the plates \cite{0668} and even virtual
electron-positron pairs \cite{0602,0611} (where the latter were found
to be negligibly small). As in the case of atom--body interactions,
various other geometries and materials have been considered such as:
Two electric \cite{0666,0652,0667,0197}, dielectric \cite{0688},
locally \cite{0626,0625,0645} and non-locally responding metal plates
\cite{0682}; two plates that are polarizable and magnetizable
\cite{0122,0123,0125,0659,0126,0129,0127}; the faces of a perfectly
conducting rectangular cavity \cite{0629,0622}; two electric
multilayer stacks \cite{0658}; a perfectly conducting plate and
cylinder \cite{0750}; two electric spheres \cite{0377}; a
perfectly conducting plate and a small electric sphere \cite{0071};
a sphere and a surrounding spherical cavity \cite{0643}. The results
qualitatively resemble the findings for the atom--atom and atom--body
interactions. In particular, retardation was found to lead to a
stronger asymptotic decrease of the forces which is softened due to
thermal effects as soon as the separations exceed the thermal
wavelengths; and the force between a polarizable object and a
magnetizable one was found to be repulsive. Perhaps a more surprising
result is the fact that two birefringent plates may exert a
non-vanishing dispersion torque on each other \cite{0639,0684}.
Moreover, the problem of Casimir energies of single
bodies\footnote{The Casimir energy of a single body can be defined as
the geometry-dependent part of the total electromagnetic energy where
the notion geometry-dependent part is subject to ambiguities, cf.~the
discussion below. For further reading on the Casimir energy of a
single body, cf. Ref.~\cite{0136}.} has been addressed, motivated by a
conjecture made by Casimir \cite{0692}, according to which an
attractive Casimir energy of an electron (modeled as a small perfectly
conducting sphere) should be able to counterbalance the repulsive
self-energy of the electron charge and thus explain its
stability.\footnote{For a further a discussion of this idea, cf.
Ref.~\cite{0487}.} However, the energy of a perfectly conducting
sphere was found to be repulsive \cite{0519,0648}, with similar
findings for a weakly dielectric sphere \cite{0525,0627,0624,0615}. On
the contrary, the Casimir energy of a weakly dielectric cylinder was
found to be attractive, in agreement with expectations
\cite{0627,0623}. The physical significance of Casimir energies of
single objects is yet unclear; in particular it was shown by
pairwise summation over microscopic dispersion energies that
dispersion energies of macroscopic objects are in fact dominated by
the always attractive volume and surface energies and may hence
never be observed \cite{0204}. In standard calculations of Casimir
energies, these volume and surface energies are either not
considered from the very beginning or discarded during
regularization procedures \cite{0615}.\vspace*{-1ex}

Normal-mode techniques have proved to be a powerful tool for studying
dispersion forces (cf.~also Refs.~\cite{0487,0636}). Nevertheless,
some principal limitations of the approach have become apparent
recently, in particular in view of the new challenges in connection
with recent improvements on the experimental side. So, normal-mode
calculations can become extremely cumbersome when applied to object
geometries relevant to practice or when a realistic description of
the electromagnetic properties of the interacting objects is required.
The limitations are also illustrated by the controversy regarding the
low-temperature behavior of dispersion forces on bodies (for a recent
account of the debate, see Ref.~\cite{0681} and references therein).
The answer to this question requires detailed knowledge of the
complicated interplay of positional, thermal and spectral factors. To
see this, one has to bear in mind that, in general, a large range of
frequencies contributes to the forces where the relative influence
of different frequency intervals is determined by the object--object
separation, temperature and the frequency dependence of the object
properties. As a consequence, approximations such as
long-/short-range, high-/low-temperature or perfect-reflectivity
limits become intrinsically intertwined. A typical material property
relevant to dispersion forces is the permittivity which is a complex
function of frequency, with the real and the imaginary part being
responsible for dispersion and absorption, respectively. In
particular, absorption which introduces additional noise into a
system, inhibits the application of normal-mode expansion on a
macroscopic level. This point was first taken into account by Lifshitz
in his calculation of the dispersion force between two electric half
spaces at finite temperature \cite{0057,0264} where he derived the
force from the average of the stress tensor of the fluctuating
electromagnetic field at the surfaces of the half spaces, with the
source of the field being the fluctuating noise current within the
dielectric matter. The required average was obtained by noting that
the current fluctuations are linked to the imaginary part of the
permittivity via the fluctuation--dissipation theorem. In this way,
Lifshitz could express the force per unit area in terms of the
permittivities $\varepsilon(\omega)$, $\varepsilon'(\omega)$ of the
two half spaces where in particular, in the non-retarded
(zero-temperature) limit the force per unit area, was obtained to
be\vspace*{-1ex}
\begin{equation}
\label{1.9}
\bar{F}=\frac{\hbar}{8\pi^2z^3}\int_0^\infty\dif\xi\,
 \frac{\varepsilon(\mi\xi)-1}{\varepsilon(\mi\xi)+1}\,
 \frac{\varepsilon'(\mi\xi)-1}{\varepsilon'(\mi\xi)+1}\,.
\end{equation}
The Lifshitz theory has been applied and extended by a number of
authors (for an overview, see Refs.~\cite{0378,0136}), who studied the
influence of different frequency ranges \cite{0644,0132,0676,0690},
effects of finite temperatures \cite{0646,0691,0689,0683} and surface
roughness \cite{0607}, and other planar structures such as
electrolytic half spaces separated by a dielectric \cite{0657},
magneto-electric half spaces \cite{0134}, metal plates of finite
thickness \cite{0612}, metal half spaces exhibiting non-local
properties \cite{0689,0680} and electric multilayer systems
\cite{0655} (for further reading, cf., e.g., Ref.~\cite{0322}). A
typical approximation for treating small deviations from planar
structures (like a sphere that is sufficiently close by a plate
\cite{0625}) is the proximity force approximation where it is assumed
that the interaction of two objects with gently curved surfaces can be
obtained by simply integrating the (Lifshitz) force per unit area
along the surfaces \cite{0601}.\footnote{Recently, validity limits for
the proximity force approximation in the case of perfectly conducting
objects have been discussed on the basis of numerical calculations
\cite{0695}.} While the debate regarding the temperature dependence of
the force between realistic metal half spaces still seems unsettled,
general consensus is reached that inclusion or neglect of material
absorption (i.e., use of a Drude-type or a plasma-type permittivity)
leads to the disagreeing results \cite{0681}. It is worth noting that
the forces in a planar structure can be reexpressed in terms of
(frequency-dependent) reflection coefficients directly accessible from
experiments. This formulation of the theory has been applied to
metal \cite{0628,0621,0742} and electric half spaces \cite{0678},
metal half spaces with non-local properties \cite{0604}, electric
multilayer stacks \cite{0741,0664} and, in some approximation, to
rough perfectly conducting \cite{0677,0674} and metal half spaces
\cite{0675,0672} where the surface roughness can give rise to a
tangential force component \cite{0673} and a torque \cite{0743}.

Lifshitz's idea of expressing dispersion forces in terms of response
functions is of course not restricted to planar systems, but can be
extended to arbitrary geometries. This can be achieved by expressing
the results obtained by normal-mode expansion in terms of the Green
tensor of the (classical) electromagnetic field \cite{0616,0528,0124}.
Alternatively, the Green tensor which contains all the necessary
information on the shape and the relevant electromagnetic properties
of the objects, can be introduced by applying path-integral techniques
\cite{0050,0069} or employing the fluctuation--dissipation theorem
\cite{0665}. The theory has been used to study the forces between two
perfectly conducting plates \cite{0616,0637,0744}, a perfectly
conducting plate and a perfectly permeable plate \cite{0124}, two
dielectric half spaces \cite{0688,0687}, two electric plates
\cite{0665} and, in some approximation, two perfectly conducting
spheres \cite{0124,0637}, a perfectly conducting sphere and a
perfectly conducting plate \cite{0637,0744}; the force on a
rectangular piston \cite{0745}; and the force and the torque between
weakly dielectric objects of arbitrary shapes \cite{0528}. In
particular, it was shown that the force between two mirror-symmetric
electric objects is always attractive \cite{0503}. Casimir energies of
a perfectly conducting sphere \cite{0616}, a dielectric sphere
\cite{0069}, a magneto-dielectric sphere \cite{0614} and a perfectly
conducting cylinder \cite{0617} have also been studied in this way.

The concept of linear-response theory has also been widely used to
study dispersion forces on atoms. In particular, it can be shown that
the (position-dependent part of the) interaction energy between a
ground-state atom and the (body-assisted) electromagnetic vacuum in
leading-order perturbation theory can be expressed in terms of the
linear response functions of the two systems,
\begin{equation}
\label{1.10}
U(\vect{r})=\frac{\hbar\mu_0}{2\pi}
 \int_{0}^{\infty}\dif\xi\,\xi^2 \alpha(\mi\xi)
 \,\trace\ten{G}{^{(1)}}(\vect{r},\vect{r},\mi\xi),
\end{equation}
i.e., the atomic polarizability $\alpha(\omega)$ on the one hand and
the scattering Green tensor
$\ten{G}{^{(1)}}(\vect{r},\vect{r},\omega)$ of the electromagnetic
field on the other, thus rendering a general expression for the force
on an atom in the presence of arbitrary bodies
\cite{0035,0039,0275,0041}\footnote{Note that the method is the
natural extension of the approach based on the electrostatic response
function which is now replaced by the response function for the
complete electromagnetic field including its transverse part.} (for an
alternative, semiclassical approach based on finding the eigenenergies
of the classical electromagnetic field interacting with a
harmonic-oscillator atom, see Ref.~\cite{0375}). The method which can
easily be extended to thermal fields \cite{0037,0046,0394}, also
applies to the dispersion interaction of two ground-state atoms in
free space \cite{0035,0037} or in the presence of bodies \cite{0036}
(cf. also Refs.~\cite{0375,0092}). However, it cannot directly be
applied to atoms in excited energy eigenstates where it is necessary
to again start from the leading-order interaction energy and only
express the field contribution in terms of the respective response
function \cite{0235,0042}. Linear response theory has been used to
study the dispersion interaction of a single ground-state atom with a
multitude of bodies such as: Perfectly conducting plates
\cite{0035,0039,0041,0037}; dielectric \cite{0077}, electric
\cite{0039,0041,0046,0653} and magneto-electric half spaces
\cite{0043}; metal half spaces exhibiting non-local properties
\cite{0275,0400} and/or surface roughness \cite{0275,0044} or being
covered by a thin overlayer \cite{0280}; perfectly conducting
\cite{0346} and dielectric spheres \cite{0069,0077,0349,0372};
dielectric cylinders \cite{0077,0395,0316,0442}; perfectly conducting
and electric planar cavities \cite{0314}; dielectric spherical
\cite{0313}, cylindrical \cite{0316} and perfectly conducting
wedge-shaped cavities \cite{0069,0372}. Moreover, the force
on an atom in an excited energy eigenstate in front of a perfectly
conducting \cite{0235,0042}, dielectric \cite{0042} and birefringent
dielectric half space \cite{0045} has been considered; and the
interaction of two ground-state atoms embedded in a non-locally
responding electrolyte \cite{0351}, placed near a perfectly conducting
\cite{0093} and a electric half space \cite{0653,0523} or inside a
perfectly conducting \cite{0092,0093} and dielectric planar cavity
\cite{0108} has been studied.

Finally, relations between microscopic and macroscopic dispersion
forces have been established whose validity is no longer restricted
to the non-retarded limit. Modeling macroscopic bodies as collections
of harmonic-oscillator atoms interacting with the electromagnetic
field and calculating the total energy of the interacting system, both
the the force on a single ground-state atom in the presence of a
dielectric half space and the force between two dielectric half spaces
were derived from microscopic atom--atom interactions \cite{0048}
where the former result was later extended beyond the
harmonic-oscillator model \cite{0087}. A harmonic-oscillator model of
atoms with the atoms being coupled to a heat bath, was used to derive
the force between absorbing dielectric half spaces, confirming the
result of Lifshitz theory \cite{0203}. The microscopic-model
calculations show that only in the limit of weakly polarizable bodies,
i.e., small values of the susceptibilities, a pairwise sum over
two-atom forces is sufficient to obtain the total force, stressing
once more the importance of many-atom interactions in the context of
body-assisted dispersion forces. Pairwise summation of two-atom
interactions can be used to obtain an approximate description of the
interaction between intricately shaped objects, e.g., bodies with
rough surfaces \cite{0638},\footnote{For bodies with small deviations
from the planar geometry, the result of pairwise summation can be
improved by introducing a correction factor obtained from Lifshitz
theory \cite{0606,0603}.} or of atom--body/body--body interactions
involving excited atoms and/or bodies \cite{0333,0522,0443}.
Conversely, from well-known formulas for the body--body interaction,
formulas for the atom--body interaction
\cite{0057,0264,0050,0069,0305,0388,0399,0392,0391,0670,0669} as
well as the atom--atom interaction
\cite{0101,0057,0264,0392,0669,0102} can be obtained in the limit of
the respective susceptibilities being asymptotically small.

As we have seen, various concepts have been developed to describe
dispersion forces---an overview over the different scenarios which
have been studied theoretically, is given in App.~\ref{app1} in
tabular form. These concepts, to some extent, are based on different
basic assumptions and hence impose different limitations upon the
applicability. The QED concepts based on normal-mode expansion of
the quantized electromagnetic field typically suffer from the fact
that when macroscopic bodies come into play, these bodies should be
regarded as non-absorbing and hence also non-dispersing. To overcome
this difficulty, arguments from other theories, such as the
fluctua\-tion--dissipation theorem of statistical physics, must be
borrowed. On the contrary, the concepts based on linear response
theory abandon an explicit field quantization and employ the
fluctuation--dissipation theorem from the beginning. However, the
applicability of methods that make use of the fluctuation--dissipation
theorem by some means or other is limited to equilibrium systems ---a
disadvantage when dynamical aspects of excited atoms are to be
considered. All concepts have in common that macroscopic bodies are
typically described in terms of macroscopic electrodynamics, i.e.,
boundary conditions at surfaces of discontinuity and/or constitutive
relations.

In this article we show that by following the formalism of
macroscopic QED in media (as developed, e.g., in
Refs.~\cite{0003,0002,0605}) from the very beginning, one can obtain
a unified approach to dispersion forces which does not only combine
the benefits of normal-mode QED and linear-response theory in a
natural way, but also accentuates the common origin of and intimate
relations between the different types of forces and extends the range
of application. In particular, the approach can be used to study
dispersion forces for a wide class of different scenarios, including
many of those listed in the above overview which have originally been
studied by means of a variety of different methods.

The further contents of the article are organized as follows. In
Sec.~\ref{sec2}, the main features of the quantization of the
electromagnetic field in linear, dispersing and absorbing media and
the interaction of the medium-assisted field with atoms is outlined,
with special emphasis on magneto-electric media described in terms of
spatially varying permittivities and permeabilities which are complex
functions of frequency. On this basis, in Sec.~\ref{sec3}, very
general formulas for the force on a macroscopic body due to its
interaction with other macroscopic bodies are presented which are
valid for arbitrarily shaped bodies as all the relevant properties
of the bodies are fully expressed in terms of the Green tensor of
the associated macroscopic Maxwell equations. Both Casimir stress and
Casimir force are introduced, and a very general relation to many-atom
van der Waals forces is established. In particular, it is shown that
both the force on a single ground-state atom interacting with a body
and the force between two ground-state atoms, can be obtained as
limiting cases of the general formulas. In Sec.~\ref{sec4}, forces on
individual atoms interacting with the body-assisted electromagnetic
field are studied in more detail, with special emphasis on explicitly
solving the corresponding quantum-mechanical interaction problem. It
is demonstrated how the force on one or two ground-state atoms in the
presence of magneto-electric bodies can be calculated by leading-order
perturbation theory, the results agreeing with those obtained in
Sec.~\ref{sec3}. A number of examples is studied where it is shown
that dispersion forces are often given by simple asymptotic power laws
in the retarded and non-retarded limits. The force on a single atom
initially prepared in an arbitrary excited quantum state is calculated
by solving the atom--field dynamics, leading to explicitly
time-dependent results. Some concluding are given in Sec.~\ref{sec5}.

%%%%%%%%%%%%%%%%%%%%%%%%%%%%%%%%%%%%%%%%%%%%%%%%%%%%%%%%%%%%%%%%%%%%%%

\section{Elements of QED in linearly responding media}
\label{sec2}

It is well known that the properties of the electromagnetic field in
media can significantly differ from those observed in free space,
and hence, the interaction of the field with atoms can strongly be
influenced by the presence of media. In classical electrodynamics,
linear media are commonly described in terms of phenomenologically
introduced macroscopic electric and magnetic susceptibilities (or
permittivities and permeabilities, respectively) available from
measurable data. This concept which can be transferred to quantum
electrodynamics, has the benefit of being universally valid because
it uses only very general physical properties, without the need for
specific microscopic matter models and involved ab initio
calculations.

%%%%%%%%%%%%%%%%%%%%%%%%%%%%%%%%%%%%%%%%%%%%%%%%%%%%%%%%%%%%%%%%%%%%%%

\subsection{The medium-assisted electromagnetic field}
\label{sec2.1}

The medium-assisted electromagnetic field in the absence of free
charges or currents obeys the macroscopic Maxwell equations which
in the Fourier domain read
\begin{align}
\label{2.1}
&\bm{\nabla}\sprod\underline{\hat{\vect{B}}}(\vect{r},\omega)=0,
 \\[.5ex]
\label{2.3}
&\bm{\nabla}\vprod\underline{\hat{\vect{E}}}(\vect{r},\omega)
 -\mi\omega\underline{\hat{\vect{B}}}(\vect{r},\omega)=\vect{0},
 \\[.5ex ]
\label{2.2b} &\varepsilon_0\bm{\nabla}\sprod
 \underline{\hat{\vect{E}}}(\vect{r},\omega)
 =\underline{\hat{\rho}}_\mathrm{in}(\vect{r},\omega),\\[.5ex]
\label{2.4b} &\kappa_0
 \bm{\nabla}\vprod\underline{\hat{\vect{B}}}(\vect{r},\omega)
 +\mi\omega\varepsilon_0
 \underline{\hat{\vect{E}}}(\vect{r},\omega)
 =\underline{\hat{\vect{j}}}_\mathrm{in}(\vect{r},\omega)
\end{align}
($\kappa_0$ $\!=$ $\!\mu_0^{-1}$) where the internal charge and
current densities of the magneto-electric media
$\underline{\hat{\rho}}_\mathrm{in}(\vect{r},\omega)$ and
$\underline{\hat{\vect{j}}}_\mathrm{in}(\vect{r},\omega)$ are the
sources for the electric field
$\underline{\hat{\vect{E}}}(\vect{r},\omega)$ and the induction
field $\hat{\underline{\vect{B}}}(\vect{r},\omega)$. Note that the
picture-independent Fourier components
$\hat{\underline{O}}(\vect{r},\omega)$ of an operator field
$\hat{O}(\vect{r})$ are defined according to
\begin{equation}
\label{2.0}
 \hat{O}(\vect{r})=\int_0^{\infty}\dif\omega\,
 \hat{\underline{O}}(\vect{r},\omega)+\mathrm{H.c.}
\end{equation}
so that $\underline{\hat{O}}(\vect{r},\omega,t)$ $\!=$
$\!\me^{-\mi\omega(t-t')} \underline{\hat{O}}(\vect{r},\omega,t')$
in the Heisenberg picture. Since the internal charge and current
densities are subject to the continuity equation
\begin{equation}
\label{2.0-1}
-\mi\omega\hat{\underline{\rho}}_\mathrm{in}(\vect{r},\omega)
 +\bm{\nabla}\sprod
 \hat{\underline{\vect{j}}}_\mathrm{in}(\vect{r},\omega)
 =0,
\end{equation}
they may be related to polarization and magnetization fields
$\hat{\underline{\vect{P}}}(\vect{r},\omega)$ and
$\hat{\underline{\vect{M}}}(\vect{r},\omega)$ as follows:
\begin{align}
\label{2.0-2} \hat{\underline{\rho}}_\mathrm{in}(\vect{r},\omega)
 =&-\bm{\nabla}\sprod
 \hat{\underline{\vect{P}}}(\vect{r},\omega),\\[.5ex]
\label{2.0-3}
 \hat{\underline{\vect{j}}}_\mathrm{in}(\vect{r},\omega)
 =&-\mi\omega\hat{\underline{\vect{P}}}(\vect{r},\omega)
 +\bm{\nabla}\vprod
 \hat{\underline{\vect{M}}}(\vect{r},\omega).
\end{align}
Upon introducing the displacement field
\begin{equation}
\label{2.5} \hat{\underline{\vect{D}}}(\vect{r},\omega) =
\varepsilon_0\hat{\underline{\vect{E}}}(\vect{r},\omega)
 +\hat{\underline{\vect{P}}}(\vect{r},\omega)
\end{equation}
and the magnetic field
\begin{equation}
\label{2.6} \hat{\underline{\vect{H}}}(\vect{r},\omega) =
\kappa_0\hat{\underline{\vect{B}}}(\vect{r},\omega)
 -\hat{\underline{\vect{M}}}(\vect{r},\omega),
\end{equation}
the inhomogeneous Maxwell equations (\ref{2.2b}) and (\ref{2.4b})
can hence be written in the familiar equivalent form
\begin{align}
\label{2.2}
&\bm{\nabla}\sprod\underline{\hat{\vect{D}}}(\vect{r},\omega)=0,
 \\[.5ex]
\label{2.4}
&\bm{\nabla}\vprod\underline{\hat{\vect{H}}}(\vect{r},\omega)
 +\mi\omega\underline{\hat{\vect{D}}}(\vect{r},\omega) = \vect{0}
\end{align}
where the source terms associated with the internal charge and
current densities are now contained in the displacement and magnetic
fields.

In particular in the case of linearly and locally responding
magneto-electric media, Eqs.~(\ref{2.5}) and (\ref{2.6}) take the form
\begin{align}
\label{2.7} &\hat{\underline{\vect{P}}}(\vect{r},\omega)
=\varepsilon_0[\varepsilon(\vect{r},\omega)-1]
 \hat{\underline{\vect{E}}}(\vect{r},\omega)
 +\hat{\underline{\vect{P}}}_\mathrm{N}(\vect{r},\omega),\\[.5ex]
\label{2.8} &\hat{\underline{\vect{M}}}(\vect{r},\omega)
=\kappa_0[1-\kappa(\vect{r},\omega)]
 \hat{\underline{\vect{B}}}(\vect{r},\omega)
 +\hat{\underline{\vect{M}}}_\mathrm{N}(\vect{r},\omega)
\end{align}
[$\kappa(\vect{r},\omega)$ $\!=$ $\!\mu^{-1}(\vect{r},\omega)$] with
$\varepsilon(\vect{r},\omega)$ and $\mu(\vect{r},\omega)$ being the
(relative) electric permittivity and magnetic permeability of the
media, respectively. Causality implies that
$\varepsilon(\vect{r},\omega)$ and $\mu(\vect{r},\omega)$ which
vary with space in general, are complex-valued functions of
frequency with the Kramers--Kronig relations being satisfied
\cite{0001}.\footnote{Note that both metals and dielectrics can be
described in terms of their permittivity with the main difference
being that the permittivity of a dielectric is analytic in the whole
upper half of the complex frequency plane whereas that of a metal
is commonly assumed to exhibit a simple pole at $\omega$ $\!=$
$\!0$ \cite{0001}.} According to the fluctuation--dissipation theorem,
$\hat{\underline{\vect{P}}}_\mathrm{N}(\vect{r},\omega)$ and
$\hat{\underline{\vect{M}}}_\mathrm{N}(\vect{r},\omega)$ are the
(linear) noise polarization and magnetization, respectively,
associated with the (linear) absorption described by the imaginary
parts of $\varepsilon(\vect{r},\omega)$
\mbox{[$\mathrm{Im}\,\varepsilon(\vect{r}, \omega)$ $\!>$ $\!0$]} and
$\mu(\vect{r},\omega)$ \mbox{[$\mathrm{Im}\,\mu(\vect{r},\omega)$
$\!>$ $\!0$]}. For simplicity, in Eqs.~(\ref{2.7}) and (\ref{2.8})
the material is assumed to be isotropic.\footnote{The theory can be
extended to arbitrary media, by starting from the general linear
response relation between the current density and the electric
field. Formulas in this article which do not explicitly refer to
material properties (but solely via the Green tensor) are valid for
arbitrary linear media \cite{1019}.}

Substituting Eqs.~(\ref{2.5}), (\ref{2.6}), (\ref{2.7}) and
(\ref{2.8}) into Eq.~(\ref{2.4}) and making use of Eq.~(\ref{2.3}),
one can verify that the electric field obeys the inhomogeneous
Helmholtz equation
\begin{equation}
\label{2.9}
\biggl[\bm{\nabla}\vprod\kappa(\vect{r},\omega)\bm{\nabla}\vprod
 \,-\,\frac{\omega^2}{c^2}\varepsilon(\vect{r},\omega)\biggr]
 \underline{\hat{\vect{E}}}(\vect{r},\omega)
 =\mi\omega\mu_0
 \underline{\hat{\vect{j}}}_\mathrm{N}(\vect{r},\omega)
\end{equation}
where the source term is determined by the noise current density
\begin{equation}
\label{2.10} \hat{\underline{\vect{j}}}_\mathrm{N}(\vect{r},\omega)
 =-\mi\omega\hat{\underline{\vect{P}}}_\mathrm{N}(\vect{r},\omega)
 +\bm{\nabla}\vprod
 \hat{\underline{\vect{M}}}_\mathrm{N}(\vect{r},\omega).
\end{equation}
Note that noise current density and noise charge density
\begin{equation}
\label{2.11} \hat{\underline{\rho}}_\mathrm{N}(\vect{r},\omega)
 =-\bm{\nabla}\sprod
 \hat{\underline{\vect{P}}}_\mathrm{N}(\vect{r},\omega)
\end{equation}
fulfill the continuity equation
\begin{equation}
\label{2.12}
-\mi\omega\hat{\underline{\rho}}_\mathrm{N}(\vect{r},\omega)
 +\bm{\nabla}\sprod
 \hat{\underline{\vect{j}}}_\mathrm{N}(\vect{r},\omega)
 =0
\end{equation}
[recall Eqs.~(\ref{2.0-1})--(\ref{2.0-3})]. The solution to
Eq.~(\ref{2.9}) can be given in the form
\begin{equation}
\label{2.15} \hat{\underline{\vect{E}}}(\vect{r},\omega)
 =\mi\omega\mu_0 \int \dif^3 r'\,
 \ten{G}(\vect{r},\vect{r}',\omega)
 \sprod\hat{\underline{\vect{j}}}_\mathrm{N}(\vect{r}',\omega)
\end{equation}
which, according to Eq.~(\ref{2.3}), implies that
\begin{equation}
\label{2.15-1}
\hat{\underline{\vect{B}}}(\vect{r},\omega)
 =\mu_0\int\dif^3 r'\,\bm{\nabla}\vprod
 \ten{G}(\vect{r},\vect{r}',\omega)
 \sprod\hat{\underline{\vect{j}}}_\mathrm{N}(\vect{r}',\omega).
\end{equation}
Here, $\ten{G}(\vect{r},\vect{r}',\omega)$ is the (classical) Green
tensor which is defined by the equation
\begin{equation}
\label{2.13}
\biggl[\bm{\nabla}\vprod\kappa(\vect{r},\omega)\bm{\nabla}\vprod
 \,-\,\frac{\omega^2}{c^2}\,\varepsilon(\vect{r},\omega)\biggr]
 \ten{G}(\vect{r},\vect{r}',\omega)=\delta(\vect{r}-\vect{r}')\ten{I}
\end{equation}
($\ten{I}$: unit tensor) together with the boundary condition
\begin{equation}
\label{2.14} \ten{G}(\vect{r},\vect{r}',\omega)\to \ten{0}
\quad\mbox{for
}\quad |\vect{r}-\vect{r}'|\to \infty.
\end{equation}
It should be pointed out that the Green tensor is uniquely defined
by Eqs.~(\ref{2.13}) and (\ref{2.14}) provided that the strict
inequalities $\mathrm{Im}\,\varepsilon(\vect{r},\omega)$ $\!>$ $\!0$
and \mbox{$\mathrm{Im}\,\mu(\vect{r},\omega)$ $\!>$ $\!0$} hold. Note
that it is an analytic function of $\omega$ in the upper complex half
plane and and has the following useful properties
($\cten{A}{}_{ij}^\trans$
$\!=$ $\!\cten{A}_{ji}$):
\begin{gather}
\label{2.17}
\ten{G}^{\ast}(\vect{r},\vect{r}',\omega)
 =\ten{G}(\vect{r},\vect{r}',-\omega^{\ast}),
\\[.5ex]
\label{2.18}
\ten{G}(\vect{r},\vect{r}',\omega)
 =\ten{G}^\trans(\vect{r}',\vect{r},\omega),
\end{gather}
\begin{multline}
\label{2.19} \int\!\dif^3 s\,\Bigl\{
 -\mathrm{Im}\,\kappa(\vect{s},\omega)
 \bigl[{\bm{\nabla}}_{\!\vect{s}}\vprod
 \ten{G}(\vect{s},\vect{r},\omega)\bigr]^\trans\sprod
 \bigl[{\bm{\nabla}}_{\!\vect{s}}\vprod
 \ten{G}^\ast(\vect{s},\vect{r}',\omega)\bigr]
\\[.5ex]
 +\frac{\omega^2}{c^2}\,
 \mathrm{Im}\,\varepsilon(\vect{s},\omega)
 \,\ten{G}(\vect{r},\vect{s},\omega)
 \sprod\ten{G}^\ast(\vect{s},\vect{r}',\omega)\Bigr\}
 =\mathrm{Im}\ten{G}(\vect{r},\vect{r}',\omega).
\end{multline}

Noise polarization and magnetization and hence noise current density,
can be related to dynamical variables
$\hat{\vect{f}}_\lambda(\vect{r},\omega)$ and
$\hat{\vect{f}}_\lambda^\dagger(\vect{r},\omega)$ \mbox{($\lambda$
$\!\in$ $\!\{{e},{m}\}$)} of the system which consists of the
electromagnetic field and the magneto-electric matter, including the
dissipative system responsible for absorption, as follows
\cite{0003,0002}:
\begin{align}
\label{2.22} &\hat{\underline{\vect{P}}}_\mathrm{N}(\vect{r},\omega)
=\mi\sqrt{\frac{\hbar\varepsilon_0}{\pi}
 \mathrm{Im}\,\varepsilon(\vect{r},\omega)}
 \,\hat{\vect{f}}_{e}(\vect{r},\omega),\\[.5ex]
\label{2.23} &\hat{\underline{\vect{M}}}_\mathrm{N}(\vect{r},\omega)
=\sqrt{-\frac{\hbar\kappa_0}{\pi}
 \mathrm{Im}\,\kappa(\vect{r},\omega)}
 \,\hat{\vect{f}}_{m}(\vect{r},\omega)
 =\sqrt{\frac{\hbar}{\pi\mu_0}\,
 \frac{\mathrm{Im}\,\mu(\vect{r},\omega)}
 {|\mu(\vect{r},\omega)|^2}}
 \,\hat{\vect{f}}_{m}(\vect{r},\omega)
\end{align}
with the $\hat{\vect{f}}_\lambda(\vect{r},\omega)$ and
$\hat{\vect{f}}_\lambda^\dagger(\vect{r},\omega)$ being attributed to
the collective Bosonic excitations of the system,
\begin{align}
\label{2.20} &\left[\hat{f}_{\lambda i}(\vect{r},\omega),
 \hat{f}_{\lambda'i'}^\dagger(\vect{r}',\omega')\right]
= \delta_{\lambda\lambda'}
 \delta_{ii'}\delta(\vect{r}-\vect{r}')
 \delta(\omega-\omega'),\\[.5ex]
\label{2.21} &\left[\hat{f}_{\lambda i}(\vect{r},\omega),
 \hat{f}_{\lambda'i'}(\vect{r}',\omega')\right] = 0.
\end{align}
By substituting Eqs.~(\ref{2.22}) and (\ref{2.23}) into
Eq.~(\ref{2.15}), on recalling Eq.~(\ref{2.10}), we may express the
medium-assisted electric field in terms of the dynamical variables
$\hat{\vect{f}}_\lambda(\vect{r},\omega)$ and
$\hat{\vect{f}}_\lambda^\dagger(\vect{r},\omega)$ to obtain
\begin{equation}
\label{2.24}
 \underline{\hat{\vect{E}}}(\vect{r},\omega)
 =\sum_{\lambda={e},{m}}
 \int\dif^3r'\,
 \ten{G}_\lambda(\vect{r},\vect{r}',\omega)
 \sprod\hat{\vect{f}}_\lambda(\vect{r}',\omega)
\end{equation}
where
\begin{align}
\label{2.25} &\ten{G}_{e}(\vect{r},\vect{r}',\omega)
 =\mi\,\frac{\omega^2}{c^2}
 \sqrt{\frac{\hbar}{\pi\varepsilon_0}\,
 \mathrm{Im}\,\varepsilon(\vect{r}',\omega)}\;
 \ten{G}(\vect{r},\vect{r}',\omega),\\[.5ex]
\label{2.30} &\ten{G}_{m}(\vect{r},\vect{r}',\omega)
 =\mi\,\frac{\omega}{c}
 \sqrt{-\frac{\hbar}{\pi\varepsilon_0}\,
 \mathrm{Im}\,\kappa(\vect{r}',\omega)}\,
 \left[\bm{\nabla}'
 \!\vprod\ten{G}(\vect{r}',\vect{r},\omega)
 \right]^\trans,
\end{align}
so that, according to Eq.~(\ref{2.0}),
\begin{align}
\label{2.24-1}
 \hat{\vect{E}}(\vect{r})
&=\int_0^\infty\dif\omega\,
 \underline{\hat{\vect{E}}}(\vect{r},\omega)
 +\mathrm{H.c.}\nonumber\\[.5ex]
&=\sum_{\lambda={e},{m}}\int\dif^3r'
 \int_0^\infty\dif\omega\,
 \ten{G}_\lambda(\vect{r},\vect{r}',\omega)
 \sprod\hat{\vect{f}}_\lambda(\vect{r}',\omega) +\mathrm{H.c.}.
\end{align}
Note that the relation (\ref{2.19}) can be written in the more
compact form
\begin{equation}
\label{2.30b} \sum_{\lambda={e},{m}}\int\dif^3s\,
 \ten{G}_\lambda(\vect{r},\vect{s},\omega)\sprod
 \ten{G}^{\ast\trans}_\lambda(\vect{r}',\vect{s},\omega)
 =\frac{\hbar\mu_0}{\pi}\,\omega^2\mathrm{Im}
 \ten{G}(\vect{r},\vect{r}',\omega).
\end{equation}
By starting from Eq.~(\ref{2.24}) and making use of the Maxwell
equations in the Fourier domain, Eqs.~(\ref{2.3}) and (\ref{2.4})
together with Eqs.~(\ref{2.5}), (\ref{2.6}), (\ref{2.7}), (\ref{2.8}),
(\ref{2.22}) and (\ref{2.23}), the other electromagnetic-field
quantities such as $\hat{\vect{B}}(\vect{r})$,
$\hat{\vect{D}}(\vect{r})$ and $\hat{\vect{H}}(\vect{r})$, can be
expressed in terms of the dynamical variables
$\hat{\vect{f}}_\lambda(\vect{r},\omega)$ and
$\hat{\vect{f}}_\lambda^\dagger(\vect{r},\omega)$ in a
straightforward way. In particular, one derives
\begin{equation}
\label{2.31}
 \underline{\hat{\vect{B}}}(\vect{r},\omega)
 =\frac{1}{\mi\omega}\sum_{\lambda={e},{m}}
 \int\dif^3r'\,
 \bm{\nabla}\vprod\ten{G}_\lambda(\vect{r},\vect{r}',\omega)
 \sprod\hat{\vect{f}}_\lambda(\vect{r}',\omega),
\end{equation}
and hence,
\begin{align}
\label{2.31-1}
 \hat{\vect{B}}(\vect{r})
&=\int_0^\infty\dif\omega\,
 \underline{\hat{\vect{B}}}(\vect{r},\omega)
 +\mathrm{H.c.}\nonumber\\[.5ex]
&=\sum_{\lambda={e},{m}}\int\dif^3r'
 \int_0^\infty\frac{\dif\omega}{\mi\omega}\,
 \bm{\nabla}\vprod\ten{G}_\lambda(\vect{r},\vect{r}',\omega)
 \sprod\hat{\vect{f}}_\lambda(\vect{r}',\omega)
 +\mathrm{H.c.}
\end{align}

In view of the treatment of the interaction of the medium-assisted
electromagnetic field with atoms, it may be expedient to express the
electric and induction fields in terms of potentials,
\begin{align}
\label{2.45} &\hat{\vect{E}}(\vect{r}) =
-\bm{\nabla}\hat{\varphi}(\vect{r}) -\dot{\hat{\vect{A}}}(\vect{r}),
 \\[.5ex]
\label{2.46} &\hat{\vect{B}}(\vect{r}) =
\bm{\nabla}\vprod\hat{\vect{A}}(\vect{r}).
\end{align}
In Coulomb gauge, $\bm{\nabla}\sprod\hat{\vect{A}}(\vect{r})$ $\!=$
$\!0$, the first and second terms on the r.h.s. of Eq.~(\ref{2.45})
are equal to the longitudinal ($\parallel$) and transverse ($\perp$)
parts of the electric field, respectively. {F}rom Eq.~(\ref{2.24-1})
it then follows that $\bm{\nabla}\hat{\varphi}(\vect{r})$ and
$\hat{\vect{A}}(\vect{r})$ can be expressed in terms of the
dynamical variables $\hat{\vect{f}}_\lambda(\vect{r},\omega)$ and
$\hat{\vect{f}}_\lambda^\dagger(\vect{r},\omega)$ as
\begin{align}
\label{2.49} &\bm{\nabla}
 \hat{\varphi}(\vect{r})
 =-\hat{\vect{E}}{}^\parallel(\vect{r})
 =-\sum_{\lambda={e},{m}}\int\dif^3r'
 \int_0^\infty \dif\omega\,
 {}^\parallel\ten{G}_\lambda(\vect{r},\vect{r}',\omega)
 \sprod\hat{\vect{f}}_\lambda(\vect{r}',\omega)
 + \mathrm{H.c.},
\\[.5ex]
\label{2.50} & \hat{\vect{A}}(\vect{r})
 =\sum_{\lambda={e},{m}}\int\dif^3r'
 \int_0^\infty\frac{\dif\omega}{\mi\omega}\,
 {}^\perp\ten{G}_\lambda(\vect{r},\vect{r}',\omega)
 \sprod\hat{\vect{f}}_\lambda(\vect{r}',\omega)
  + \mathrm{H.c.}
\end{align}
Note that the longitudinal (transverse) part of a vector field is
given by
\begin{equation}
\label{2.47} \vect{F}^{\parallel(\perp)}(\vect{r})
 =\int\dif^3 r'\,
 \bm{\delta}^{\parallel(\perp)}(\vect{r}-\vect{r}')
 \sprod\vect{F}(\vect{r}')
\end{equation}
where
\begin{equation}
\label{2.48} \bm{\delta}^\parallel(\vect{r})
  =-\bm{\nabla}\tprod\bm{\nabla}\left(\frac{1}{4\pi r}\right),
\quad\bm{\delta}^\perp(\vect{r})
 =\delta(\vect{r})\ten{I}-\bm{\delta}^\parallel(\vect{r})
\end{equation}
(and for a tensor field accordingly).

The commutation relations for the electromagnetic fields can be
deduced from the commutation relations for the dynamical variables
$\hat{\vect{f}}_\lambda(\vect{r},\omega)$ and
$\hat{\vect{f}}_\lambda^\dagger(\vect{r},\omega)$ as given by
Eqs.~(\ref{2.20}) and (\ref{2.21}). In particular, it can be shown
\cite{0003,0002} that the electric and induction fields obey the
well-known (equal-time) commutation relations
\begin{align}
\label{2.35}
&\left[\hat{E}_i(\vect{r}),\hat{B}_{i'}(\vect{r}')\right]
 =-\mi\hbar\varepsilon_0^{-1}\epsilon_{ii'k}\partial_k
 \delta(\vect{r}-\vect{r}'),\\[.5ex]
\label{2.34}
&\left[\hat{E}_i(\vect{r}),\hat{E}_{i'}(\vect{r}')\right]=0
 =\left[\hat{B}_i(\vect{r}),\hat{B}_{i'}(\vect{r}')\right].
\end{align}
Introducing the canonical momentum field associated with the
(transverse) vector potential,
\begin{equation}
\label{2.51b} \hat{\vect{\Pi}}(\vect{r})
 =-\varepsilon_0\hat{\vect{E}}{^\perp}(\vect{r}),
\end{equation}
one can also prove that the (equal-time) commutation relations
\begin{align}
\label{2.53} &
\left[\hat{A}_i(\vect{r}),\hat{\Pi}_{i'}(\vect{r}')\right]
 =\mi\hbar\delta^\perp_{ii'}(\vect{r}-\vect{r}'),
\\[.5ex]
\label{2.52} &
\left[\hat{A}_i(\vect{r}),\hat{A}_{i'}(\vect{r}')\right]=0
 =\left[\hat{\Pi}_i(\vect{r}),\hat{\Pi}_{i'}(\vect{r}')\right]
\end{align}
are fulfilled.

It is an almost trivial consequence of the quantization scheme that
upon choosing the Hamiltonian of the combined system to be
\begin{equation}
\label{2.39} \hat{H}_\mathrm{mf} =\sum_{\lambda={e},{m}}\int\dif^3r
\int_0^\infty
 \dif\omega\,\hbar\omega\,
 \hat{\vect{f}}_{\lambda}^\dagger(\vect{r},\omega)
 \sprod\hat{\vect{f}}_{\lambda}(\vect{r},\omega),
\end{equation}
the Heisenberg equation of motion
\begin{equation}
\label{2.40}
\dot{\hat{O}}=\frac{\mi}{\hbar}
 \bigl[\hat{H}_\mathrm{mf},\hat{O}\bigr]
\end{equation}
generates the correct Maxwell equations in the time domain,
\begin{align}
&\bm{\nabla}\vprod\hat{\vect{E}}(\vect{r})
 +\dot{\hat{\vect{B}}}(\vect{r})=\vect{0},\\[.5ex]
\label{2.44} &\bm{\nabla}\vprod\hat{\vect{H}}(\vect{r})
 -\dot{\hat{\vect{D}}}(\vect{r})=\vect{0}.
\end{align}
Note that the Maxwell equations
$\bm{\nabla}\sprod\hat{\vect{B}}(\vect{r})$ $\!=$ $\!0$ and
$\bm{\nabla}\sprod\hat{\vect{D}}(\vect{r})$ $\!=$ $\!0$ are
fulfilled by construction.

The Hilbert space can be spanned by Fock states obtained in the
usual way by repeated application of the creation operators
$\hat{\vect{f}}_\lambda^\dagger(\vect{r},\omega)$ on the ground
state $|\{0\}\rangle$ which is defined by
\begin{equation}
\label{2.36} \hat{\vect{f}}_\lambda(\vect{r},\omega)|\{0\}\rangle
 =\vect{0}\  \quad\forall\ \lambda,\vect{r},\omega.
\end{equation}
Note that the ground state refers to the combined system of the
electromagnetic field and the medium. In particular, from
Eq.~(\ref{2.24}) together with the commutation relations (\ref{2.20})
and (\ref{2.21}) one derives, on using the integral relation
(\ref{2.30b}),
\begin{equation}
\label{2.36-1} \langle\{0\}|\underline{\hat{E}}_i(\vect{r},\omega)
\underline{\hat{E}}^\dagger_{i'}(\vect{r}',\omega')|\{0\}\rangle =
\pi^{-1}\hbar\mu_0\omega^2 \mathrm{Im}\,
\cten{G}_{ii'}(\vect{r},\vect{r}',\omega)\delta(\omega-\omega'),
\end{equation}
which reveals that the ground-state fluctuations of the electric field
are determined by the imaginary part of the Green tensor---in full
agreement with the fluctuation--dissipation theorem.

It should be stressed that $\mathrm{Im}\,\varepsilon(\vect{r},\omega)$
$\!>$ $\!0$ and $\mathrm{Im}\,\mu(\vect{r},\omega)$ $\!>$ $\!0$
are assumed to hold everywhere. Even in almost empty regions or
regions where absorption is very small and can be neglected in
practice, the imaginary parts of the permittivity and permeability
must not be set equal to zero in the integrands of expressions of the
type (\ref{2.24-1}). To allow for empty-space regions, the limits
\mbox{$\mathrm{Im}\,\varepsilon(\vect{r},\omega)$ $\!\to$ $\!0$} and
$\mathrm{Im}\,\mu(\vect{r},\omega)$ $\!\to$ $\!0$ may be performed
\emph{a posteriori}, i.e., after taking the desired expectation values
and having carried out all spatial integrals. In this sense, the
theory provides the quantized electromagnetic field in the presence of
an arbitrary arrangement of linear, causal magneto-electric bodies
characterized by their permittivities and permeabilities where
$\mathrm{Im}\,\varepsilon(\vect{r},\omega)$ $\!\ge$ $\!0$ and
\mbox{$\mathrm{Im}\,\mu(\vect{r},\omega)$ $\!\ge$
$\!0$}.

As outlined above, quantization of the electromagnetic field in the
presence of dispersing and absorbing magneto-electric bodies can be
performed by starting from the macroscopic Maxwell equations including
noise terms associated with absorption, expressing the electromagnetic
field in terms of these noise terms and relating them to Bosonic
dynamical variables in an appropriate way (cf. also
Refs.~\cite{0716,0717,0718}). Alternatively, absorption can be
accounted for by expressing the electromagnetic field in terms of
auxiliary fields with the dynamics of the auxiliary fields being such
that the Maxwell equations are fulfilled \cite{0710}. It has been
shown that the two approaches are equivalent \cite{0711}. It is worth
noting that the quantization scheme is in full agreement with the
results of (quasi-)microscopic models of dielectric matter where the
polarization is modeled by harmonic-oscillator fields and damping is
accounted for by introducing a bath of additional harmonic oscillators
\cite{1000}. After a Fano diagonalization \cite{0252} of the total
Hamiltonian of the system (which consists of the electromagnetic
field, the polarization and the bath), an expression of the
form~(\ref{2.39}) is obtained. A model of this type was first
developed for homogeneous dielectrics \cite{1000} and later extended
to inhomogeneous dielectric bodies \cite{0713,1001,0714}, including
bodies exhibiting non-local properties \cite{0715}. In particular in
the latter case, the differential equation (\ref{2.13}) for the Green
tensor obviously changes to an integro-differential equation and
Eqs.~(\ref{2.22}) and (\ref{2.23}) must be modified accordingly.

%%%%%%%%%%%%%%%%%%%%%%%%%%%%%%%%%%%%%%%%%%%%%%%%%%%%%%%%%%%%%%%%%%%%%%

\subsection{Atom--field interaction}
\label{sec2.2}

Let us consider a system of nonrelativistic particles of masses
$m_\alpha$ and charges $q_\alpha$ which form an atomic system, e.g.,
an atom or a molecule, interacting with the medium-assisted
electromagnetic field. The Hamiltonian governing the dynamics of the
atomic system (briefly referred to as atom in the following) in the
absence of the medium-assisted electromagnetic field is commonly
given in the form
\begin{equation}
\label{2.55} \hat{H}_\mathrm{at}
 =\sum_{\alpha}\frac{\hat{\vect{p}}_{\alpha}^2}{2m_{\alpha}}
  +{\textstyle\frac{1}{2}}\int\dif^3r\,
 \hat{\rho}_\mathrm{at}(\vect{r})
 \hat{\varphi}_\mathrm{at}(\vect{r})
\end{equation}
where $\hat{\rho}_\mathrm{at}(\vect{r})$ and
$\hat{\varphi}_\mathrm{at}(\vect{r})$, respectively, are the charge
density and the scalar potential which are attributed to the atom,
\begin{align}
\label{2.56} &\hat{\rho}_\mathrm{at}(\vect{r})
 =\sum_{\alpha}q_{\alpha}\delta(\vect{r}-\hat{\vect{r}}_{\alpha}),
  \\[.5ex]
\label{2.57} &\hat{\varphi}_\mathrm{at}(\vect{r})
 =\int\dif^3r'
 \frac{\hat{\rho}_\mathrm{at}(\vect{r}')}
 {4\pi\varepsilon_0|\vect{r}-\vect{r}'|}
 =\sum_{\alpha}\frac{q_{\alpha}}
 {4\pi\varepsilon_0|\vect{r}-\hat{\vect{r}}_{\alpha}|}\,,
\end{align}
and the standard commutation relations
\begin{align}
\label{2.54} &\left[\hat{r}_{\alpha i},\hat{p}_{\alpha'i'}\right]
 =\mi\hbar\delta_{\alpha\alpha'}\delta_{ii'},
\\[.5ex]
\label{2.54-1} &\left[\hat{r}_{\alpha i},\hat{r}_{\alpha'i'}\right]
 =0=\left[\hat{p}_{\alpha i},\hat{p}_{\alpha'i'}\right]
\end{align}
hold. Obviously, $\hat{\varphi}_\mathrm{at}(\vect{r})$ and
$\hat{\rho}_\mathrm{at}(\vect{r})$ obey the Poisson equation
\begin{equation}
\label{2.57b}
\varepsilon_0\Delta\hat{\varphi}_\mathrm{at}(\vect{r})
 =-\hat{\rho}_\mathrm{at}(\vect{r}),
\end{equation}
and the continuity equation
\begin{equation}
\label{2.57-1} \dot{\hat{\rho}}_\mathrm{at}(\vect{r})
+\bm{\nabla}\sprod\hat{\vect{j}}_\mathrm{at}(\vect{r})=0
\end{equation}
is fulfilled where the atomic current density
$\hat{\vect{j}}_\mathrm{at}(\vect{r})$ reads
\begin{equation}
\label{2.63a} \hat{\vect{j}}_\mathrm{at}(\vect{r})
 ={\textstyle\frac{1}{2}}\sum_\alpha
 q_\alpha\left[\dot{\hat{\vect{r}}}_\alpha
 \delta(\vect{r}-\hat{\vect{r}}_\alpha)
 +\delta(\vect{r}-\hat{\vect{r}}_\alpha)
 \dot{\hat{\vect{r}}}_\alpha\right].
\end{equation}

It may be useful \cite{0005,0008} to introduce center-of-mass and
relative coordinates
\begin{equation}
\label{2.58} \hat{\vect{r}}_{A}=\sum_\alpha\frac{m_\alpha}{m_{A}}
 \,\hat{\vect{r}}_\alpha,\qquad
\hat{\overline{\vect{r}}}_\alpha=\hat{\vect{r}}_\alpha
 -\hat{\vect{r}}_{A}
\end{equation}
($m_{A}=\sum_\alpha m_\alpha$) with the associated momenta being
\begin{equation}
\label{2.59} \hat{\vect{p}}_{A}=\sum_\alpha\hat{\vect{p}}_\alpha,
\qquad \hat{\overline{\vect{p}}}_\alpha=\hat{\vect{p}}_\alpha
 -\frac{m_\alpha}{m_{A}}\,\hat{\vect{p}}_{A}.
\end{equation}
Combining Eqs.~(\ref{2.55}) and (\ref{2.59}), the atomic Hamiltonian
may be written in the form
\begin{align}
\label{2.63} \hat{H}_\mathrm{at}
 =&\;\frac{\hat{\vect{p}}_{A}^2}{2m_{A}}
 +\sum_{\alpha}
 \frac{\hat{\overline{\vect{p}}}{}_{\alpha}^2}{2m_{\alpha}}
 +{\textstyle\frac{1}{2}}\int\dif^3r\,
 \hat{\rho}_\mathrm{at}(\vect{r})
 \hat{\varphi}_\mathrm{at}(\vect{r})
 \nonumber\\[.5ex]
 =&\;\frac{\hat{\vect{p}}_{A}^2}{2m_{A}}
 +\sum_n E_n |n\rangle\langle n|
\end{align}
where $E_n$ and $|n\rangle$ are the eigenenergies and eigenstates of
the internal Hamiltonian. {F}rom the commutation relations
(\ref{2.54}) and (\ref{2.54-1}) it then follows that the
non-vanishing commutators of the new variables are
\begin{align}
\label{2.60} &\left[\hat{r}_{{A}i},\hat{p}_{{A}i'}\right]
 =\mi\hbar\delta_{ii'},\\[.5ex]
\label{2.61} & \left[\hat{\overline{r}}_{\alpha i},
 \hat{\overline{p}}_{\alpha'i'}\right]
 =\mi\hbar\delta_{ii'}
 \left(\delta_{\alpha\alpha'}
 -\frac{m_{\alpha'}}{m_{A}}\right).
\end{align}
In particular, when $m_{\alpha'}/m_{A}$ $\!\ll$ $\!1$, then
\begin{equation}
\label{2.61-1}
 \left[\hat{\overline{r}}_{\alpha i},
 \hat{\overline{p}}_{\alpha'i'}\right]
 \simeq \mi\hbar\delta_{ii'}\delta_{\alpha\alpha'}.
\end{equation}

Further atomic quantities that will be of interest are the atomic
polarization $\hat{\vect{P}}_\mathrm{at}(\vect{r})$ and
magnetization $\hat{\vect{M}}_\mathrm{at}(\vect{r})$ \cite{0006},
\begin{align}
\label{2.64}
&\hspace{-1ex}\hat{\vect{P}}_\mathrm{at}(\vect{r})
 =\sum_\alpha q_\alpha
 \hat{\overline{\vect{r}}}_\alpha\int _0^1\dif\sigma\,
 \delta\bigl(\vect{r}-\hat{\vect{r}}_{A}
  -\sigma\hat{\overline{\vect{r}}}_\alpha\bigr),\\[.5ex]
\label{2.65}
&\hspace{-1ex}\hat{\vect{M}}_\mathrm{at}(\vect{r})
 ={\textstyle\frac{1}{2}}\sum_\alpha
 q_\alpha\!\int _0^1\!\!\dif\sigma\,\sigma
 \left[\delta\bigl(\vect{r}\!-\!\hat{\vect{r}}_{A}
 \!-\!\sigma\hat{\overline{\vect{r}}}_\alpha\bigr)
 \hat{\overline{\vect{r}}}_\alpha\vprod
 \dot{\hat{\overline{\vect{r}}}}_\alpha
 -\dot{\hat{\overline{\vect{r}}}}_\alpha\vprod
 \hat{\overline{\vect{r}}}_\alpha
 \delta\bigl(\vect{r}\!-\!\hat{\vect{r}}_{A}
 \!-\!\sigma\hat{\overline{\vect{r}}}_\alpha\bigr)\right]\!;
\end{align}
it can be shown that for neutral atoms, the atomic charge and current
densities are related to the atomic polarization and magnetization
according to
\begin{equation}
\label{2.66c}
 \hat{\rho}_\mathrm{at}(\vect{r})
 =-\bm{\nabla}\sprod\hat{\vect{P}}_\mathrm{at}(\vect{r})
\end{equation}
and
\begin{equation}
\label{2.66d} \hat{\vect{j}}_\mathrm{at}(\vect{r})
 =\dot{\hat{\vect{P}}}_\mathrm{at}(\vect{r})
 +\bm{\nabla}\vprod\hat{\vect{M}}_\mathrm{at}(\vect{r})
 +\hat{\vect{j}}_\mathrm{r}(\vect{r})
\end{equation}
where
\begin{equation}
\label{2.66d2} \hat{\vect{j}}_\mathrm{r}(\vect{r})
 ={\textstyle\frac{1}{2}}\bm{\nabla}\vprod\left[
 \hat{\vect{P}}_\mathrm{at}(\vect{r})\vprod
 \dot{\hat{\vect{r}}}_{A}
 -\dot{\hat{\vect{r}}}_{A}
 \vprod\hat{\vect{P}}_\mathrm{at}(\vect{r})\right]
\end{equation}
which is due to the center-of-mass motion, is known as the
R\"{o}ntgen current density \cite{0007,0005}. Note that
Eqs.~(\ref{2.57b}) and (\ref{2.66c}) imply
\begin{equation}
\label{2.66e}
\varepsilon_0\bm{\nabla}\hat{\varphi}_\mathrm{at}(\vect{r})
 = \hat{\vect{P}}_\mathrm{at}^\parallel(\vect{r}).
\end{equation}

Expanding the delta functions in Eqs.~(\ref{2.64}) and (\ref{2.65})
in powers of the relative coordinates
$\hat{\overline{\vect{r}}}_\alpha$, we see that the leading-order
terms are the electric and magnetic dipole densities associated with
the atom,
\begin{align}
\label{2.64-1} &\hat{\vect{P}}_\mathrm{at}(\vect{r})
= \hat{\vect{d}}\delta(\vect{r}-\hat{\vect{r}}_{A}),\\[.5ex]
\label{2.65-1} &\hat{\vect{M}}_\mathrm{at}(\vect{r}) =
\hat{\vect{m}}\delta(\vect{r}-\hat{\vect{r}}_{A})
\end{align}
where the electric and magnetic atomic dipole moments read
\begin{align}
\label{2.66} &\hat{\vect{d}}
 =\sum_\alpha q_\alpha\hat{\overline{\vect{r}}}_\alpha
 =\sum_\alpha q_\alpha\hat{\vect{r}}_\alpha,\\[0.5ex]
\label{2.66b}
&\hat{\vect{m}}
 ={\textstyle\frac{1}{2}}\sum_\alpha q_\alpha
  \hat{\overline{\vect{r}}}_\alpha\vprod
  \dot{\hat{\overline{\vect{r}}}}_\alpha.
\end{align}
Note that the second equality in Eq.~(\ref{2.66}) only holds for
neutral atoms. Using the atomic Hamiltonian (\ref{2.63}) together
with the commutation relation (\ref{2.61}) and the definition
(\ref{2.59}), one can easily verify the useful relation
\begin{equation}
\label{2.66f} \sum_\alpha\frac{q_\alpha}{m_\alpha}\,
 \langle m|\hat{\overline{\vect{p}}}_\alpha|n\rangle
 =\mi\omega_{mn}\vect{d}_{mn}
\end{equation}
[$\omega_{mn}$ $\!=$ $\!(E_m$ $\!-$ $\!E_n)/\hbar$, $\vect{d}_{mn}$
$\!=$ $\!\langle m|\hat{\vect{d}}|n\rangle$] which in turn implies
the well-known sum rule
\begin{equation}
\label{2.66i} \frac{1}{2\hbar}\sum_m\omega_{mn}
 (\vect{d}_{nm}\tprod\vect{d}_{mn}
 +\vect{d}_{mn}\tprod\vect{d}_{nm})
 =\sum_\alpha\frac{q_\alpha^2}{2m_\alpha}\,\ten{I}.
\end{equation}

%%%%%%%%%%%%%%%%%%%%%%%%%%%%%%%%%%%%%%%%%%%%%%%%%%%%%%%%%%%%%%%%%%%%%%

\subsubsection{Minimal coupling}
\label{sec2.2.1}

Having established the Hamiltonians of the medium-assisted
field and the atom, we next consider the atom--field interaction.
According to the minimal coupling scheme (cf.,~e.g.,
Ref.~\cite{0007}), this may be done by making the replacement
$\hat{\vect{p}}_\alpha \mapsto \hat{\vect{p}}_\alpha$ $\!-$
$\!q_\alpha\hat{\vect{A}}(\hat{\vect{r}}_\alpha)$ in the atomic
Hamiltonian (\ref{2.55}), summing the Hamiltonians of the
medium-assisted field and the atom and adding the Coulomb
interaction of the atom with the medium-assisted field, leading to
\cite{0003,0008}
\begin{align}
\label{2.67} \hat{H}
=&\,\sum_{\lambda={e},{m}}\int\dif^3r
 \int_0^\infty\dif\omega\,
 \hbar\omega\,\hat{\vect{f}}_\lambda^{\dagger}(\vect{r},\omega)
 \sprod\hat{\vect{f}}_\lambda(\vect{r},\omega)
 +{\textstyle\frac{1}{2}}\sum_\alpha  m_\alpha^{-1}
 \left[\hat{\vect{p}}_\alpha
 -q_{\alpha}\hat{\vect{A}}(\hat{\vect{r}}_\alpha)\right]^2
 \nonumber\\[.5ex]
&\,+{\textstyle\frac{1}{2}}\int\dif^3r\,
 \hat{\rho}_\mathrm{at}(\vect{r})
 \hat{\varphi}_\mathrm{at}(\vect{r})
 +\int\dif^3r\,\hat{\rho}_\mathrm{at}(\vect{r})
 \hat{\varphi}(\vect{r})\nonumber\\[.5ex]
=&\;\hat{H}_\mathrm{mf}+\hat{H}_\mathrm{at}+\hat{H}_\mathrm{int}
\end{align}
where $\hat{\varphi}(\vect{r})$ and $\hat{\vect{A}}(\vect{r})$ must
be thought of as being expressed in terms of the dynamical variables
$\hat{\vect{f}}_\lambda(\vect{r},\omega)$ and
$\hat{\vect{f}}_\lambda^\dagger(\vect{r},\omega)$, according to
Eqs.~(\ref{2.49}) and (\ref{2.50}). Hence, the atom--field
interaction energy reads
\begin{equation}
\label{2.68} \hat{H}_\mathrm{int}=
 \sum_\alpha q_\alpha\hat{\varphi}(\hat{\vect{r}}_\alpha)
 -\sum_\alpha\frac{q_\alpha}{m_\alpha}\,
 \hat{\vect{p}}_\alpha\sprod
 \hat{\vect{A}}(\hat{\vect{r}}_\alpha)
 +\sum_\alpha\frac{q_\alpha^2}{2m_\alpha}\,
 \hat{\vect{A}}^2(\hat{\vect{r}}_\alpha).
\end{equation}
Note that the scalar product of $\hat{\vect{p}}_\alpha$ and
$\hat{\vect{A}}(\hat{\vect{r}}_\alpha)$ commutes in the
Coulomb gauge used.

The total electromagnetic field in the presence of the atom reads
\begin{alignat}{4}
\label{2.69} &
\hat{\bm{\mathcal{E}}}(\vect{r})
 =\hat{\vect{E}}(\vect{r})
 -\bm{\nabla}\hat{\varphi}_\mathrm{at}(\vect{r}),
&\qquad&\hat{\bm{\mathcal{B}}}(\vect{r})
 =\hat{\vect{B}}(\vect{r}),\\[.5ex]
\label{2.70} 
&\hat{\bm{\mathcal{D}}}(\vect{r})
=\hat{\vect{D}}(\vect{r})
 -\varepsilon_0\bm{\nabla}\hat{\varphi}_\mathrm{at}(\vect{r}),
&\qquad&\hat{\bm{\mathcal{H}}}(\vect{r})
 =\hat{\vect{H}}(\vect{r}).
\end{alignat}
Obviously, $\hat{\bm{\mathcal{B}}}(\vect{r})$ and
$\hat{\bm{\mathcal{D}}}(\vect{r})$ obey the Maxwell equations
\begin{align}
\label{2.71} 
&\bm{\nabla}\sprod\hat{\bm{\mathcal{B}}}(\vect{r})
=0,\\[.5ex]
\label{2.72} 
&\bm{\nabla}\sprod\hat{\bm{\mathcal{D}}}(\vect{r})
 =\hat{\rho}_\mathrm{at}(\vect{r}),
\end{align}
and it is a straightforward calculation \cite{0003,0008} to verify
that the Hamiltonian (\ref{2.67}) generates the remaining two Maxwell
equations
\begin{align}
\label{2.73}
&\bm{\nabla}\vprod\hat{\bm{\mathcal{E}}}(\vect{r})
 +\dot{\hat{\bm{\mathcal{B}}}}(\vect{r})
 =\vect{0},\\[.5ex]
\label{2.74}
&\bm{\nabla}\vprod\hat{\bm{\mathcal{H}}}(\vect{r})
 -\dot{\hat{\bm{\mathcal{D}}}}(\vect{r})
 =\hat{\vect{j}}_\mathrm{at}(\vect{r})
\end{align}
and the Newton equations of motion for the charged particles,
\begin{equation}
\label{2.76} m_\alpha\ddot{\hat{\vect{r}}}_\alpha =
q_\alpha\hat{\bm{\mathcal{E}}}(\vect{r}_\alpha)
 +{\textstyle\frac{1}{2}}q_\alpha
 \left[\dot{\hat{\vect{r}}}_\alpha
 \vprod\hat{\bm{\mathcal{B}}}(\vect{r}_\alpha)
 -\hat{\bm{\mathcal{B}}}(\vect{r}_\alpha)
 \vprod\dot{\hat{\vect{r}}}_\alpha\right]
\end{equation}
where
\begin{equation}
\label{2.75} \dot{\hat{\vect{r}}}_\alpha = m_\alpha^{-1}
 \left[\hat{\vect{p}}_\alpha
 -q_\alpha\hat{\vect{A}}(\hat{\vect{r}}_\alpha)\right]\!.
\end{equation}

In many cases of practical interest one may assume that the atom is
small compared to the wavelength of the relevant electromagnetic
field. It is hence useful to employ center-of-mass and relative
coordinates [Eqs.~(\ref{2.58}) and (\ref{2.59})] and apply the
long-wavelength approximation by performing a leading-order
expansion of the interaction Hamiltonian (\ref{2.68}) in terms of
the relative particle coordinates $\hat{\overline{\vect{r}}}_\alpha$.
Considering a neutral atom and recalling Eq.~(\ref{2.49}), one
finds
\begin{equation}
\label{2.77} \hat{H}_\mathrm{int}
 =-\hat{\vect{d}}\sprod\hat{\vect{E}}{}^\parallel(\hat{\vect{r}}_{A})
 -\sum_{\alpha}\frac{q_\alpha}{m_\alpha}\,
 \hat{\overline{\vect{p}}}_\alpha\sprod
 \hat{\vect{A}}(\hat{\vect{r}}_{A})
 +\sum_\alpha\frac{q_\alpha^2}{2m_\alpha}
 \,\hat{\vect{A}}^2(\hat{\vect{r}}_{A}).
\end{equation}
Note that the last term on the r.h.s. of Eq.~(\ref{2.77}) is
independent of the relative particle coordinates and hence does not
act on the internal state of the atom. When considering processes
caused by strong resonant transitions between different internal
states of the atom, it may therefore be neglected.

%%%%%%%%%%%%%%%%%%%%%%%%%%%%%%%%%%%%%%%%%%%%%%%%%%%%%%%%%%%%%%%%%%%%%%

\subsubsection{Multipolar coupling}
\label{sec2.2.2}

An equivalent description of the atom--field interaction that is
widely used is based on the multipolar-coupling
Hamiltonian.\footnote{For an extension of the formulas given below
to the case of more than one atoms, see Ref.~\cite{0009}.} For a
neutral atom, the transition from the minimal-coupling Hamiltonian
(\ref{2.67}) to the multipolar-coupling Hamiltonian is a canonical
transformation of the dynamical variables, corresponding to a unitary
transformation with the transformation operator being given by
\begin{equation}
\label{2.79} \hat{U}=\exp\left[\frac{\mi}{\hbar}\int\dif^3r\,
 \hat{\vect{P}}_\mathrm{at}(\vect{r})\sprod
 \hat{\vect{A}}(\vect{r})\right]
\end{equation}
where $\hat{\vect{A}}(\vect{r})$ and
$\hat{\vect{P}}_\mathrm{at}(\vect{r})$ are defined by
Eqs.~(\ref{2.50}) and (\ref{2.64}), respectively. This
transformation is commonly known as the Power--Zienau--Woolley
transformation \cite{0013,0014}; obviously it does not change
$\hat{\vect{r}}_\alpha$,
\begin{equation}
\label{2.79-1} \hat{\vect{r}}_\alpha' =
\hat{U}\hat{\vect{r}}_\alpha\hat{U}^\dagger = \hat{\vect{r}}_\alpha
\end{equation}
and a straightforward calculation yields \cite{0003,0008}
\begin{equation}
\label{2.79-2} \hat{\vect{p}}'_\alpha
=\hat{U}\hat{\vect{p}}_\alpha\hat{U}^\dagger = \hat{\vect{p}}_\alpha
 -q_\alpha\hat{\vect{A}}(\hat{\vect{r}}_\alpha)
 -\int\dif^3 r\,\hat{\bm{\Xi}}_\alpha(\vect{r})
 \vprod\hat{\vect{B}}(\vect{r})
\end{equation}
and
\begin{equation}
\label{2.81} \hat{\vect{f}}_\lambda'(\vect{r},\omega)
=\hat{U}\hat{\vect{f}}_\lambda(\vect{r},\omega)\hat{U}^\dagger
 =\hat{\vect{f}}_\lambda(\vect{r},\omega)
 +\frac{1}{\hbar\omega}\int\dif^3 r'\,
 \hat{\vect{P}}_\mathrm{at}^\perp(\vect{r}')
 \sprod\ten{G}_\lambda^\ast(\vect{r}',\vect{r},\omega)
\end{equation}
where
\begin{multline}
\label{2.86} \hat{\bm{\Xi}}_\alpha(\vect{r})
= q_\alpha\hat{\overline{\vect{r}}}_\alpha
 \int _0^1\dif\sigma\, \sigma
 \delta\bigl(\vect{r}-\hat{\vect{r}}_{A}
 -\sigma\hat{\overline{\vect{r}}}_\alpha\bigr)\\[.5ex]
-\frac{m_\alpha}{m_{A}}
 \sum_\beta q_\beta
 \hat{\overline{\vect{r}}}_\beta
 \int _0^1\dif\sigma\, \sigma
 \delta\bigl(\vect{r}-\hat{\vect{r}}_{A}
 -\sigma\hat{\overline{\vect{r}}}_\beta\bigr)
 +\frac{m_\alpha}{m_{A}}\,\hat{\vect{P}}_\mathrm{at}(\vect{r}).
\end{multline}

Now we may express the minimal-coupling Hamiltonian~(\ref{2.67}) in
terms of the transformed variables to obtain the multipolar-coupling
Hamiltonian in the form
\begin{multline}
\label{2.82} \hat{H}=\sum_{\lambda=e,m}\int\dif^3r
 \int_0^\infty\dif\omega\,\hbar\omega
 \hat{\vect{f}}_\lambda^{\prime\dagger}(\vect{r},\omega)
 \sprod\hat{\vect{f}}_\lambda'(\vect{r},\omega)
 +\frac{1}{2\varepsilon_0}\int\dif^3r\,
 \hat{\vect{P}}^{\prime 2}_\mathrm{at}(\vect{r})\\[.5ex]
 +\sum_\alpha\frac{1}{2 m_\alpha}
 \left[\hat{\vect{p}}'_\alpha
 +\int\dif^3 r\,\hat{\bm{\Xi}}'_\alpha(\vect{r})
 \vprod\hat{\vect{B}}'(\vect{r})\right]^2
 -\int\dif^3r\,\hat{\vect{P}}'_\mathrm{at}(\vect{r})
 \sprod\hat{\vect{E}}'(\vect{r}).
\end{multline}
Here, $\hat{\vect{E}}'(\vect{r})$ and $\hat{\vect{B}}'(\vect{r})$,
respectively, are given by Eqs.~(\ref{2.24-1}) and (\ref{2.31-1})
with $\hat{\vect{f}}_\lambda(\vect{r},\omega)$
[$\hat{\vect{f}}^\dagger_\lambda(\vect{r},\omega)$] being replaced
with $\hat{\vect{f}}'_\lambda(\vect{r},\omega)$
[$\hat{\vect{f}}^{\prime\dagger}_\lambda(\vect{r},\omega)$]. Note
that $\hat{\vect{r}}'_\alpha$ $\!=$ $\!\hat{\vect{r}}_\alpha$,
\mbox{$\hat{\overline{\vect{r}}}{}'_\alpha$ $\!=$
$\!\hat{\overline{\vect{r}}}_\alpha$},
$\hat{\overline{\vect{r}}}{}'_{A}$ $\!=$
$\!\hat{\overline{\vect{r}}}_{A}$,
$\hat{\vect{P}}'_\mathrm{at}(\vect{r})$ $\!=$
$\!\hat{\vect{P}}_\mathrm{at}(\vect{r})$,
$\hat{\vect{\bm{\Xi}}}'_\alpha(\vect{r})$ $\!=$
$\!\hat{\vect{\bm{\Xi}}}_\alpha(\vect{r})$,
$\hat{\vect{B}}'(\vect{r})$ $\!=$ $\!\hat{\vect{B}}(\vect{r})$, but
\begin{equation}
\label{2.93} \hat{\vect{E}}'(\vect{r})=\hat{\vect{E}}(\vect{r})
 +\varepsilon_0^{-1}
 \hat{\vect{P}}^{\perp}_\mathrm{at}(\vect{r})
\end{equation}
which means that the transformed (medium-assisted) electric field
$\hat{\vect{E}}'(\vect{r})$ has the physical meaning of a
displacement field, in contrast to $\hat{\vect{E}}(\vect{r})$ which
has the physical meaning of an electric field.

Hamiltonian (\ref{2.82}) can be decomposed into three parts,
\begin{equation}
\label{2.82-1}
\hat{H}=\hat{H}_\mathrm{mf'}+\hat{H}_\mathrm{at'}
 +\hat{H}_\mathrm{int'}
\end{equation}
where $\hat{H}_\mathrm{mf'}$ is given by Eq.~(\ref{2.39}) with the
primed variables in place of the unprimed ones,
\begin{equation}
\label{2.83} \hat{H}_\mathrm{mf'} =
 \sum_{\lambda={e},{m}}\int\dif^3r \int_0^\infty
 \dif\omega\,\hbar\omega\,
 \hat{\vect{f}}_{\lambda}^{\prime\dagger}(\vect{r},\omega)
 \sprod\hat{\vect{f}}'_{\lambda}(\vect{r},\omega),
\end{equation}
$\hat{H}_\mathrm{at'}$ is the atomic Hamiltonian,
\begin{align}
\label{2.84} \hat{H}_\mathrm{at'} &=
\frac{\hat{\vect{p}}_{A}^{\prime 2}}{2m_{A}} + \sum_{\alpha}
\frac{\hat{\overline{\vect{p}}}{}_{\alpha}^{\prime 2}}{2m_{\alpha}}
 +\frac{1}{2\varepsilon_0}\int\dif^3r\,
\hat{\vect{P}}^{\prime 2}_\mathrm{at}(\vect{r})
\nonumber\\[.5ex]
&= \frac{\hat{\vect{p}}_{A}^{\prime 2}}{2m_{A}}
 +\sum_n E'_n |n'\rangle\langle n'|
\end{align}
and $\hat{H}_\mathrm{int'}$ is the coupling term,
\begin{multline}
\label{2.85} \hat{H}_\mathrm{int'}=
 -\int\dif^3r\,\hat{\vect{P}}'_\mathrm{at}(\vect{r})
 \sprod\hat{\vect{E}}'(\vect{r})
 -\int\dif^3r\,\hat{\widetilde{\vect{M}}}{}'_\mathrm{at}(\vect{r})
 \sprod\hat{\vect{B}}'(\vect{r})
\\[.5ex]
 +\sum_\alpha\frac{1}{2 m_\alpha}
 \left[\int\dif^3 r\,\hat{\bm{\Xi}}'_\alpha(\vect{r})
 \vprod\hat{\vect{B}}'(\vect{r})\right]^2
 + \frac{1}{m_{A}}\int\dif^3 r\,
 \hat{\vect{p}}'_{A}\sprod
 \hat{\vect{P}}'_\mathrm{at}(\vect{r})
 \vprod\hat{\vect{B}}'(\vect{r})
\end{multline}
where
\begin{equation}
\label{2.85b} \hat{\widetilde{\vect{M}}}{}'_\mathrm{at}(\vect{r})
=\sum_\alpha \frac{q_\alpha}{2m_\alpha}\int_0^1\dif\sigma\,\sigma
 \left[\delta\bigl(\vect{r}\!-\!\hat{\vect{r}}'_{A}
 \!-\!\sigma\hat{\overline{\vect{r}}}{}'_\alpha\bigr)
 \hat{\overline{\vect{r}}}{}'_\alpha\vprod
 \hat{\overline{\vect{p}}}{}'_\alpha
 -\hat{\overline{\vect{p}}}{}'_\alpha\vprod
 \hat{\overline{\vect{r}}}{}'_\alpha
 \delta\bigl(\vect{r}\!-\!\hat{\vect{r}}'_{A}
 \!-\!\sigma\hat{\overline{\vect{r}}}{}'_\alpha\bigr)\right].
\end{equation}
Note that in contrast to the physical magnetization
$\hat{\vect{M}}_\mathrm{at}(\vect{r})$ [Eq.~(\ref{2.65})],
$\hat{\widetilde{\vect{M}}}_\mathrm{at}(\vect{r})$ is defined in
terms of the canonically conjugated momenta rather than the
velocities, as is required in a canonical formalism. The Hamiltonian
(\ref{2.82}) implies the relation
\begin{equation}
\label{2.92} m_\alpha\dot{\hat{\vect{r}}}'_\alpha
 =\hat{\vect{p}}'_\alpha
 +\int\dif^3r\,\hat{\bm{\Xi}}'_\alpha(\vect{r})
 \vprod\hat{\vect{B}}'(\vect{r})
\end{equation}
and it is not difficult to see [recall Eqs.~(\ref{2.75}) and
(\ref{2.79-2})] that $m_\alpha\dot{\hat{\vect{r}}}'_\alpha$ $\!=$
$\!m_\alpha\dot{\hat{\vect{r}}}_\alpha$. It should be pointed out
that the eigenenergies $E'_n$ of the internal Hamiltonian in
Eq.~(\ref{2.84}) may be different from the corresponding ones of the
internal Hamiltonian in Eq.~(\ref{2.63}), because of the additional
term contained in
\begin{equation}
\label{2.84-1} \frac{1}{2\varepsilon_0}\int\dif^3r\,
\hat{\vect{P}}^{\prime 2}_\mathrm{at}(\vect{r}) =
{\textstyle\frac{1}{2}}\int\dif{r}\,
\hat{\rho}'_\mathrm{at}(\vect{r})\varphi'_\mathrm{at}(\vect{r}) +
\frac{1}{2\varepsilon_0}\int\dif^3r
\left[\hat{\vect{P}}^{\prime\perp}_\mathrm{at}(\vect{r})\right]^2.
\end{equation}
Accordingly, the eigenstates of the two internal Hamiltonians are
not related to each other via the unitary transformation $\hat{U}$
[Eq.~(\ref{2.79})] in general.

One of the advantages of the multipolar coupling scheme is the fact
that it allows for a systematic expansion in terms of the electric
and magnetic multipole moments of the atom. In particular, in the
long-wavelength approximation, by retaining only the leading-order
terms in the relative coordinates
$\hat{\overline{\vect{r}}}{}'_\alpha$, the interaction energy
(\ref{2.85}) reads
\begin{align}
\label{2.88} \hat{H}_\mathrm{int'}=&\;
 -\hat{\vect{d}}'\sprod
 \hat{\vect{E}}'(\hat{\vect{r}}'_{A})
 -\hat{\widetilde{\vect{m}}}{}'\sprod
 \hat{\vect{B}}'(\hat{\vect{r}}'_{A})
 +\sum_\alpha\frac{q_\alpha^2}{8m_\alpha}
 \left[\hat{\bar{\vect{r}}}{}'_\alpha\vprod
 \hat{\vect{B}}'(\hat{\vect{r}}'_{A})\right]^2
 \nonumber\\[.5ex]
&\;+\frac{3}{8m_{A}}\left[\hat{\vect{d}}'\vprod
 \hat{\vect{B}}'(\hat{\vect{r}}_{A})\right]^2
 +\frac{1}{m_{A}}\,\hat{\vect{p}}'_{A}
 \sprod\hat{\vect{d}}'\vprod
 \hat{\vect{B}}'(\hat{\vect{r}}'_{A})
\end{align}
where
\begin{equation}
\label{2.89} \hat{\widetilde{\vect{m}}}{}'
 =\sum_\alpha\frac{q_\alpha}{2m_\alpha}\,
  \hat{\overline{\vect{r}}}{}'_\alpha\vprod
  \hat{\overline{\vect{p}}}{}'_\alpha.
\end{equation}
Note that, in contrast to $\hat{\vect{m}}$ [Eq.~(\ref{2.66b})],
$\hat{\widetilde{\vect{m}}}$ is defined in terms of the canonical
momenta. The first two terms on the r.h.s. of Eq.~(\ref{2.88})
represent electric and magnetic dipole interactions, respectively; the
next two terms describe the (generalized) diamagnetic interaction; and
the last term is the R\"{o}ntgen interaction due to the center-of-mass
motion. For non-magnetic atoms, Eq.~(\ref{2.89}) reduces to the
interaction Hamiltonian in electric-dipole approximation,
\begin{equation}
\label{2.90}
\hat{H}_\mathrm{int'}=
 -\hat{\vect{d}}'\sprod
 \hat{\vect{E}}'(\hat{\vect{r}}'_{A})
 +\frac{\hat{\vect{p}}'_{A}}{m_{A}}
 \sprod\hat{\vect{d}}'\vprod
 \hat{\vect{B}}'(\hat{\vect{r}}'_{A})
\end{equation}
which in cases where the influence of the center-of-mass motion on
the atom--field interaction does not need to be taken into account,
reduces to
\begin{equation}
\label{2.90-1}
\hat{H}_\mathrm{int'}= -\hat{\vect{d}}'\sprod
 \hat{\vect{E}}'(\hat{\vect{r}}'_{A}).
\end{equation}

%%%%%%%%%%%%%%%%%%%%%%%%%%%%%%%%%%%%%%%%%%%%%%%%%%%%%%%%%%%%%%%%%%%%%%

\section{Forces on bodies}
\label{sec3}

Electromagnetic forces are Lorentz forces. As known, the total Lorentz
force $\hat{\vect{F}}_\mathrm{L}$ acting on the matter contained in a
volume $V$ is given by
\begin{equation}
\label{3.4}
\hat{\vect{F}}_\mathrm{L}
 =\int_{V}\dif^3r\,\left[
 \hat{\rho}(\vect{r})\hat{\vect{E}}(\vect{r})
 +\hat{\vect{j}}(\vect{r})\vprod\hat{\vect{B}}(\vect{r})\right]\!.
\end{equation}
Here, the electromagnetic field acts on the the total charge and
current densities $\hat{\rho}(\vect{r})$ and
$\hat{\vect{j}}(\vect{r})$, respectively, which in general include the
internal charge and current densities
$\hat{\rho}_\mathrm{in}(\vect{r})$ and
$\hat{\vect{j}}_\mathrm{in}(\vect{r})$, respectively,
which are attributed to a medium [Eqs.~(\ref{2.0-2}) and
(\ref{2.0-3})] as well as those due to the presence of
additional sources, such as $\hat{\rho}_\mathrm{at}(\vect{r})$ and
$\hat{\vect{j}}_\mathrm{at}(\vect{r})$ [Eqs.~(\ref{2.56}) and
(\ref{2.63a})]. With the help of the Maxwell equations [as given by
Eqs.~(\ref{2.1})--(\ref{2.4b}) with $\hat{\rho}_\mathrm{in}(\vect{r})$
$\!\mapsto$ $\hat{\rho}(\vect{r})$,
\mbox{$\hat{\vect{j}}_\mathrm{in}(\vect{r})$ $\!\mapsto$
$\!\hat{\vect{j}}(\vect{r})$}] one easily finds
\begin{equation}
\label{3.2}
\hat{\rho}(\vect{r})\hat{\vect{E}}(\vect{r})
 +\hat{\vect{j}}(\vect{r})\vprod\hat{\vect{B}}(\vect{r})
=\bm{\nabla}\sprod\hat{\ten{T}}(\vect{r})
 -\varepsilon_{0}\frac{\partial}{\partial t}
 \left[\hat{\vect{E}}(\vect{r})\vprod\hat{\vect{B}}(\vect{r})\right]
\end{equation}
so that
\begin{equation}
\label{3.5}
\hat{\vect{F}}_\mathrm{L}
=\int_{\partial V}\dif\vect{a}\sprod\hat{\ten{T}}(\vect{r})
 -\varepsilon_0\,\frac{\dif}{\dif t}
 \int_V \dif^3r\,\hat{\vect{E}}(\vect{r})
 \vprod\hat{\vect{B}}(\vect{r})
\end{equation}
where the Maxwell stress tensor
\begin{equation}
\label{3.3}
\hat{\ten{T}}(\vect{r})
=\varepsilon_0\hat{\vect{E}}(\vect{r})
 \tprod\hat{\vect{E}}(\vect{r})
 +\mu_0^{-1}\hat{\vect{B}}(\vect{r})
 \tprod\hat{\vect{B}}(\vect{r})
 -\textstyle{\frac{1}{2}}\left[\varepsilon_0
 \hat{\vect{E}}^2(\vect{r})
 +\mu_0^{-1}\hat{\vect{B}}^2(\vect{r})\right]\ten{I}
\end{equation}
has been introduced. In particular, if the volume integral in the
second term on the r.h.s. of Eq.~(\ref{3.5}) does not depend on time,
then the total force reduces to the surface integral
\begin{equation}
\label{3.6}
\hat{\vect{F}}_\mathrm{L}
 =\int_{\partial V}\dif\hat{\vect{F}}_\mathrm{L}
\end{equation}
where
\begin{equation}
\label{3.7}
\dif\hat{\vect{F}}_\mathrm{L}
 =\dif\vect{a}\sprod \hat{\ten{T}}(\vect{r})
 =\hat{\ten{T}}(\vect{r})\sprod\dif\vect{a}
\end{equation}
may be regarded as the infinitesimal force element acting on an
infinitesimal surface element $\dif\vect{a}$. Note that a constant
term in the stress tensor does not contribute to the integral in
Eq.~(\ref{3.6}) and can therefore be omitted.

If the Minkowski stress tensor
\begin{multline}
\label{3.8}
\hat{\ten{T}}{}^\mathrm{(M)}(\vect{r})
=\hat{\vect{D}}(\vect{r})\tprod\hat{\vect{E}}(\vect{r})
 +\hat{\vect{H}}(\vect{r})\tprod\hat{\vect{B}}(\vect{r})
 -\textstyle{\frac{1}{2}}\left[
 \hat{\vect{D}}(\vect{r})\sprod\hat{\vect{E}}(\vect{r})
 +\hat{\vect{H}}(\vect{r})\sprod\hat{\vect{B}}(\vect{r})
 \right]\ten{I}
\\[.5ex]
=\hat{\ten{T}}(\vect{r})
 +\hat{\vect{P}}(\vect{r})\tprod\hat{\vect{E}}(\vect{r})
 -\hat{\vect{M}}(\vect{r})\tprod\hat{\vect{B}}(\vect{r})
 -\textstyle{\frac{1}{2}}\left[
 \hat{\vect{P}}(\vect{r})\sprod\hat{\vect{E}}(\vect{r})
 -\hat{\vect{M}}(\vect{r})\sprod\hat{\vect{B}}(\vect{r})\right]\ten{I}
\end{multline}
(which agrees with Abraham's stress tensor \cite{1003}) is used in
Eq.~(\ref{3.6}) [together with Eq.~(\ref{3.7})] instead of the Maxwell
stress tensor $\hat{\ten{T}}(\vect{r})$ to calculate the force, one
finds
\begin{multline}
\label{3.9}
\dif\hat{\vect{F}}^\mathrm{(M)}
=\dif\vect{a}\sprod\hat{\ten{T}}{}^\mathrm{(M)}(\vect{r})
 =\dif\hat{\vect{F}}_\mathrm{L}\\[.5ex]
+\dif\vect{a}\sprod\left\{
 \hat{\vect{P}}(\vect{r})\tprod\hat{\vect{E}}(\vect{r})
 -\hat{\vect{M}}(\vect{r})\tprod\hat{\vect{B}}(\vect{r})
 -\textstyle{\frac{1}{2}}\left[
 \hat{\vect{P}}(\vect{r})\sprod\hat{\vect{E}}(\vect{r})
 -\hat{\vect{M}}(\vect{r})\sprod\hat{\vect{B}}(\vect{r})\right]
 \ten{I}\right\}\!,
\end{multline}
and it is seen that in general
\begin{equation}
\label{3.10}
\dif\hat{\vect{F}}_\mathrm{L}
 \neq\dif\vect{a}\sprod\hat{\ten{T}}{}^\mathrm{(M)}(\vect{r}).
\end{equation}
That is to say, the use of the Minkowski stress tensor is expected not
to yield the Lorentz force, in general. Indeed, a careful analysis and
interpretation of classical electromagnetic force experiments
\cite{1010,1011,1007,1006,1008,1009,1005,1012} shows that the
(energy--momentum four-tensor associated with the) Lorentz force
passes the theoretical and experimental tests and qualifies for a
correct description of the energy--momentum properties of the
electromagnetic field in macroscopic electrodynamics \cite{1004} (also
see Secs.~\ref{sec3.1} and \ref{sec3.2}).

In classical electrodynamics, electrically neutral material bodies at
zero temperature which do not carry a permanent polarization and/or
magnetization are not subject to a Lorentz force in the absence of
external electromagnetic fields. As already noted in
Sec.~\ref{sec1.1}, the situation changes in quantum electrodynamics,
since the ground-state fluctuations of the body-assisted
electromagnetic field and the body's polarization/magnetization charge
and current densities can give rise to a non-vanishing ground-state
expectation value of the Lorentz force---the Casimir force
\cite{0198}
\begin{equation}
\label{3.10-2}
\vect{F}=\int_{V}\dif^3r\,\left\{\langle\{0\}|\left[
 \hat{\rho}(\vect{r})\hat{\vect{E}}(\vect{r}')
 +\hat{\vect{j}}(\vect{r})\vprod\hat{\vect{B}}(\vect{r}')\right]
 |\{0\}\rangle\right\}_{\vect{r'}\to\vect{r}}.
\end{equation}
Here, the coincidence limit $\vect{r'}\to\vect{r}$ must be performed
in such a way that unphysical (divergent) self-force contributions are
discarded after the vacuum expectation value has been calculated for
$\vect{r}'$ $\neq$ $\!\vect{r}$, an explicit prescription will
be given below Eq.~(\ref{3.19}).

To calculate the Casimir force, let us consider linear media that
locally respond to the electromagnetic field and can be characterized
by a spatially varying complex permittivity
$\varepsilon(\vect{r},\omega)$ and a spatially varying complex
permeability $\mu(\vect{r},\omega)$. Following Sec.~\ref{sec2.1}, we
may write the medium-assisted electric and induction fields in the
form of Eqs.~(\ref{2.24-1}) and (\ref{2.31-1}). Provided that the
volume of interest $V$ does not contain any additional charges or
currents, the charge and current densities that are subject to the
Lorentz force~(\ref{3.4}) are the internal ones,
$\hat{\rho}(\vect{r})$ $\!=$ $\!\hat{\rho}_\mathrm{in}(\vect{r})$ and
$\hat{\vect{j}}(\vect{r})$ $\!=$
$\!\hat{\vect{j}}_\mathrm{in}(\vect{r})$ [recall
Eqs.~(\ref{2.0-2}) and (\ref{2.0-3})]. Making use of Eqs.~(\ref{2.7})
and (\ref{2.8}) together with Eqs.~(\ref{2.11})--(\ref{2.13}), one can
easily see that
\begin{equation}
\label{3.15}
\hat{\underline{\rho}}(\vect{r},\omega)
 =-\varepsilon_{0}
 \bm{\nabla}\sprod\left\{[\varepsilon(\vect{r},\omega)-1]
 \hat{\underline{\vect{E}}}(\vect{r},\omega)\right\}
 +(\mi\omega)^{-1}\bm{\nabla}\sprod
 \hat{\underline{\vect{j}}}_\mathrm{N}(\vect{r},\omega)
\end{equation}
and
\begin{align}
\label{3.16}
\hat{\underline{\vect{j}}}(\vect{r},\omega)
=&\;-\mi\omega \varepsilon_{0}
 [\varepsilon(\vect{r},\omega)-1]
 \hat{\underline{\vect{E}}}(\vect{r},\omega)
 \nonumber\\[.5ex]
&\;+\,\bm{\nabla}\vprod\left\{\kappa_{0}
 [1-\kappa(\vect{r},\omega)]
 \hat{\underline{\vect{B}}}(\vect{r},\omega)\right\}
 +\hat{\underline{\vect{j}}}_\mathrm{N}(\vect{r},\omega).
\end{align}
Taking into account that $\hat{\underline{\vect{E}}}(\vect{r},\omega)$
and $\hat{\underline{\vect{B}}}(\vect{r},\omega)$ can be given in the
forms (\ref{2.15}) and (\ref{2.31}), respectively, and that the Green
tensor $\ten{G}(\vect{r},\vect{r'},\omega)$ obeys the differential
equation (\ref{2.13}), one may perform Eqs.~(\ref{3.15}) and
(\ref{3.16}) to obtain
\begin{align}
\label{3.17}
&\hat{\underline{\rho}}(\vect{r},\omega)
 =\frac{\mi\omega}{c^2}\int\dif^3r'\,
 \bm{\nabla}\sprod\ten{G}(\vect{r},\vect{r'},\omega)\sprod
 \hat{\underline{\vect{j}}}_\mathrm{N}(\vect{r}',\omega),\\[.5ex]
\label{3.18}
&\hat{\underline{\vect{j}}}(\vect{r},\omega)
 =\int \dif^3r'\biggl[\bm{\nabla}\vprod\bm{\nabla}\vprod\,
 -\frac{\omega^2}{c^2}\biggr]
 \ten{G}(\vect{r},\vect{r'},\omega)\sprod
 \hat{\underline{\vect{j}}}_\mathrm{N}(\vect{r}',\omega).
\end{align}
In this way, the fields $\hat{\underline{\rho}}(\vect{r},\omega)$,
$\hat{\underline{\vect{j}}}(\vect{r},\omega)$,
$\hat{\underline{\vect{E}}}(\vect{r},\omega)$ [Eq.~(\ref{2.15})]
and $\hat{\underline{\vect{B}}}(\vect{r},\omega)$ [Eq.~(\ref{2.15-1})]
are expressed in terms of the noise current density
$\hat{\underline{\vect{j}}}_\mathrm{N}(\vect{r},\omega)$. Making use
of Eq.~(\ref{2.10}) together with Eqs.~(\ref{2.22}) and (\ref{2.23})
and recalling the commutation relations (\ref{2.20}) and (\ref{2.21}),
one can easily calculate the ground-state correlation function
$\langle\{0\}|\underline{\hat{\vect{j}}}_\mathrm{N}(\vect{r},\omega)
\tprod
\underline{\hat{\vect{j}}}_\mathrm{N}^\dagger(\vect{r}',\omega')
|\{0\}\rangle$ which can then be used, on recalling
Eqs.~(\ref{2.15}), (\ref{2.15-1}), (\ref{3.17}) and (\ref{3.18}),
to calculate all the correlation functions relevant to the Casimir
force, as given by Eq.~(\ref{3.10-2}). The result is \cite{0663}
\begin{align}
\label{3.19}
\vect{F}
=&\;
 \frac{\hbar}{\pi}\int_{V}\dif^3r\int_{0}^\infty\dif\omega\,
 \biggl(\frac{\omega^2}{c^2}\bm{\nabla}\sprod
 \mathrm{Im}\ten{G}(\vect{r},\vect{r'},\omega)
 \nonumber\\[.5ex]
&\hspace{19ex}+\trace\biggl\{\ten{I}\vprod
 \biggl[\bm{\nabla}\vprod\bm{\nabla}\vprod\,
 -\frac{\omega^2}{c^2}\biggr]
 \mathrm{Im}\ten{G}(\vect{r},\vect{r'},\omega)\vprod
 \overleftarrow{\bm{\nabla}}'\biggr\}
 \biggr)_{\vect{r'}\to\vect{r}}\nonumber\\[.5ex]
=&\;-\frac{\hbar}{\pi}\int_{V}\dif^3r\int_{0}^\infty\dif\xi\,
 \biggl(\frac{\xi^2}{c^2}\bm{\nabla}\sprod
 \ten{G}(\vect{r},\vect{r'},\mi\xi)
 \nonumber\\[.5ex]
&\hspace{17ex}
-\trace\biggl\{\ten{I}\vprod
\biggl[\bm{\nabla}\vprod\bm{\nabla}\vprod\,
 +\frac{\xi^2}{c^2}\biggr]
 \ten{G}(\vect{r},\vect{r'},\mi\xi)\vprod
 \overleftarrow{\bm{\nabla}}'\biggr\}
 \biggr)
 _{\vect{r'}\to\vect{r}}
\end{align}
[$(\trace\ten{T})_j$ $\!=$ $\!\cten{T}_{ljl}$,
$\overleftarrow{\bm{\nabla}}$ introduces differentiation to the left]
where it is now apparent that in the coincidence limit
$\vect{r'}\to\vect{r}$ the Green tensor has to be replaced with its
scattering part at each space point. In particular, when the material
in the space region $V$ is homogeneous, then the Green tensor therein
can be globally decomposed into a bulk part
$\ten{G}^{(0)}(\vect{r},\vect{r'},\omega)$ and a scattering part
$\ten{G}^{(1)}(\vect{r},\vect{r'},\omega)$,
\begin{equation}
\label{3.22}
\ten{G}(\vect{r},\vect{r'},\omega)
 =\ten{G}^{(0)}(\vect{r},\vect{r'},\omega)
 +\ten{G}^{(1)}(\vect{r},\vect{r'},\omega)\qquad(\vect{r}\in V).
\end{equation}
In this case, the coincidence limit $\vect{r'}\to\vect{r}$ simply
means that the Green tensor $\ten{G}(\vect{r},\vect{r'},\omega)$ can
be globally replaced by its well-behaved scattering part
$\ten{G}^{(1)}(\vect{r},\vect{r'},\omega)$.

According to Eqs.~(\ref{3.3})--(\ref{3.7}), the Casimir force can be
equivalently rewritten as a surface integral over a stress tensor
\cite{0198},
\begin{equation}
\label{3.23}
\vect{F}=\int_{\partial V}\dif\vect{a}
 \sprod\ten{T}(\vect{r},\vect{r}')_{\vect{r'}\to\vect{r}}
\end{equation}
where
\begin{multline}
\label{3.24}
\ten{T}(\vect{r},\vect{r}')
=\langle\{0\}|\bigl\{\varepsilon_0
 \hat{\vect{E}}(\vect{r})\tprod\hat{\vect{E}}(\vect{r'})
 +\mu_0^{-1}\hat{\vect{B}}(\vect{r})\tprod\hat{\vect{B}}(\vect{r'})
 \\[.5ex]
-\textstyle{\frac{1}{2}}\bigl[\varepsilon_0
 \hat{\vect{E}}(\vect{r})\sprod\hat{\vect{E}}(\vect{r'})
 +\mu_0^{-1}\hat{\vect{B}}(\vect{r})\sprod\hat{\vect{B}}(\vect{r'})
 \bigr]\ten{I}\bigr\}|\{0\}\rangle
\end{multline}
which leads to
\begin{equation}
\label{3.25}
\ten{T}(\vect{r},\vect{r}')
 =\ten{S}(\vect{r},\vect{r}')
 -{\textstyle\frac{1}{2}}
 \bigl[\trace\ten{S}(\vect{r},\vect{r}')\bigr]\ten{I}
\end{equation}
with
\begin{align}
\label{3.26}
\ten{S}(\vect{r},\vect{r}')
=&\;\frac{\hbar}{\pi}\int_{0}^{\infty} \dif\omega
 \biggl[\frac{\omega^2}{c^2}\,
 \mathrm{Im}\ten{G}(\vect{r},\vect{r}',\omega)
 -\bm{\nabla}\vprod
 \mathrm{Im}\ten{G}(\vect{r},\vect{r}',\omega)
 \vprod\overleftarrow{\bm{\nabla}}'\biggr]\nonumber\\[.5ex]
=&\,-\frac{\hbar}{\pi}\int_{0}^{\infty} \dif\xi
 \biggl[\frac{\xi^2}{c^2}\,\ten{G}(\vect{r},\vect{r}',\mi\xi)
 +\bm{\nabla}\vprod\ten{G}(\vect{r},\vect{r}',\mi\xi)
 \vprod\overleftarrow{\bm{\nabla}}'\biggr].
\end{align}

Both Eq.~(\ref{3.19}) and Eq.~(\ref{3.23}) [together with
Eqs.~(\ref{3.25}) and (\ref{3.26})] are valid for arbitrary bodies
that linearly respond to the electromagnetic field, since the force is
fully determined by the Green tensor of the classical, macroscopic
Maxwell equations with the material properties entering the force
formulas only via the Green tensor. Moreover, Eqs.~(\ref{3.19}) and
(\ref{3.26}) reveal that the force is proportional to $\hbar$ and
hence represents a pure quantum effect. The results can be generalized
to finite temperatures $T$ in a straightforward way, by averaging in
Eqs.~(\ref{3.10-2}) and (\ref{3.24}) over the thermal state instead of
the vacuum state. As a consequence, the r.h.s. of the first equalities
of Eqs.~(\ref{3.19}) and (\ref{3.26}) are modified according to
\cite{0198}
\begin{equation}
\label{3.26-2}
\int_{0}^{\infty}\dif\omega\,\ldots\
 \mapsto\ \int_{0}^{\infty}\dif\omega\,
 \coth{\left(\frac{\hbar\omega}{2k_\mathrm{B}T}\right)}\ldots,
\end{equation}
($k_\mathrm{B}$, Boltzmann constant) so that the final forms of these
equations change as
\begin{equation}
\label{3.26-3}
\frac{\hbar}{\pi}\int_{0}^{\infty}\dif\xi\,f(\mi\xi)\
 \mapsto\ 2k_\mathrm{B}T\sum_{n=0}^\infty
 \bigl(1-{\textstyle\frac{1}{2}}\delta_{n0}\bigr)f(\mi\xi_n)
\end{equation}
with
\begin{equation}
\label{3.26-4}
\xi_n=\frac{2\pi k_\mathrm{B}T}{\hbar}\,n
\end{equation}
being the Matsubara frequencies.

%%%%%%%%%%%%%%%%%%%%%%%%%%%%%%%%%%%%%%%%%%%%%%%%%%%%%%%%%%%%%%%%%%%%%%

\subsection{Casimir stress in planar structures}
\label{sec3.1}

Let us apply the theory to a planar magneto-electric structure defined
according to
\begin{equation}
\label{3.27}
\varepsilon(\vect{r},\omega)
 =\begin{cases}
 \varepsilon_{-}(z,\omega)&\quad z<0,\\
 \varepsilon(\omega)&\quad 0<z<d,\\
 \varepsilon_{+}(z,\omega)&\quad z>d,
 \end{cases}
\end{equation}
\begin{equation}
\label{3.28}
\mu(\vect{r},\omega)
 =\begin{cases}
 \mu_{-}(z,\omega)&\quad z<0,\\
 \mu(\omega) & \quad 0<z<d,\\
 \mu_{+}(z,\omega)&\quad z>d
 \end{cases}
\end{equation}
and restrict our attention to the zero-temperature limit. To determine
the Casimir stress in the interspace $0$ $\!<$ $\!z$ $\!<$ $\!d$, we
need the scattering part of the Green tensor in Eq.~(\ref{3.26}) for
both spatial arguments within the interspace ($0\!<\!z\!=\!z'\!<\!d$).
Since the component $\vect{q}$ of the wave vector parallel to the
interfaces is conserved and the polarizations $\sigma$ $\!=$ $\!s,p$
decouple, the required Green tensor (as given in App.~\ref{appA}) can
be expressed in terms of reflection coefficients $r_{\sigma\pm}$ $\!=$
$\!r_{\sigma\pm}(\omega,q)$ ($q$ $\!=$ $\!|\vect{q}|$) referring to
reflection of waves at the right ($+$) and left ($-$) wall,
respectively, as seen from the interspace. Explicit (recurrence)
expressions for the reflection coefficients are available if the walls
are multi-slab magneto-electrics, cf. Eqs.~(\ref{A.7}) and
(\ref{A.8}).\footnote{For continuous wall profiles, Riccati-type
equations have to be solved \cite{0217}.} In the simplest case of two
homogeneous, semi-infinite half spaces, the coefficients
$r_{\sigma\pm}$ reduce to the well-known Fresnel amplitudes,
Eq.~(\ref{A.10}).

In order to determine the Casimir force, it is clear for symmetry
reasons that one requires the $z$ component
$\cten{T}_{\!zz}(\vect{r})$ $\!\equiv$
$\cten{T}_{\!zz}(\vect{r},\vect{r})_{\vect{r}'\to\vect{r}}$ of the
stress tensor in the interspace \mbox{$0$ $\!<$ $\!z$ $\!<$ $\!d$}
which, upon using the Green tensor from App.~\ref{appA}, can be given
in the form \cite{0198}
\begin{equation}
\label{3.29}
\cten{T}_{\!zz}(\vect{r})
=-\frac{\hbar}{8\pi^2}\int_{0}^{\infty}\dif\xi
 \int_{0}^{\infty}\dif q\,\frac{q}{b}\,\mu(\mi\xi)
 g(z,\mi\xi,q)
\end{equation}
where the function $g(z,\xi,q)$ is defined by
\begin{align}
\label{3.30}
g(z,\mi\xi,q)
=&\;-2\left[b^2 (1+n^{-2})+q^2(1-n^{-2})\right]
 \me ^{-2bd}\,r_{s+}r_{s-}D_s^{-1}\nonumber\\[.5ex]
&\;-2\left[b^2 (1+n^{-2})-q^2 (1-n^{-2})\right]
 \me^{-2bd}\,r_{p+}r_{p-}D_p^{-1}\nonumber\\[.5ex]
&\;+(b^2-q^2)(1-n^{-2})\left[\me^{-2b z}r_{s-}+
 \me^{-2b(d-z)}r_{s+}\right]D_s^{-1}\nonumber\\[.5ex]
&\;-(b^2-q^2)(1-n^{-2})\left[\me ^{-2bz}r_{p-}+
 \me^{-2b(d-z)}r_{p+}\right]D_p^{-1}
\end{align}
with
\begin{gather}
\label{3.31}
n=n(\mi\xi)
 =\sqrt{\varepsilon(\mi\xi)\mu(\mi\xi)}\,,\\[.5ex]
\label{3.32}
b=b(\mi\xi,q)
 =\sqrt{n^2(\mi\xi)\frac{\xi^2}{c^2}+q^2}\,,\\[.5ex]
\label{3.33}
D_\sigma=D_\sigma(\mi\xi,q)
 =1-r_{\sigma +} r_{\sigma -} \me^{-2bd}.
\end{gather}
According to Eq.~(\ref{3.23}), Eq.~(\ref{3.29}) [together with
Eqs.~(\ref{3.30})--(\ref{3.33})] gives the force per unit area between
two arbitrary planar multilayer stacks of (locally responding)
dispersing and absorbing magneto-electric material where the
interspace between them may contain an additional magneto-electric
medium. It is worth noting that many specific planar systems that can
be addressed by means of Eq.~(\ref{3.29}) have been studied by using
alternative methods: The most prominent example is the case of two
electric half spaces separated by vacuum, as first considered by
Lifshitz \cite{0057,0264} and later readdressed
\cite{0644,0132,0676,0690,0628,0678}, inter alia based on normal-mode
QED \cite{0652,0667,0197,0626}. Extended scenarios range from electric
half spaces separated by an electric medium (as studied by means of
electrostatic theory \cite{0650,0649,0640,0651}, normal-mode QED
\cite{0668} and an extended Lifshitz theory \cite{0657}) over
electric plates of finite thickness (as addressed on the basis of the
Lifshitz theory \cite{0612} and linear-response theory \cite{0665}),
electric multilayer stacks (as treated by generalizing the results of
the Lifshitz theory \cite{0655,0741,0664} as well as normal-mode QED
\cite{0658}) to magneto-electric half spaces (as studied by means of
Lifshitz theory \cite{0124} as well as normal-mode QED
\cite{0122,0123,0125,0659,0126,0134}). For purely electric multilayer
systems, various effects not being taken into account by
Eq.~(\ref{3.29}), have also been addressed in a number of works, such
as the influence on the force of finite temperature
\cite{0666,0688,0625,0645,0682,0125,0126,0057,0264,0646,0691,%
0689,0683,0621,0742,0616}, surface roughness
\cite{0747,0748,0749,0630,0631,0607,0677,0674,0675,0672,0673} and
non-locally responding materials \cite{0341,0682,0689,0680}. As
outlined above, finite temperature can be easily included in
Eq.~(\ref{3.29}) by applying Eqs.~(\ref{3.26-3}) and (\ref{3.26-4})
\cite{0198}, whereas surface roughness as well as non-local material
response can be taken into account by returning to the more general
formula (\ref{3.23}) [together with Eqs.~(\ref{3.25}) and
(\ref{3.26})] and specifying the Green tensor appropriately.

To further (numerically) evaluate Eq.~(\ref{3.29}), knowledge of the
$\xi$- and $q$-depen\-dence of the reflection coefficients which
depend on the respective planar system (see, e.g., the examples
studied in Refs.~\cite{0670,0669,0647,0660,0133}), is required. Let us
here restrict our attention to the limit of perfectly conducting
surfaces, i.e., $r_{p\pm}$ $\!=$ $\!-r_{s\pm}$ $\!=$ $\!1$ and assume
that the wall separation $d$ is sufficiently large, so that the
permittivity and the permeability of the medium in the interspace can
be replaced by their static values $\varepsilon$ $\!\equiv$
$\varepsilon(0)$ and $\mu$ $\!\equiv$ $\mu(0)$. It is then not
difficult to calculate the simplified integrals in Eq.~(\ref{3.29})
analytically to obtain the attractive Casimir force per unit area,
$\bar{F}$ $\!=$ $\!\cten{T}_{\!zz}(d)$ [recall Eq.~(\ref{3.23})] which
acts between two perfectly reflecting plates as\footnote{Note that
the terms proportional to $\me^{-2b(d-z)}$ in Eq.~(\ref{3.30}) give
rise to divergent integrals in Eq.~(\ref{3.29}) in the limit \mbox{$z$
$\!\to$ $\!d$} which obviously results from the assumptions of
frequency-independent response of the plates and the intervening
medium and infinite lateral extension of the plates. Since in a
realistic system these contributions are canceled by similar
contributions occurring at the backs of the plates \cite{0198}, they
can be discarded.}
\begin{equation}
\label{3.34}
\bar{F}
 =\frac{\pi^2\hbar c}{240}\,\sqrt{\frac{\mu}{\varepsilon}}
 \biggl(\frac{2}{3}+\frac{1}{3\varepsilon\mu}\biggr)
 \frac{1}{d^4}\,.
\end{equation}
In particular, when the interspace is empty
($\varepsilon$ $\!=$ $\!1$, $\mu$ $\!=$
$\!1$), then Eq.~(\ref{3.34}) reduces to
\begin{equation}
\label{3.35}
\bar{F}=\frac{\pi^2\hbar c}{240}\,\frac{1}{d^4}\,,
\end{equation}
in agreement with the famous result~(\ref{1.6}) first obtained by
Casimir on the basis of a normal-mode expansion \cite{0373} and
subsequently rederived \cite{0131,0632}, inter alia based on classical
orbits \cite{0124,0637} or dimensional arguments \cite{0068}.

In the same approximation, the force that acts on a perfectly
conducting plate in a planar cavity bounded by perfectly conducting
walls and filled with a magneto-electric medium obviously reads
\begin{equation}
\label{3.36}
\bar{F}=\frac{\pi^2\hbar c}{240}\,\sqrt{\frac{\mu}{\varepsilon}}\,
 \biggl(\frac{2}{3}+\frac{1}{3\varepsilon\mu}\biggr)
 \biggl(\frac{1}{d_\mathrm{r}^4}-\frac{1}{d_\mathrm{l}^4}\biggr)
\end{equation}
where $d_\mathrm{l}$ ($d_\mathrm{r}$) is the left (right) plate--wall
separation. On the contrary, use of the Minkowski stress tensor
(\ref{3.8}) leads, for $\mu$ $\!=$ $\!1$, to \cite{0647}
\begin{equation}
\label{3.37}
\bar{F}^\mathrm{(M)}
=\frac{\pi^2\hbar c }{240}\,\frac{1}{\sqrt{\varepsilon}}
\biggl(\frac{1}{d_\mathrm{r}^{4}}
-\frac{1}{d_\mathrm{l}^{4}}\biggr).
\end{equation}
Comparing the two results for $\mu$ $\!=$ $\!1$, we see that
$|\bar{F}|$ $\!\leq$ $|\bar{F}^\mathrm{(M)}|$. Introduction of a
(polarizable) medium into the interspace between the plate and the
cavity walls is obviously associated with some screening, thereby
reducing the force acting on the plate. Since the internal charges and
currents of the interspace medium are fully included only in the
Lorentz force [recall Eqs.~(\ref{3.4})--(\ref{3.7})], the force based
on the Minkowski stress tensor or an equivalent quantity
underestimates the screening effect and is hence larger than the
Lorentz force in general.

%%%%%%%%%%%%%%%%%%%%%%%%%%%%%%%%%%%%%%%%%%%%%%%%%%%%%%%%%%%%%%%%%%%%%%

\subsection{Macro- and micro-objects}
\label{sec3.2}

Let us return to the general formula (\ref{3.19}) and consider the
Casimir force acting on electric matter of susceptibility
$\chi(\vect{r},\omega)$ $\!=$ $\!\varepsilon(\vect{r},\omega)\!-\!1$
in some particular space region $V$ in the presence of arbitrary
linearly responding bodies (outside~$V$) in more detail. If
$\overline{\ten{G}}(\vect{r},\vect{r}',\omega)$ and
$\ten{G}(\vect{r},\vect{r}',\omega)$, respectively, denote the Green
tensors in the absence and presence of the electric matter in $V$
with both of them taking into account the bodies in the remaining
space, the differential equation (\ref{2.13}) for
$\ten{G}(\vect{r},\vect{r}',\omega)$ can be converted into the
Dyson-type integral equation
\begin{equation}
\label{3.38}
\ten{G}(\vect{r},\vect{r}',\omega)
=\overline{\ten{G}}(\vect{r},\vect{r}',\omega)
 +\frac{\omega^2}{c^2}\int_V \dif^3s\,\chi(\vect{s},\omega)
 \overline{\ten{G}}(\vect{r},\vect{s},\omega)\sprod
 \ten{G}(\vect{s},\vect{r}',\omega)
\end{equation}
where, for $\vect{r}$ $\!\in$ $\!V$, the Green tensor
$\overline{\ten{G}}(\vect{r},\vect{r}',\omega)$ satisfies the same
differential equation as the free-space Green tensor,
\begin{equation}
\label{3.39}
\left[\bm{\nabla}\vprod\bm{\nabla}\vprod\,
 -\frac{\omega^2}{c^2}\right]
 \overline{\ten{G}}(\vect{r},\vect{r}',\omega)
 =\delta(\vect{r}-\vect{r}')\ten{I},
\end{equation}
from which it follows that
\begin{equation}
\label{3.40}
\frac{\omega^2}{c^2}\,\bm{\nabla}\sprod
 \overline{\ten{G}}(\vect{r},\vect{r}',\omega)
=-\bm{\nabla}\delta(\vect{r}-\vect{r}')
\end{equation}
for $\vect{r}$ $\!\in$ $\!V$. Equations~(\ref{3.38}) and (\ref{3.40})
imply ($\vect{r}\in V$, $\omega$ real)
\begin{equation}
\label{3.41}
\bm{\nabla}\sprod\mathrm{Im}\ten{G}(\vect{r},\vect{r}',\omega)
 =-\bm{\nabla}\sprod\mathrm{Im}[
 \chi(\vect{r},\omega)\ten{G}(\vect{r},\vect{r}',\omega)].
\end{equation}
In a similar way, one finds that ($\vect{r}\in V$, $\omega$ real)
\begin{equation}
\label{3.43}
\left[\bm{\nabla}\vprod\bm{\nabla}\vprod\,
 -\frac{\omega^2}{c^2}\right]
 \mathrm{Im}\ten{G}(\vect{r},\vect{r'},\omega)\vprod
 \overleftarrow{\bm{\nabla}}'
=\frac{\omega^2}{c^2}\,\mathrm{Im}[
\chi(\vect{r},\omega)\ten{G}(\vect{r},\vect{r}',\omega)]
\vprod\overleftarrow{\bm{\nabla}}'.
\end{equation}
Substituting Eqs.~(\ref{3.41}) and (\ref{3.43}) into Eq.~(\ref{3.19})
one can then show that the Casimir force acting on an electric body of
volume $V_\mathrm{I}$ which is an inner part of a larger electric
body (occupying volume $V$) reads \cite{0663}
\begin{multline}
\label{3.44}
\vect{F}=-\frac{\hbar}{2\pi}\int_{0}^{\infty}\dif\xi\,
 \frac{\xi^2}{c^2}
 \Bigl\{\int_{V_\mathrm{I}}\dif^3r\,\chi(\vect{r},\mi\xi)\bm{\nabla}
 \trace[\ten{G}(\vect{r},\vect{r}',\mi\xi)]_{\vect{r'}\to\vect{r}}
 \\[.5ex]
-2\int_{\partial V_\mathrm{I}}\dif\vect{a}\sprod
 \chi(\vect{r},\mi\xi)[\ten{G}(\vect{r},\vect{r}',\mi\xi)
 ]_{\vect{r'}\to\vect{r}}\Bigr\}.
\end{multline}
In particular, in the case of an isolated body, i.e., when the region
$V\supset V_\mathrm{I}$ is empty apart from the electric matter
contained in $V_\mathrm{I}$, then the surface integral can be dropped,
hence
\begin{equation}
\label{3.45}
\vect{F}=-\frac{\hbar}{2\pi}\int_{V_\mathrm{I}}\dif^3r
 \int_{0}^{\infty}\dif\xi\,
 \frac{\xi^2}{c^2}\,\chi(\vect{r},\mi\xi)\bm{\nabla}
 \trace[\ten{G}(\vect{r},\vect{r}',\mi\xi)]_{\vect{r'}\to\vect{r}}.
\end{equation}

Let us briefly compare Eq.~(\ref{3.44}) with the equation obtained on
the basis of the Minkowski stress tensor,
\begin{equation}
\label{3.44-1}
\vect{F}^{\mathrm{(M)}}=\frac{\hbar}{2\pi}\int_{V_\mathrm{I}}\dif^3r
 \int_{0}^{\infty}\dif\xi\,\frac{\xi^2}{c^2}
 \left[\bm{\nabla}\chi(\vect{r},\mi\xi)\right]\,
 \trace[\ten{G}(\vect{r},\vect{r},\mi\xi)]_{\vect{r}'\to\vect{r}}.
\end{equation}
It differs from Eq.~(\ref{3.44}) by a surface integral, in general
\cite{0663}. Hence, the two force formulas agree in the case of an
isolated body where the surface integrals do not contribute to the
force and both equations reduce to Eq.~(\ref{3.45}). In contrast to
Eq.~(\ref{3.44}), application of Eq.~(\ref{3.44-1}) to any inner,
homogeneous part of a body leads to the paradoxical result that the
force identically vanishes, because of
$\bm{\nabla}\chi(\vect{r},\mi\xi)$ $\!=$ $\!0$. In other words, the
only atoms that are subject to a force are those at the surface of the
body. On the contrary, it is known that the van der Waals forces on
all atoms of a body contribute to the Casimir force
(Sec.~\ref{sec3.2.1}).

We have seen that within the framework of macroscopic QED, Casimir
forces on linearly responding bodies can be expressed in terms of the
respective Green tensor of the Maxwell equations for the body-assisted
electromagnetic field. Hence, the main problem to be solved in
practice is the determination of the Green tensors for the specific
systems of interest. Since closed formulas for Green tensors are only
available for highly symmetric systems (see, e.g., Ref.~\cite{0217}),
approximative and numerical methods are required. For example, one can
start from an appropriately chosen Green tensor as zeroth-order
approximation to the exact one and perform a Born expansion of the
exact Green tensor by iteratively solving the corresponding Dyson-type
equation. In particular, iteratively solving Eq.~(\ref{3.38}) yields
the Born series
\begin{multline}
\label{3.46}
\ten{G}(\vect{r},\vect{r}',\omega)
 =\overline{\ten{G}}(\vect{r},\vect{r}',\omega)
 +\sum_{K=1}^\infty\Bigl(\frac{\omega}{c}\Bigr)^{2K}
 \Biggl[\prod_{J=1}^K\int_V\dif^3s_J\,
 \chi(\vect{s}_J,\omega)\Biggr]\\[.5ex]
 \times\overline{\ten{G}}(\vect{r},\vect{s}_1,\omega)\sprod
 \overline{\ten{G}}(\vect{s}_1,\vect{s}_2,\omega)\cdots
 \overline{\ten{G}}(\vect{s}_K,\vect{r}',\omega).
\end{multline}

%%%%%%%%%%%%%%%%%%%%%%%%%%%%%%%%%%%%%%%%%%%%%%%%%%%%%%%%%%%%%%%%%%%%%%

\subsubsection{Weakly polarizable bodies, micro-objects and atoms}
\label{sec3.2.1}

The force formulas (\ref{3.44}) and (\ref{3.45}) which follow from
macroscopic QED without involved microscopic considerations, do not
only apply to electric macro-objects but also to micro-objects.
Moreover they also allow for studying the limiting case of individual
atoms and determining in this way even the dispersion forces with
which bodies act on atoms and atoms act on each other in the presence
of bodies. To see this, let us consider dielectric bodies which
may be typically thought of as consisting of distinguishable
(electrically neutral but polarizable) micro-constituents (again
briefly referred to as atoms), so that the Clausius--Mossotti relation
\cite{0001,1018}
\begin{align}
\label{3.47}
\chi(\vect{r},\omega)
&=\varepsilon_{0}^{-1}\eta(\vect{r})\alpha(\omega)
 [1-\eta(\vect{r})\alpha(\omega)/(3\varepsilon_{0})]^{-1}
 \nonumber\\[.5ex]
&=\varepsilon_{0}^{-1} \eta(\vect{r})\alpha(\omega)\,
 [1+\chi(\vect{r},\omega)/3]
\end{align}
may be assumed to be valid where $\alpha(\omega)$ is the atomic
polarizability and $\eta(\vect{r})$ their number
density.\footnote{Note that Eq.~(\ref{3.47}) is consistent with the
requirement that both $\alpha(\omega)$ and $\chi(\vect{r},\omega)$
be Fourier transforms of response functions iff
$\eta(\vect{r})\alpha(0)/(3\varepsilon_{0})$ $\!<$ $\!1$.} Let
$V_\mathrm{I}$ be the volume of an isolated dielectric body of
susceptibility $\chi(\vect{r},\omega)$. The Born series (\ref{3.46})
and the Clausius--Mossotti relation (\ref{3.47}) imply that when the
body is sufficiently small and/or weakly polarizable, then the force,
as given by Eq.~(\ref{3.45}), is essentially determined by the
leading-order term proportional to $\chi(\vect{r},\omega)$ $\!\simeq$
$\!\varepsilon_{0}^{-1}\eta(\vect{r})\alpha(\omega)$, so that
Eq.~(\ref{3.45}) approximates to\footnote{For a small and/or weakly
polarizable body that is an inner part of a larger body, see
Refs.~\cite{0392,0663,0661}.}
\begin{equation}
\label{3.48}
\vect{F}=-\frac{\hbar\mu_0}{2\pi}\int_{V_\mathrm{I}}\dif^3r\,
 \eta(\vect{r})
 \int_{0}^{\infty} \dif\xi\,\xi^2\alpha(\mi\xi)
 \bm{\nabla}\,\trace
 \overline{\ten{G}}{^{(1)}}(\vect{r},\vect{r},\mi\xi)
\end{equation}
where, according to the decomposition
\begin{equation}
\label{3.49}
\overline{\ten{G}}(\vect{r},\vect{r'},\omega)
 =\overline{\ten{G}}{^{(0)}}(\vect{r},\vect{r'},\omega)
 +\overline{\ten{G}}{^{(1)}}(\vect{r},\vect{r'},\omega)
 \quad(\vect{r}\in V)
\end{equation}
[cf.~Eq.~(\ref{3.22})],
$\overline{\ten{G}}{^{(1)}}(\vect{r},\vect{r}',\omega)$
is simply the scattering part of the Green tensor
$\overline{\ten{G}}(\vect{r},\vect{r}',\omega)$
of the system without the dielectric body under consideration.

It can be easily seen that Eq.~(\ref{3.48}) may be rewritten as
\cite{0663}
\begin{equation}
\label{3.50}
\vect{F}
 =\int_{V_\mathrm{I}}\dif^3r\,\eta(\vect{r})\vect{F}(\vect{r})
\end{equation}
where
\begin{equation}
\label{3.51}
\vect{F}(\vect{r})=-\bm{\nabla}U(\vect{r})
\end{equation}
with
\begin{equation}
\label{3.52}
U(\vect{r})=\frac{\hbar\mu_0}{2\pi}
 \int_{0}^{\infty}\dif\xi\,\xi^2 \alpha(\mi\xi)
 \,\trace\overline{\ten{G}}{^{(1)}}(\vect{r},\vect{r},\mi\xi)
\end{equation}
being nothing but the van der Waals potential of a single ground-state
atom of polarizability $\alpha(\omega)$ at position $\vect{r}$ in the
presence of arbitrary linearly responding bodies at zero temperature
(Sec.~\ref{sec4}). Note that in the limiting case when
$V_\mathrm{I}\to 0$ and $\eta\to\infty$ but $\eta V_\mathrm{I}$ $\!=$
$\!1$, such that $V_\mathrm{I}$ covers a single atom at position
$\vect{r}_{A}$, then $\vect{F}$ reduces to
 $\vect{F}(\vect{r}_{A})$---the force acting on a single atom. Note
that a relation of the kind~(\ref{3.50}) was already used by Lifshitz
to deduce the dispersion force between a single atom and an electric
half space from that between a dielectric and an electric half space
\cite{0057,0264}.

Equation (\ref{3.50}) reveals that the force acting on a weakly
polarizable dielectric body is the sum of the forces acting on all
body atoms due to their interaction with other bodies giving rise to
the Green tensor $\overline{\ten{G}}(\vect{r},\vect{r}',\omega)$. Let
us consider in more detail the interaction of a weakly polarizable
body with a second isolated dielectric body of volume $V'_\mathrm{I}$
which is also weakly polarizable [$\chi'(\vect{r},\omega)$ $\!\simeq$
$\!\varepsilon_{0}^{-1}\eta'(\vect{r})\alpha'(\omega)$]. Denoting the
Green tensor associated with all remaining bodies except for the two
under consideration by
$\widetilde{\ten{G}}(\vect{r},\vect{r}',\omega)$, expanding
$\overline{\ten{G}}(\vect{r},\vect{r}',\omega)$, by starting from
$\widetilde{\ten{G}}(\vect{r},\vect{r}',\omega)$ in the Born series
[i.e., using Eq.~(\ref{3.46}) \mbox{with $\ten{G}$ $\!\mapsto$
$\!\overline{\ten{G}}$}, $\overline{\ten{G}}$ $\!\mapsto$
$\!\widetilde{\ten{G}}$ and $V$ $\!\mapsto$ $\!V'_\mathrm{I}$] and
again omitting terms of higher than linear order in the
susceptibility, Eq.~(\ref{3.48}) leads to \cite{0663}
\begin{equation}
\label{3.54}
\vect{F}=\int_{V_\mathrm{I}}\dif^3r\,\eta(\vect{r})
 \int_{V'_\mathrm{I}}\dif^3r'\,\eta'(\vect{r})
 \vect{F}(\vect{r},\vect{r}')
\end{equation}
where
\begin{equation}
\label{3.55}
\vect{F}(\vect{r},\vect{r}')
= - \bm{\nabla}U(\vect{r},\vect{r}'),
\end{equation}
is the force with which an atom of polarizability $\alpha'(\omega)$
at position $\vect{r}'$ acts on an atom of polarizability
$\alpha(\omega)$ at position $\vect{r}$ with
\begin{equation}
\label{3.56}
U(\vect{r},\vect{r}')=-\frac{\hbar\mu_0^2}{2\pi}
 \int_{0}^{\infty}\dif\xi\,\xi^4\alpha(\mi\xi)\alpha'(\mi\xi)
 \trace\bigl[\widetilde{\ten{G}}(\vect{r},\vect{r}',\mi\xi)
 \sprod\widetilde{\ten{G}}(\vect{r}',\vect{r},\mi\xi)\bigr]
\end{equation}
being the two-atom van der Waals potential \cite{0009,0113}.
Equation~(\ref{3.54}) clearly shows that the Casimir force is a volume
force and not a surface force as could be suggested on the basis of
the Minkowski stress tensor. According to Eq.~(\ref{3.54}), the
Casimir force between weakly polarizable dielectric bodies is the sum
of all two-atom van der Waals forces between the body atoms---a
result which was already obtained by Lifshitz for the special case of
two dielectric half spaces \cite{0057,0264}. In fact, such a relation
formed the basis of early calculations of dispersion forces between
bodies \cite{0642,0641} and it is still used for treating bodies
exhibiting surface roughness \cite{0638} or containing excited media
\cite{0333,0522}.

The force between a polarizable atom and a magnetizable one can be
obtained in a similar way \cite{0491}. For this purpose, we consider
the interaction of polarizable atoms contained in the first, weakly
polarizable body (volume $V_\mathrm{I}$) with a second, weakly
magnetizable body of volume $V'_\mathrm{I}$ and magnetic
susceptibility $\zeta(\vect{r},\omega)$ $\!=$
$\!\mu(\vect{r},\omega)\!-\!1$ $\!=$
$\!\mu_0\eta'(\vect{r})\beta'(\omega)$ where $\beta'(\omega)$ denotes
the magnetizability of the atoms contained in $V'_\mathrm{I}$.
Again expanding the Green tensor
$\overline{\ten{G}}(\vect{r},\vect{r}',\omega)$ associated with all
the bodies except for the first one by starting from the Green tensor
$\widetilde{\ten{G}}(\vect{r},\vect{r}',\omega)$ (associated with all
the bodies except for the two under consideration) and retaining only
the linear order in $\zeta(\vect{r},\omega)$, one obtains
\begin{equation}
\label{3.67}
\overline{\ten{G}}(\vect{r},\vect{r}',\omega)
 =\widetilde{\ten{G}}(\vect{r},\vect{r}',\omega)
 -\int_{V'_\mathrm{I}}\dif^3s\,\zeta(\vect{s},\omega)
 \left[\widetilde{\ten{G}}(\vect{r},\vect{s},\omega)\vprod
 \overleftarrow{\bm{\nabla}}_{\!\vect{s}}\right]\sprod
 \bm{\nabla}_{\!\vect{s}}\vprod
 \widetilde{\ten{G}}(\vect{s},\vect{r}',\omega).
\end{equation}
Upon substitution of Eq.~(\ref{3.67}), the force on the first body
(\ref{3.48}) can again be written as a sum over two-atom forces,
Eqs.~(\ref{3.54}) and (\ref{3.55}) where the potential of a
polarizable atom interacting with a magnetizable one reads \cite{0491}
\begin{multline}
\label{3.68}
U(\vect{r},\vect{r}')=-\frac{\hbar\mu_0^2}{2\pi}
 \int_0^\infty\dif\xi\,\xi^2
 \alpha(\mi\xi)\beta'(\mi\xi)\\[.5ex]
 \times\trace\!\left\{\left[
 \widetilde{\ten{G}}(\vect{r},\vect{s},\mi\xi)
 \vprod\overleftarrow{\bm{\nabla}}_{\!\vect{r}}\right]
 \sprod\bm{\nabla}_{\!\vect{s}}\vprod
 \widetilde{\ten{G}} (\vect{s},\vect{r},\mi\xi)
 \right\}_{\vect{s}=\vect{r}'}.
\end{multline}

%%%%%%%%%%%%%%%%%%%%%%%%%%%%%%%%%%%%%%%%%%%%%%%%%%%%%%%%%%%%%%%%%%%%%%

\subsubsection{Many-atom van der Waals interactions}
\label{sec3.2.2}

In general, not only two-atom interactions but all many-atom
interactions must be taken into account to obtain exact dispersion
forces involving macroscopic bodies. To illustrate this point, let us
return to Eq.~(\ref{3.52}) and consider the interaction of a single
atom with a dielectric body (volume $V$), whose susceptibility
$\chi(\vect{r},\omega)$ is given by the Clausius--Mossotti relation
(\ref{3.47}). Recall from Sec.~\ref{sec3.2.1} that
$\overline{\ten{G}}(\vect{r},\vect{r}',\omega)$ denotes the Green
tensor of the whole arrangement of bodies and
$\widetilde{\ten{G}}(\vect{r},\vect{r}',\omega)$ is the Green tensor
of all (background) bodies except for the one under consideration.
Substituting the Born series (\ref{3.46}) [\mbox{$\ten{G}$ $\!\mapsto$
$\!\overline{\ten{G}}$}, $\!\overline{\ten{G}}$ $\!\mapsto$
$\!\widetilde{\ten{G}}$] into Eq.~(\ref{3.52}), one can write
the atom--body potential in the form
\begin{equation}
\label{3.59}
U(\vect{r})=\sum_{K=0}^\infty U_K(\vect{r})
\end{equation}
where
\begin{equation}
\label{3.60}
U_0(\vect{r})
 =\frac{\hbar\mu_0}{2\pi}\int_0^\infty\dif\xi\,\xi^2\alpha(\mi\xi)
 \trace\widetilde{\ten{G}}^{(1)}(\vect{r},\vect{r},\mi\xi)
\end{equation}
is the atomic potential due to the background bodies\footnote{Note
that $U_0(\vect{r})$ $\!=$ $\!0$ for free-space background, i.e., if
there are no further bodies.} and ($K$ $\!\ge$ $\!1$)
\begin{multline}
\label{3.61}
U_K(\vect{r})=\frac{(-1)^K\hbar\mu_0}{2\pi c^{2K}}
 \Biggl[\prod_{J=1}^K\int_V\dif^3s_J\,\chi(\vect{s}_J,\mi\xi)\Biggr]
 \int_0^{\infty}\dif \xi\,\xi^{2K+2}\alpha(\mi\xi)
\\[.5ex]
 \times\trace\bigl[
 \widetilde{\ten{G}}(\vect{r},\vect{s}_1,\mi\xi)\sprod
 \widetilde{\ten{G}}(\vect{s}_1,\vect{s}_2,\mi\xi)\cdots
 \widetilde{\ten{G}}(\vect{s}_K,\vect{r},\mi\xi)\bigr]
\end{multline}
is the contribution to the potential that is of $K$th order in the
susceptibility of the dielectric body. To further treat the sum on the
right-hand side of Eq.~(\ref{3.61}), the Green tensors therein
are decomposed into singular and regular parts according to
\begin{equation}
\label{3.62}
\widetilde{\ten{G}}(\vect{r},\vect{r}',\omega)
 =-\frac{1}{3}\Bigl(\frac{c}{\omega}\Bigr)^2
 \delta(\vect{r}-\vect{r}')\ten{I}
 +\widetilde{\ten{G}}{}'(\vect{r},\vect{r}',\omega)
\end{equation}
and use is made of the Clausius--Mossotti relation (\ref{3.47}). A
somewhat lengthy calculation then leads to the result that
\cite{0113,0020}
\begin{equation}
\label{3.63}
U(\vect{r})
 =U_0(\vect{r})+\sum_{K=1}^\infty\frac{1}{K!}
 \Biggl[\prod_{J=1}^K\int\dif^3s_J\,\eta(\vect{s}_J)\Biggr]
 U(\vect{r},\vect{s}_1,\ldots,\vect{s}_K)
\end{equation}
where
\begin{multline}
\label{3.65}
U(\vect{r}_1,\vect{r}_2,\ldots,\vect{r}_N)
 =\frac{(-1)^{N-1}\hbar\mu_0^N}{(1+\delta_{2N})\pi}
 \int_0^\infty\dif\xi\,\xi^{2N}
 \alpha_1(\mi\xi)\ldots\alpha_N(\mi\xi)\\[.5ex]
\times\mathcal{S}\,\trace\!\bigl[
 \widetilde{\ten{G}}{'}(\vect{r}_1,\vect{r}_2,\mi\xi)\sprod
 \widetilde{\ten{G}}{'}(\vect{r}_2,\vect{r}_3,\mi\xi)\cdots
 \widetilde{\ten{G}}{'}(\vect{r}_N,\vect{r}_1,\mi\xi)\bigr]
\end{multline}
is the $N$-atom van der Waals potential in the presence of arbitrary
linearly responding (background) bodies at zero temperature and hence
generalizes the free-space result given in Refs.~\cite{0090,0091}. In
Eq.~(\ref{3.65}), the symbol $\mathcal{S}$ introduces the
symmetrization prescription
\begin{equation}
\label{3.66}
\mathcal{S}f(\vect{r}_1,\vect{r}_2,\ldots,\vect{r}_N)
= \frac{1}{(2-\delta_{2N})N}\sum_{\Pi\in P(N)}
 f(\vect{r}_{\Pi(1)},\vect{r}_{\Pi(2)},\ldots,\vect{r}_{\Pi(N)})
\end{equation}
where $P(N)$ denotes the permutation group of the numbers
$1,2,\ldots,N$. Note that
$\widetilde{\ten{G}}{'}(\vect{r},\vect{r}',\omega)$
$\!=$ $\!\widetilde{\ten{G}}(\vect{r},\vect{r}',\omega)$ for
$\vect{r}$ $\!\neq$ $\!\vect{r}'$. Clearly, for $N$ $\!=$ $\!2$,
Eq.~(\ref{3.65}) reduces to the two-atom potential already given,
Eq.~(\ref{3.56}).

Equation~(\ref{3.63}) reveals that under very general conditions the
interaction of a single ground-state atom with a dielectric body of
Clausius--Mossotti susceptibility may be regarded as being due to all
many-atom interactions of the atom in question with the body atoms.
A relation of this type was first derived for the special case of a
homogeneous half space filled with harmonic-oscillator atoms
\cite{0119,0120,0048} and later extended to homogeneous dielectric
bodies of arbitrary shapes with vacuum background by means of the
Ewald--Oseen extinction theorem \cite{0087}. Note that in close
analogy to Eq.~(\ref{3.63}), the dispersion force between two
dielectric bodies is due to all many-atom interactions of atoms in the
first body with atoms in the second one; this was explicitly verified
for the cases of two homogeneous half spaces \cite{0120,0048} and
spheres \cite{0343} and was also shown for two homogeneous bodies of
arbitrary shapes \cite{0342}.

%%%%%%%%%%%%%%%%%%%%%%%%%%%%%%%%%%%%%%%%%%%%%%%%%%%%%%%%%%%%%%%%%%%%%%

\section{Forces on atoms}
\label{sec4}

As demonstrated in Secs.~\ref{sec3.2.1} and \ref{sec3.2.2}, forces
on individual ground-state atoms in the presence of linearly
responding bodies can be deduced from the forces between dielectric
bodies of Clausius--Mossotti type in the limiting case of the bodies
being weakly polarizable. Alternatively, these forces can be derived
by explicitly studying the interaction of atoms with the body-assisted
electromagnetic field according to Sec.~\ref{sec2.2}. This approach to
dispersion forces on atoms allows for studying the influence of the
internal atomic dynamics on the forces; in particular, excited atoms
can also be considered.

%%%%%%%%%%%%%%%%%%%%%%%%%%%%%%%%%%%%%%%%%%%%%%%%%%%%%%%%%%%%%%%%%%%%%%

\subsection{Ground-state atoms}
\label{sec4.1}

Atoms initially prepared in their ground state will remain in this
state provided that the body-assisted field is also initially prepared
in the ground state. In addition, the atom--field interaction in this
case will involve only virtual, off-resonant transitions of the atoms
and the body-assisted field. Consequently, dispersion forces on
ground-state atoms may adequately be described within the framework of
time-independent perturbation theory.

%%%%%%%%%%%%%%%%%%%%%%%%%%%%%%%%%%%%%%%%%%%%%%%%%%%%%%%%%%%%%%%%%%%%%%

\subsubsection{Single-atom force}
\label{sec4.1.1}

Following the idea of Casimir and Polder \cite{0030}, one can derive
the force on a single atom at zero temperature from the shift $\Delta
E$ of the system's ground-state energy $E$ for given center-of-mass
position of the atom which arises from the atom--field coupling. The
potential $U(\vect{r}_{A})$, whose negative gradient gives the sought
van der Waals force, is the position-dependent part of this
energy shift,\footnote{The position-independent part $\Delta^{(0)}E$
is a contribution to the Lamb shift in free space; for a discussion of
the Lamb shift within the multipolar coupling scheme, see,~e.g.,
Ref.~\cite{0013}.}
\begin{equation}
\label{4.1}
\Delta E=\Delta^{(0)}E+U(\vect{r}_{A}).
\end{equation}
It can be seen that the effective center-of-mass Hamiltonian
\begin{equation}
\label{4.2}
\hat{H}_\mathrm{eff}
 =\frac{\hat{\vect{p}}_{A}^2}{2m_{A}}+U(\hat{\vect{r}}_{A})
\end{equation}
leads to the equation of motion
\begin{equation}
\label{4.3}
\vect{F}(\hat{\vect{r}}_{A})
=m_{A}\ddot{\hat{\vect{r}}}_{A}
 =-\frac{1}{\hbar^2}\bigl[\hat{H}_\mathrm{eff},
 \bigl[\hat{H}_\mathrm{eff},m_{A}\hat{\vect{r}}_{A}\bigr]\bigr]
 =-\bm{\nabla}_{\!\!{A}}U(\hat{\vect{r}}_{A}).
\end{equation}

Making use of the interaction Hamiltonian (\ref{2.90-1}), one obtains
the ground-state energy shift for a non-magnetic atom in second-order
(i.e., leading-order) perturbation theory\footnote{In the following,
the multipolar coupling scheme will be employed and the primes
discriminating the respective atomic and field variables from
the minimal coupling ones will be dropped, for notational convenience.
Equation~(\ref{2.90-1}) can be employed, because the second term in
Eq.~(\ref{2.90}) gives rise to a contribution of order $v/c$ ($v$,
center-of-mass speed) \cite{0008} which can be neglected.} as
\begin{equation}
\label{4.4}
\Delta E
 =\sum_k\sum_{\lambda=e,m}\int\dif^3r\,
 \mathcal{P}\!\int_0^\infty\dif\omega\,
 \frac{\bigl|\langle 0|\langle\{0\}|
 -\!\hat{\vect{d}}\sprod\hat{\vect{E}}(\vect{r}_{A})
 |\bm{1}_\lambda(\vect{r},\omega)\rangle
 |k\rangle\bigr|^2}{-\hbar(\omega_{k0}+\omega)}
\end{equation}
[$\mathcal{P}$, principal part;
$|\vect{1}_\lambda(\vect{r},\omega)\rangle$ $\!=$
$\!\hat{\vect{f}}^\dagger_\lambda(\vect{r},\omega)|\{0\}\rangle$].
Using Eqs.~(\ref{2.20})--(\ref{2.24}), (\ref{2.24-1}) and
(\ref{2.30b}), one derives
\begin{equation}
\label{4.5}
\Delta E
 =-\frac{\mu_0}{\pi}\sum_k
 \mathcal{P}\int_0^\infty
 \frac{\dif\omega}{\omega_{k0}+\omega}\,\omega^2
 \vect{d}_{0k}\sprod\mathrm{Im}\,\ten{G}
 (\vect{r}_{A},\vect{r}_{A},\omega)\sprod\vect{d}_{k0}
\end{equation}
where $\ten{G}(\vect{r},\vect{r}',\omega)$ is the Green tensor of the
body configuration considered. By discarding the position-independent
contribution $\Delta E^{(0)}$ associated with
$\ten{G}^{(0)}(\vect{r}_{A},\vect{r}_{A},\omega)$ [recall
Eqs.~(\ref{3.22}) and (\ref{4.1})] which may be thought of as being
already included in the unperturbed energy, the van der Waals
potential can be written in the form \cite{0008,0012,0018}
\begin{equation}
\label{4.6-1}
U(\vect{r}_{A})=\frac{\hbar\mu_0}{2\pi}
 \int_0^\infty\dif\xi\,\xi^2\trace\bigl[\bm{\alpha}(\mi\xi)\sprod
 \ten{G}^{(1)}(\vect{r}_{A},\vect{r}_{A},\mi\xi)\bigr]
\end{equation}
where
\begin{equation}
\label{4.7-1}
\bm{\alpha}(\omega)
 =\lim_{\epsilon\to 0}\frac{2}{\hbar}\sum_k
 \frac{\omega_{k0}\vect{d}_{0k}\tprod\vect{d}_{k0}}
 {\omega_{k0}^2-\omega^2-\mi \omega\epsilon}
\end{equation}
is the ground-state polarizability of the atom. For isotropic atoms,
it simplifies to
\begin{equation}
\label{4.7}
\bm{\alpha}(\omega)=\alpha(\omega)\ten{I}
 =\lim_{\epsilon\to 0}\frac{2}{3\hbar}\sum_k
 \frac{\omega_{k0}|\vect{d}_{0k}|^2}
 {\omega_{k0}^2-\omega^2-\mi \omega\epsilon}\,\ten{I},
\end{equation}
so the potential simplifies to
\begin{equation}
\label{4.6}
U(\vect{r}_{A})=\frac{\hbar\mu_0}{2\pi}
 \int_0^\infty\dif\xi\,\xi^2\alpha(\mi\xi)\,\trace
 \ten{G}^{(1)}(\vect{r}_{A},\vect{r}_{A},\mi\xi).
\end{equation}
The perturbative result hence agrees with what has been inferred from
the force on weakly polarizable bodies and renders an explicit
expression for the polarizability. Note that the scattering Green
tensor $\ten{G}^{(1)}$ in Eq.~(\ref{4.6}) has exactly the same meaning
as $\overline{\ten{G}}$ in Eq.~(\ref{3.52}). {F}rom Eq.~(\ref{4.6})
together with Eq.~(\ref{4.7}) it can be seen that the potential can
be given in the equivalent form
\begin{equation}
\label{4.8}
U(\vect{r}_{A})
 =-\frac{\hbar\mu_0}{2\pi}\int_0^\infty\dif\omega\,\omega^2
 \mathrm{Im}\bigl[\alpha(\omega)\trace
 \ten{G}^{(1)}(\vect{r}_{A},\vect{r}_{A},\omega)\bigr]
\end{equation}
which allows for a simple physical interpretation of the force as
being due to correlation of the fluctuating electromagnetic field with
the corresponding induced electric dipole of the atom plus the
correlation of the fluctuating electric dipole with its induced
electric field \cite{0046}. It should be pointed out that an analogous
treatment based on the minimal-coupling Hamiltonian (\ref{2.77}) leads
to the formally same result \cite{0008} where of course the
unperturbed eigenstates and energies occurring in the polarizability
(\ref{4.7-1}) are now determined by the atomic Hamiltonian
(\ref{2.55}) in place of (\ref{2.84}). Needless to say that both
results are approximations to the same Hamiltonian of the total
system. Bearing in mind that the ground-state energy shift is entirely
due to virtual, off-resonant transitions, it is crucial to retain the
$\vect{A}^2$~term in the minimal coupling scheme which contributes to
the ground-state energy shift already in first-order perturbation
theory.

Equation~(\ref{4.6-1}) gives the potential of a single ground-state
atom in the presence of an arbitrary arrangement of linearly
responding bodies at zero temperature in terms of the polarizability
of the atom and the Green tensor of the body-assisted electromagnetic
field. A relation of this kind was first derived for arbitrary
electric bodies on the basis of linear-response theory
\cite{0035,0039,0041}, recall Eq.~(\ref{1.10}); alternatively, it was
obtained from a QED path-integral approach \cite{0050} and
semiclassical considerations \cite{0375}. A perturbative derivation
based on macroscopic QED very similar to that presented here is given
in Refs.~\cite{0186,0439,0187}. The linear-response approach has also
been applied to magneto-electric bodies \cite{0043} and finite
temperatures \cite{0037,0046,0394}. Note that in close analogy to the
case of the Casimir force, the single-atom potential at finite
temperature can be obtained from the zero-temperature
result~(\ref{4.6-1}) or (\ref{4.8}) by making the
replacements~(\ref{3.26-3}) or (\ref{3.26-2}), respectively.

%%%%%%%%%%%%%%%%%%%%%%%%%%%%%%%%%%%%%%%%%%%%%%%%%%%%%%%%%%%%%%%%%%%%%%

\subsubsection{Two-atom force}
\label{sec4.1.2}

The interaction potential of two polarizable ground-state atoms in the
presence of linearly responding bodies giving rise to the Green tensor
$\ten{G}(\vect{r},\vect{r}',\omega)$ can also be obtained by means of
time-independent perturbation theory. The leading contribution is now
of fourth order in the atom--field interaction and a somewhat lengthy
calculation yields \cite{0009,0113}
\begin{multline}
\label{4.9-1}
U(\vect{r}_{A},\vect{r}_{B})
 =-\frac{\hbar\mu_0^2}{2\pi}\int_0^\infty\dif\xi\,\xi^4\\[.5ex]
\times\trace[\bm{\alpha}_{A}(\mi\xi)\sprod
 \ten{G}(\vect{r}_{A},\vect{r}_{B},\mi\xi)
 \sprod\bm{\alpha}_{B}(\mi\xi)
 \sprod\ten{G}(\vect{r}_{B},\vect{r}_{A},\mi\xi)]
\end{multline}
which for isotropic atoms reduces to
\begin{multline}
\label{4.9}
U(\vect{r}_{A},\vect{r}_{B})
 =-\frac{\hbar\mu_0^2}{2\pi}\int_0^\infty\dif\xi\,
 \xi^4\alpha_{A}(\mi\xi)\alpha_{B}(\mi\xi)\\[.5ex]
\times\trace[\ten{G}(\vect{r}_{A},\vect{r}_{B},\mi\xi)
 \sprod\ten{G}(\vect{r}_{B},\vect{r}_{A},\mi\xi)].
\end{multline}
Note that Eq.~(\ref{4.9}) agrees with Eq.~(\ref{3.56})
[$\widetilde{\ten{G}}{'}(\vect{r}_1,\vect{r}_2,\omega)$ $\!\mapsto$
$\!\ten{G}(\vect{r}_{A},\vect{r}_{B},\omega)$ for \mbox{$\vect{r}_1$
$\!\neq$ $\!\vect{r}_2$}]. The total force acting on atom ${A}$
(${B}$)
can be obtained by supplementing the single-atom force
$\vect{F}(\vect{r}_{{A}({B})})$ with the two-atom force
\begin{equation}
\label{4.10}
\vect{F}(\vect{r}_{{A}({B})},\vect{r}_{{B}({A})})
 =-\bm{\nabla}_{\!\!{A}({B})}U(\vect{r}_{A},\vect{r}_{B})
\end{equation}
where in general $\vect{F}(\vect{r}_{A},\vect{r}_{B})$ $\!\neq$
$-\vect{F}(\vect{r}_{B},\vect{r}_{A})$, due to the presence of the
bodies.

Equations~(\ref{4.9-1})--(\ref{4.10}) form a general basis for
calculating two-atom potentials in the presence of linearly responding
bodies at zero temperature. An equation of the type~(\ref{4.9}) was
first derived for the case of electric bodies by means of
linear-response theory \cite{0036}. Derivations based on semiclassical
models of harmonic-oscillator atoms interacting in the presence of
perfectly conducting \cite{0092} and electric bodies~\cite{0375} were
given and extended to allow for finite temperatures \cite{0093}.
Equation (\ref{4.9}) has been used to study the interaction of two
atoms embedded in an electrolyte \cite{0351}, situated near perfectly
conducting \cite{0093}, non-local metallic \cite{0518}, electric
\cite{0036,0653,0523} and magneto-electric half spaces
\cite{0009,0491}, placed inside perfectly conducting \cite{0092,0093}
and electric planar cavities \cite{0108}.

Let us consider the simplest case of two atoms in free space where
the single-atom force identically vanishes and the two-atom potential
(\ref{4.9}) can be calculated by using the free-space Green tensor
$\ten{G}(\vect{r},\vect{r}',\omega)$ $\!=$
$\!\ten{G}_\mathrm{free}(\vect{r},\vect{r}',\omega)$
(App.~\ref{appA}), leading to ($r$ $\!=$ $|\vect{r}_{A}$ $\!-$
$\!\vect{r}_{B}|$)
\begin{equation}
\label{4.11}
U(\vect{r}_{A},\vect{r}_{B})
 =-\frac{\hbar}{32\pi^3\varepsilon_0^2r^6}
 \int_0^\infty\dif\xi\,
 \alpha_A(\mi\xi)\alpha_B(\mi\xi)g(\xi r/c)
\end{equation}
where
\begin{equation}
\label{4.12}
g(x)=2\me^{-2x}\bigl(3+6x+5x^2+2x^3+x^4\bigr),
\end{equation}
which shows that the interaction between two polarizable ground-state
atoms is always attractive. Equations~(\ref{4.11}) and (\ref{4.12})
are in agreement with the famous result of Casimir and Polder
\cite{0030}, whose derivation was based on fourth-order perturbation
theory and normal-mode QED. The problem has been reconsidered many
times, inter alia within the frameworks of normal-mode QED
\cite{0011,0521,0498,0056,0320,0051,0201,0200} and linear-response
theory \cite{0035}; and it has even become a common textbook
example \cite{0325,0323,0007}.

In the retarded limit where $r$ $\!\gg$ $\!c/\omega_{\mathrm{min}}$
($\omega_{\mathrm{min}}$ denoting the minimum of the relevant
resonance frequencies of atoms ${A}$ and ${B}$), the function
$g(\xi r/c)$ effectively limits the $\xi$~integral in Eq.~(\ref{4.11})
to a range where $\alpha_{{A}({B})}(\mi\xi)$ $\!\simeq$
$\!\alpha_{{A}({B})}(0)$, so the potential approaches
\begin{equation}
\label{4.13}
U(\vect{r}_{A},\vect{r}_{B})
=-\frac{23\hbar c\alpha_{A}(0)\alpha_{B}(0)}
 {64\pi^3\varepsilon_0^2r^7}\,,
\end{equation}
as already pointed out in Ref.~\cite{0030}, cf. Eq.~(\ref{1.7}). In
the non-retarded limit where \mbox{$r$ $\!\ll$
$\!c/\omega_\mathrm{max}$} ($\omega_{\mathrm{max}}$ denoting the
maximum of the relevant resonance frequencies of atoms $A$ and $B$),
the integral is limited by the polarizabilities
$\alpha_{{A}({B})}(\mi\xi)$ to a range where $g(\mi\xi)$ $\!\simeq$
$g(0)$ $\!=$ $\!6$, leading to
\begin{equation}
\label{4.14}
U(\vect{r}_{A},\vect{r}_{B})=-\frac{3\hbar}{16\pi^3\varepsilon_0^2r^6}
 \int_0^\infty\dif\xi\,\alpha_{A}(\mi\xi)\alpha_{B}(\mi\xi).
\end{equation}
Upon recalling Eq.~(\ref{4.7}), one may easily verify that this
non-retarded asymptote is nothing but the well-known London
potential~(\ref{1.2}) which was originally obtained from a
perturbative treatment of the Coulomb interaction \cite{0374}
(cf.~Refs.~\cite{0510,0507} for similar derivations).

It is illustrative to compare the potential between two polarizable
atoms with the potential between a polarizable atom $A$ of
polarizability $\alpha_{A}(\omega)$ and a magnetizable atom $B$
of magnetizability $\beta_{B}(\omega)$ which, according to
Eq.~(\ref{3.68}), is given by
\begin{multline}
\label{4.15}
U(\vect{r}_{A},\vect{r}_{B})
 =-\frac{\hbar\mu_0^2}{2\pi}
 \int_0^\infty\dif\xi\,\xi^2
 \alpha_{A}(\mi\xi)\beta_{B}(\mi\xi)\\[.5ex]
 \times\trace\bigl\{\bigl[
 \ten{G}(\vect{r}_{A},\vect{r},\mi\xi)
 \vprod\overleftarrow{\bm{\nabla}}\bigr]
 \sprod\bm{\nabla}\vprod
 \ten{G} (\vect{r},\vect{r}_{A},\mi\xi)
 \bigr\}_{\vect{r}=\vect{r}_{B}}.
\end{multline}
When the two atoms are in free space so that
$\ten{G}(\vect{r},\vect{r}',\omega)$ $\!=$
$\!\ten{G}_\mathrm{free}(\vect{r},\vect{r}',\omega)$,
then Eq.~(\ref{4.15}) reads
\begin{equation}
\label{4.16}
U(\vect{r}_{A},\vect{r}_{B}) =
\frac{\hbar\mu_0^2}{32\pi^3r^4}
 \int_0^{\infty}\dif\xi\,\xi^2
 \alpha_{A}(\mi\xi)\beta_{B}(\mi\xi) h(\xi r/c)
\end{equation}
where
\begin{equation}
\label{4.17}
 h(x) = 2\me^{-2x}\bigl(1+2x+x^2\bigr)
\end{equation}
which is in agreement with results found on the basis of normal-mode
QED \cite{0094,0096}. In contrast to the attractive interaction
between two polarizable atoms, the interaction between a polarizable
and a magnetizable atom is always repulsive, as can be easily seen
from Eq.~(\ref{4.16}) together with Eq.~(\ref{4.17}). In particular in
the retarded and non-retarded limits, respectively, Eq.~(\ref{4.16})
reduces to
\begin{equation}
\label{4.19}
U(\vect{r}_{A},\vect{r}_{B})
 =\frac{7\hbar c\mu_0\alpha_{A}(0)\beta_{B}(0)}
 {64\pi^3\varepsilon_0 r^7}
\end{equation}
and
\begin{equation}
\label{4.20}
U(\vect{r}_{A},\vect{r}_{B})
=\frac{\hbar\mu_0^2}{16\pi^3r^4}
 \int_0^\infty\dif\xi\,\xi^2
 \alpha_{A}(\mi\xi)\beta_{B}(\mi\xi)
\end{equation}
which was already given in Refs.~\cite{0089,0095} and
\cite{0499,0121}.

Comparing Eqs.~(\ref{4.13}) and (\ref{4.19}), we see that in the
retarded limit the absolute value of the force between two
polarizable atoms and that between a polarizable and a magnetizable
atom follow the same $1/r^8$~power law with the strength being weaker
in the latter case by a factor $7/23$. Comparison of Eqs.~(\ref{4.14})
and (\ref{4.20}) shows that in the non-retarded limit the absolute
value of the force between a polarizable and a magnetizable atom
which follows a $1/r^5$~power law, is more weakly diverging than that
between two polarizable atoms which follows a $1/r^7$~power law.

%%%%%%%%%%%%%%%%%%%%%%%%%%%%%%%%%%%%%%%%%%%%%%%%%%%%%%%%%%%%%%%%%%%%%%

\subsubsection{Atom in a planar structure}
\label{sec4.1.3}

Let us return to Eq.~(\ref{4.6-1}) for the potential of a single
ground-state atom in the presence of an arbitrary arrangement of
linearly responding bodies at zero temperature. It has been used
to calculate the potential for a variety of particular geometries,
including different planar structures (see below) as well as perfectly
conducting \cite{0110,0346}, non-local metallic \cite{0270,0060},
dielectric \cite{0110,0069,0077,0349,0372,0017} and magneto-electric
spheres \cite{0113}; dielectric \cite{0284,0077,0395,0316} and
non-local metallic cylinders \cite{0040,0397}; magneto-dielectric
rings \cite{0020}; electric cylindrical shells \cite{0186,0439,0187};
electric \cite{0313} and non-local metallic spherical cavities
\cite{0396}; electric cylindrical cavities \cite{0316} and perfectly
conducting wedge-shaped cavities \cite{0069,0372}.

To illustrate the theory, let us consider an atom placed in a
free-space region between two planar magneto-electric walls, as
defined by Eqs.~(\ref{3.27}) and (\ref{3.28}) with
$\varepsilon(\omega)$ $\!=$ $\!\mu(\omega)$ $\!=$ $\!1$. Inserting the
Green tensor for this system (App.~\ref{appA}) into Eq.~(\ref{4.6})
leads to the single-atom potential \cite{0012,0019}
\begin{multline}
\label{4.21}
U(z_{A})
=\frac{\hbar\mu_0}{8\pi^2}
 \int_0^\infty\dif\xi\,\xi^2\alpha(\mi\xi)
 \int_0^\infty\dif q\,\frac{q}{b}\biggl\{
 \me^{-2bz_{A}}\biggl[\frac{r_{s-}}{D_s}
 -\biggl(1+2\,\frac{q^2c^2}{\xi^2}\biggr)
 \frac{r_{p-}}{D_p}\biggr]\\[.5ex]
+\,\me^{-2b(d-z_{A})}
 \biggl[\frac{r_{s+}}{D_s}-\biggl(1+2\,\frac{q^2c^2}{\xi^2}\biggr)
 \frac{r_{p+}}{D_p}\biggr]\biggr\}
\end{multline}
where $z_{A}$ is the separation of the atom from the left wall,
$d$ is the separation of the two walls, $b$ and $D_\sigma$ are given
by Eqs.~(\ref{3.32}) and (\ref{3.33}), respectively, with $n(\mi\xi)$
$\!=$ $\!1$, and $r_{\sigma\pm}$ $\!=$ $r_{\sigma\pm}(\xi,q)$ are
again the reflection coefficients associated with the left/right
walls. If the atom is placed near a single wall, say the right wall is
missing, Eq.~(\ref{4.21}) reduces to ($r_{\sigma}$ $\!\equiv$
$\!r_{\sigma -}$)
\begin{equation}
\label{4.22}
U(z_{A})=\frac{\hbar\mu_0}{8\pi^2}
 \int_0^\infty\dif\xi\,\xi^2
 \alpha(\mi\xi)\int_0^\infty\dif q\,
 \frac{q}{b}\,\me^{-2bz_{A}}\biggl[r_s
 -\biggl(1+2\frac{q^2c^2}{\xi^2}\biggr)r_p\biggr].
\end{equation}

%%%%%%%%%%%%%%%%%%%%%%%%%%%%%%%%%%%%%%%%%%%%%%%%%%%%%%%%%%%%%%%%%%%%%%

\paragraph{Perfectly reflecting plate}
\label{sec4.1.3.1}

Consider first an atom placed near a perfectly reflecting electric
(i.e., perfectly conducting) plate, $r_p$ $\!=$ $\!-r_s$ $\!=$ $\!1$.
By changing the integration variable in Eq.~(\ref{4.22}) from $q$ to
$b$ [Eq.~(\ref{3.32})], the resulting integral can be
performed to obtain
\begin{equation}
\label{4.23}
U(z_{A})=-\frac{\hbar}{16\pi^2\varepsilon_0z_{A}^3}
 \int_0^\infty\dif\xi\,\alpha(\mi\xi)
 \,\me^{-2\xi z_{A}/c}
 \Biggl[1+2\biggl(\frac{\xi z_{A}}{c}\biggr)
 +2\biggl(\frac{\xi z_{A}}{c}\biggr)^2\Biggr],
\end{equation}
in agreement with the result found by Casimir and Polder
on the basis of normal-mode QED \cite{0030} (cf.~also
Refs.~\cite{0047,0056}) which has been reproduced by means of
linear-response theory \cite{0035,0039,0041} and dynamical
image-dipole treatments \cite{0321}. In the retarded limit, $z_{A}$
$\!\gg$ $\!c/\omega_{\mathrm{min}}$, the exponential $\me^{-2\xi
z_{A}/c}$ effectively limits the $\xi$~integral to the region where
$\alpha(\mi\xi)$ $\!\simeq$ $\alpha(0)$, so the potential approaches
\begin{equation}
\label{4.24}
U(z_A)=-\frac{3\hbar c\alpha(0)}{32\pi^2\varepsilon_0z_A^4}\,,
\end{equation}
as already noted in Refs.~\cite{0030,0325,0054}, cf. Eq.~(\ref{1.8}).
In the non-retarded limit, $z_{A}$ $\!\ll$ $\!c/\omega_\mathrm{max}$,
the polarizability $\alpha(\mi\xi)$ restricts the integration to the
region where $\xi z_{A}/c$ $\!\simeq$ $\!0$, leading to
\begin{equation}
\label{4.25}
U(z_A)=-\frac{1}{48\pi\varepsilon_0z_A^3}\sum_k|\vect{d}_{0k}|^2
 =-\frac{\langle 0|\hat{\vect{d}}^2|0\rangle}
 {48\pi\varepsilon_0z_A^3}\,,
\end{equation}
in agreement with the result~(\ref{1.3}) obtained by Lennard-Jones on
the basis of an electrostatic calculation \cite{0022} [recall
Eq.~(\ref{4.7})].

In contrast, the potential of an atom in front of an infinitely
permeable plate is repulsive, as can be seen by setting $r_p$ $\!=$
$\!-r_s$ $\!=$ $\!-1$ in Eq.~(\ref{4.22}),
\begin{equation}
\label{4.26}
U(z_{A})=\frac{\hbar}{16\pi^2\varepsilon_0z_{A}^3}
 \int_0^\infty\dif\xi\,\alpha(\mi\xi)
 \me^{-2\xi z_{A}/c}
 \Biggl[1+2\biggl(\frac{\xi z_A}{c}\biggr)
 +2\biggl(\frac{\xi z_A}{c}\biggr)^2\Biggr]\,.
\end{equation}
It approaches
\begin{equation}
\label{4.27}
U(z_A)=\frac{3\hbar c\alpha(0)}{32\pi^2\varepsilon_0z_A^4}
\end{equation}
in the retarded limit (cf. also Ref.~\cite{0330}) and
\begin{equation}
\label{4.28}
U(z_A)=
\frac{\langle 0|\hat{\vect{d}}^2|0\rangle}
{48\pi\varepsilon_0z_A^3}
\end{equation}
in the non-retarded limit.

\begin{figure}[!t!]
\begin{center}
\includegraphics[width=\linewidth]{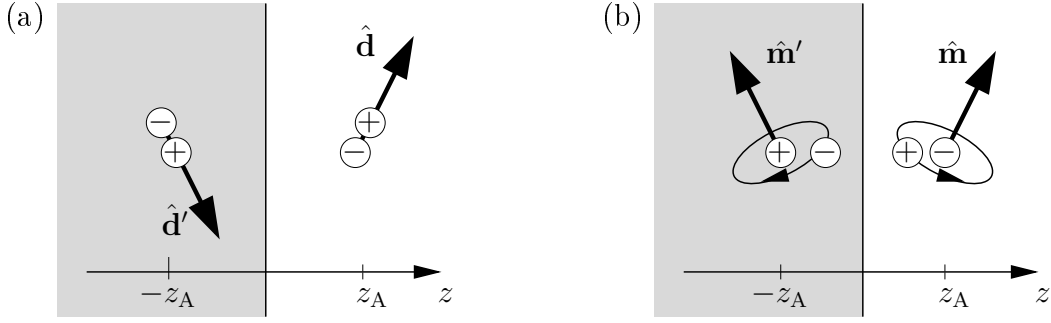}
\end{center}
\caption{
\label{fig1}
The image dipole construction for an (a) electric (b) magnetic dipole
in front of a perfectly reflecting electric plate, is shown.
}
\end{figure}

The different signs of the non-retarded potentials (\ref{4.25}) and
(\ref{4.28}) in the cases of a perfectly reflecting electric and a
perfectly reflecting magnetic plate, respectively, can be understood
from an image-dipole model \cite{0022}. The non-retarded potential
can be regarded as being due to the Coulomb interaction of an electric
dipole $\hat{\vect{d}}$ $\!=$ $(\hat{d}_x,\hat{d}_y,\hat{d}_z)$
situated at distance $z_{A}$ from the plate with its image
$\hat{\vect{d}}'$ $\!=$ $(-\hat{d}_x,-\hat{d}_y,\hat{d}_z)$
in the plate [Fig~\ref{fig1}(a)].
%%%%%%%%%%%%%%% F I G U R E %%%%%%%%%%%%%%%%%%%%%%%%%%%%%%%%%%%%%%%%%
%%%%%%%%%%%%%%%%%%%%%%%%%%%%%%%%%%%%%%%%%%%%%%%%%%%%%%%%%%%%%%%%%%%%%%
The average interaction energy of the dipole and its image hence reads
\cite{0001}\footnote{The factor $1/2$ in Eq.~(\ref{4.29}) accounts for
the fact that the second dipole is induced by the first one.}
\begin{equation}
\label{4.29}
U(z_{A})
 =\frac{1}{2}\,\frac{\langle 0|\hat{\vect{d}}\sprod\hat{\vect{d}}'
 -3\hat{d}_z\hat{d'_z}|0\rangle}
 {4\pi\varepsilon_0(2z_{A})^3}
 =-\frac{\langle 0|\hat{\vect{d}}^2+\hat{d}_z^2|0\rangle}
 {64\pi\varepsilon_0z_{A}^3}
 =-\frac{\langle 0|\hat{\vect{d}}^2|0\rangle}
 {48\pi\varepsilon_0z_{A}^3}
\end{equation}
[$\langle 0|\hat{d}_x^2|0\rangle$ $\!=$
$\!\langle 0|\hat{d}_y^2|0\rangle$ $\!=$
$\!\langle 0|\hat{d}_z^2|0\rangle$ $\!=$
$(1/3)\langle 0|\hat{\vect{d}}^2|0\rangle$],
in agreement with Eq.~(\ref{4.25}).

Since the interaction of a polarizable atom with an infinitely
permeable plate is equivalent to the interaction of a magnetizable
atom with a perfectly reflecting electric plate by virtue of the
duality of electric and magnetic fields, a magnetic dipole
$\hat{\vect{m}}$ $\!=$ $(\hat{m}_x,\hat{m}_y,\hat{m}_z)$ in front of a
perfectly reflecting electric plate can be considered. A magnetic
dipole behaves like a pseudo-vector under reflection, so its image is
given by $\hat{\vect{m}}'$ $\!=$ $(\hat{m}_x,\hat{m}_y,-\hat{m}_z)$
[Fig~\ref{fig1}(b)]. The interaction energy of the magnetic dipole and
its image reads
\begin{equation}
\label{4.30}
U(z_{A})
 =\frac{1}{2}\,\frac{\langle 0|\hat{\vect{m}}\sprod\hat{\vect{m}}'
 -3\hat{m}_z\hat{m'_z}|0\rangle}
 {4\pi\varepsilon_0(2z_{A})^3}
 =\frac{\langle 0|\hat{\vect{m}}^2|0\rangle}
 {48\pi\varepsilon_0z_{A}^3}
\end{equation}
which by means of a duality transformation is equivalent to
Eq.~(\ref{4.28}). The different signs of the potentials (\ref{4.25})
and (\ref{4.28}) can thus be understood from the different reflection
behavior of electric and magnetic dipoles.

%%%%%%%%%%%%%%%%%%%%%%%%%%%%%%%%%%%%%%%%%%%%%%%%%%%%%%%%%%%%%%%%%%%%%%

\paragraph{Magneto-electric half space}
\label{sec4.1.3.2}

To be more realistic, let us next consider an atom in front of a
semi-infinite half space of given $\varepsilon(\omega)$ and
$\mu(\omega)$. Upon substitution of the Fresnel reflection
coefficients~(\ref{A.10}), Eq.~(\ref{4.22}) takes the form
\cite{0012,0019}
\begin{multline}
\label{4.31}
U(z_{A})
=\frac{\hbar\mu_0}{8\pi^2}
 \int_0^\infty\dif\xi\,\xi^2\alpha(\mi\xi)
 \int_0^\infty\dif q\,\frac{q}{b}\,\me^{-2bz_{A}}
 \biggl[\frac{\mu(\mi\xi)b-b_1}
 {\mu(\mi\xi)b+b_1}\\[.5ex]
-\biggl(1+2\frac{q^2c^2}{\xi^2}\biggr)
 \frac{\varepsilon(\mi\xi)b-b_1}
 {\varepsilon(\mi\xi)b+b_1}\biggr]
\end{multline}
with $b_1$ $\!\equiv$ $\!b^1_-$ defined as in Eq.~(\ref{A.9}), in
agreement with the result found by means of linear-response theory
\cite{0043}. From Eq.~(\ref{4.31}), the results obtained by means of
normal-mode QED \cite{0051,0031,0330} and linear-response theory
\cite{0036,0039,0041} for an electric half space can also be
recovered.

\begin{figure}[!t!]
\begin{center}
\includegraphics[width=0.75\linewidth]{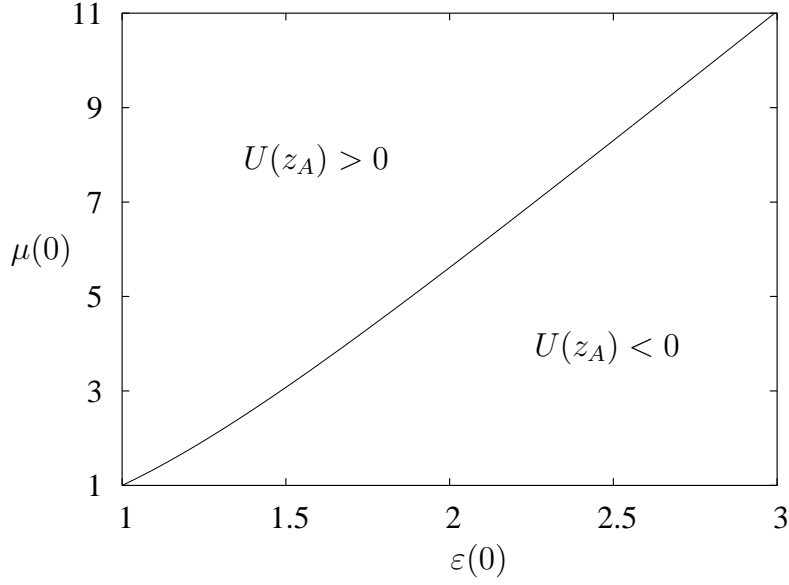}
\end{center}
\caption{
\label{fig2}
Borderline of attractive and repulsive retarded potentials
of a ground-state atom in front of a magneto-dielectric half space.
}
\end{figure}

One can show that in the retarded limit $z_A$ $\!\gg$
$\!c/\omega_\mathrm{min}$ (with $\omega_\mathrm{min}$ being the
minimum of all relevant atom and medium resonance frequencies)
the potential takes the asymptotic form
\cite{0012,0019}\footnote{Obviously this limit does not apply for
metals where the condition can never be fulfilled. Similarly,
Eqs.~(\ref{4.33}), (\ref{4.39}) and (\ref{4.40}) only hold for
dielectrics.}
\begin{multline}
\label{4.32}
U(z_{A})=-\frac{3\hbar c\alpha(0)}{64\pi^2\varepsilon_0z_{A}^4}
 \int_{1}^\infty\dif v\,
 \biggl[\Bigl(\frac{2}{v^2}-\frac{1}{v^4}\Bigr)
 \frac{\varepsilon(0)v-\sqrt{\varepsilon(0)\mu(0)-1+v^2}}
 {\varepsilon(0)v+\sqrt{\varepsilon(0)\mu(0)-1+v^2 }}\\[.5ex]
-\,\frac{1}{v^4}\,
 \frac{\mu(0)v-\sqrt{\varepsilon(0)\mu(0)-1+v^2}}
 {\mu(0)v+\sqrt{\varepsilon(0)\mu(0)-1+v^2}}\,\biggr]
\end{multline}
which can be attractive or repulsive, depending on the strengths of
the competing magnetic and electric properties of the half space.
Figure~\ref{fig2} shows the borderline between attractive and
repulsive potentials in the $\varepsilon(0)\mu(0)$-plane. In
particular, repulsion occurs iff $\mu(0)\!-\!1$ $\!>$
$\!3.29[\varepsilon(0)\!-\!1]$ or $\mu(0)$ $\!>$
$\!5.11\varepsilon(0)$ for weak and strong magneto-dielectric
properties, respectively.
%%%%%%%%%%%%%%%  F I G U R E %%%%%%%%%%%%%%%%%%%%%%%%%%%%%%%%%%%%%%%%%
%%%%%%%%%%%%%%%%%%%%%%%%%%%%%%%%%%%%%%%%%%%%%%%%%%%%%%%%%%%%%%%%%%%%%%

In the non-retarded limit, $n(0)z_A$ $\!\ll$ $\!c/\omega_\mathrm{max}$
[$\omega_\mathrm{max}$, maximum of all relevant atom and medium
resonance frequencies; $n(0)$ $\!=$ $\!\sqrt{\varepsilon(0)\mu(0)}$],
the situation is more complex, because electric and magnetic medium
properties give rise to potentials with different asymptotic power
laws. In particular, the potential approaches
\begin{equation}
\label{4.33}
U(z_{A})=-\frac{\hbar}{16\pi^2\varepsilon_0z_{A}^3}
 \int_0^\infty\dif\xi\,\alpha(\mi\xi)\,
 \frac{\varepsilon(\mi\xi)-1}{\varepsilon(\mi\xi)+1}
\end{equation}
in the case of a purely dielectric half space---in agreement with the
result~(\ref{1.4}) found by considering the Coulomb interaction and
using image-dipole methods \cite{0277} or linear-response theory
\cite{0027}---and
\begin{equation}
\label{4.34}
U(z_{A})=\frac{\hbar}{32\pi^2\varepsilon_0z_{A}}
 \int_0^\infty\dif\xi\,\biggl(\frac{\xi}{c}\biggr)^2\alpha(\mi\xi)\,
  \frac{[\mu(\mi\xi)-1][\mu(\mi\xi)+3]}{\mu(\mi\xi)+1}
\end{equation}
%
%%%%%%%%%%%%%%%  F I G U R E %%%%%%%%%%%%%%%%%%%%%%%%%%%%%%%%%%%%%%%%%
\begin{figure}[!t!]
\begin{center}
\includegraphics[width=\linewidth]{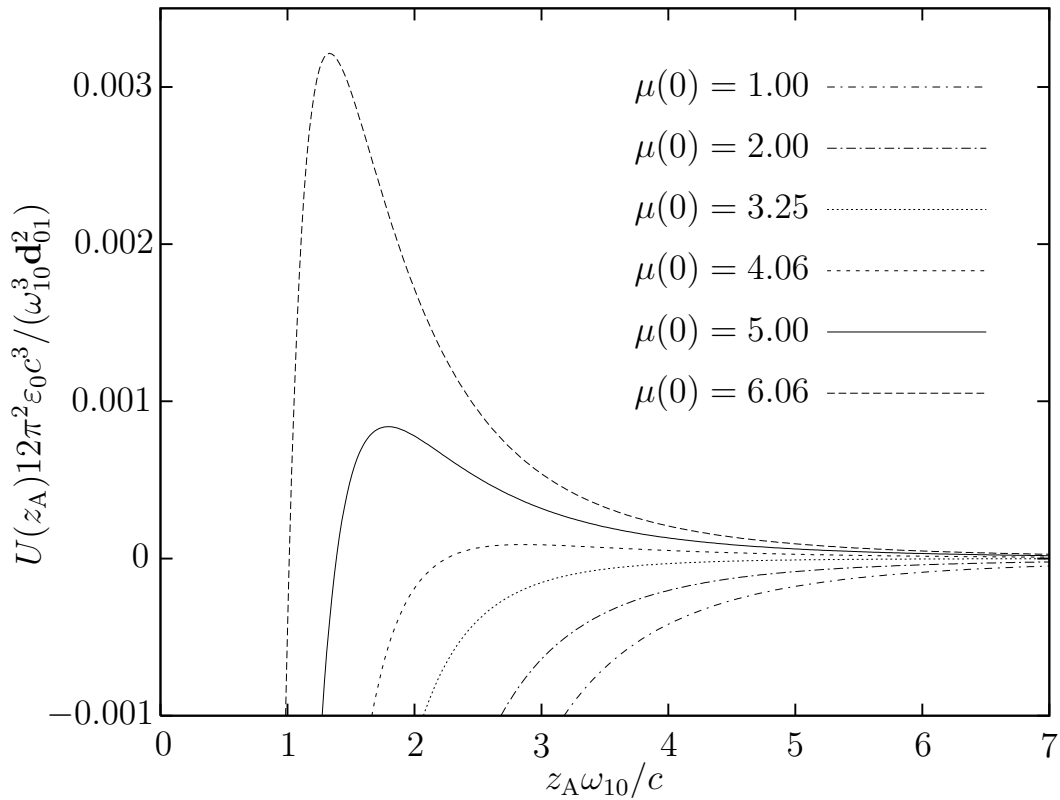}
\end{center}
\caption{
\label{fig3}
The potential of a ground-state two-level atom in front of a
magneto-dielectric half space, Eq.~(\ref{4.31}), is shown as a
function of the distance between the atom and the half space for
different values of $\mu(0)$ ($\omega_{\mathrm{P}e}/\omega_{10}$ $\!=$
$\!0.75$, $\omega_{\mathrm{T}e}/\omega_{10}$ $\!=$ $\!1.03$,
\mbox{$\omega_{\mathrm{T}m}/\omega_{10}$ $\!=$ $\!1$},
$\gamma_{e}/\omega_{10}$ $\!=$ $\!\gamma_{m}/\omega_{10}$ $\!=$
$\!0.001$).
}
\end{figure}%
%%%%%%%%%%%%%%%%%%%%%%%%%%%%%%%%%%%%%%%%%%%%%%%%%%%%%%%%%%%%%%%%%%%%%%
in the case of a purely magnetic one \cite{0012,0019}. Thus for a
magneto-dielectric half space the attractive $1/z_{A}^3$ potential
associated with the polarizability of the half space will always
dominate the repulsive $1/z_{A}$ potential related to its
magnetizability. It should be noted that the non-retarded limit is
in general incompatible with the limit of perfect reflectivity
considered in Sec.~\ref{sec4.1.3.1}. So, Eq.~(\ref{4.34}) does not
approach Eq.~(\ref{4.28}) as $\mu(\mi\xi)$ tends to infinity. On the
contrary, Eq.~(\ref{4.33}) converges to Eq.~(\ref{4.25}) as
$\varepsilon(\mi\xi)$ tends to infinity~\cite{0296}.

Combining the observations from the retarded and non-retarded
limits, one may thus expect a potential barrier, provided that the
permeability of the half space is sufficiently large \cite{0012,0019}.
This is illustrated in Fig.~\ref{fig3} where the potential of a
two-level atom as a function of its distance from the half space, is
shown for various values of the (static) permeability. In the
figure, the permittivity and permeability of the half space have been
assumed to be of Drude--Lorentz type,
\begin{equation}
\label{4.35}
\varepsilon(\omega)=1+\frac{\omega_{\mathrm{P}e}^2}
 {\omega^2_{{\mathrm{T}e}}-\omega^2-\mi\omega\gamma_e}\,,
\quad
\mu(\omega)=1+\frac{\omega_{\mathrm{P}m}^2}
 {\omega^2_{\mathrm{T}m}-\omega^2-\mi\omega\gamma_m}\,.
\end{equation}
{F}rom the figure it is seen that with increasing value of $\mu(0)$,
a potential barrier begins to form at intermediate distances, as
expected. It is shifted to smaller distances and increases in height
as the value of $\mu(0)$ is further increased.

%%%%%%%%%%%%%%%%%%%%%%%%%%%%%%%%%%%%%%%%%%%%%%%%%%%%%%%%%%%%%%%%%%%%%%

\paragraph{Plate of finite thickness}
\label{sec4.1.3.3}

Consider now an atom in front of magneto-electric plate of finite
thickness $d$. Evaluating the relevant reflection coefficients
(\ref{A.7}) and (\ref{A.8}) ($d$ $\!\equiv$ $\!d_-^1$), one finds that
Eq.~(\ref{4.22}) takes the form \cite{0012,0019}
\begin{multline}
\label{4.37}
U(z_{A})
=\frac{\hbar\mu_0}{8\pi^2}
 \int_0^\infty\dif\xi\,\xi^2\alpha(\mi\xi)
 \int_0^\infty\dif q\,\frac{q}{b}\,\me^{-2bz_{A}}\\[.5ex]
 \times\biggl\{\frac{[\mu^2(\mi\xi)b^2-b_1^2]\tanh(b_1d)}
 {2\mu(\mi\xi)b b_1+[\mu^2(\mi\xi)b^2+b_1^2]\tanh(b_1d)}\\[.5ex]
-\,\biggl(1+2\,\frac{q^2c^2}{\xi^2}\biggr)
 \frac{[\varepsilon^2(\mi\xi)b^2-b_1^2]\tanh(b_1d)}
 {2\varepsilon(\mi\xi)b b_1+
 [\varepsilon^2(\mi\xi)b^2+b_1^2]\tanh(b_1d)}
 \biggr\}
\end{multline}
which reduces to the result in Ref.~\cite{0032} in the special case of
an electric plate. For an asymptotically thick plate, $d$ $\!\gg$
$\!z_A$, the exponential factor restricts the integral in
Eq.~(\ref{4.37}) to a region where \mbox{$b_1d$ $\!\sim$ $\!d/(2z_A)$
$\!\gg$ $\!1$}. One may hence make the approximation $\tanh(b_1d)$
$\!\simeq$ $\!1$, leading back to Eq.~(\ref{4.31}) which demonstrates
that the semi-infinite half space is a good model provided that $d$
$\!\gg$ $\!z_A$.

%%%%%%%%%%%%%%%  F I G U R E %%%%%%%%%%%%%%%%%%%%%%%%%%%%%%%%%%%%%%%%%
\begin{figure}[!t!]
\begin{center}
\includegraphics[width=\linewidth]{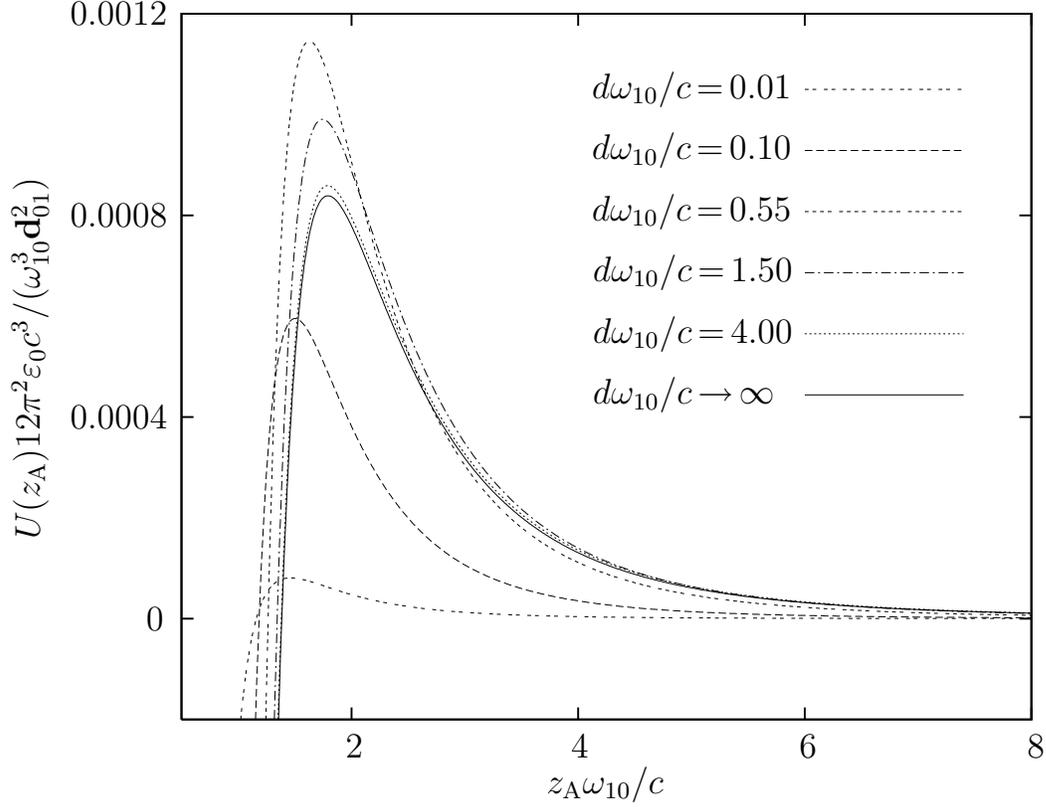}
\end{center}
\caption{
\label{fig4}
The potential of a ground-state two-level atom in front of a
magneto-dielectric plate, Eq.~(\ref{4.37}), is shown as a function of
the atom--plate separation for different values of the plate thickness
$d$ [$\mu(0)$ $\!=$ $\!5$; whereas all other parameters are the same
as in Fig.~\ref{fig2}].
}
\end{figure}%
%%%%%%%%%%%%%%%%%%%%%%%%%%%%%%%%%%%%%%%%%%%%%%%%%%%%%%%%%%%%%%%%%%%%%%
On the contrary, in the limit of an asymptotically thin plate, $n(0)d$
$\!\ll$ $\!z_A$, the approximation $b_1d$ $\!\ll$ $\!1$ results in
\cite{0012,0019}
\begin{multline}
\label{4.38}
U(z_{A})
=\frac{\hbar\mu_0d}{8\pi^2}
 \int_0^\infty\dif\xi\,\xi^2\alpha(\mi\xi)
 \int_0^\infty\dif q\,\frac{q}{b}\,\me^{-2bz_{A}}
 \biggl[\frac{\mu^2(\mi\xi)b^2-b_1^2}{2\mu(\mi\xi)b}\\[.5ex]
-\,\biggl(1+2\frac{q^2c^2}{\xi^2}\biggr)
 \frac{\varepsilon^2(\mi\xi)b^2-b_1^2}
 {2\varepsilon(\mi\xi)b}\biggr]
\end{multline}
which in the retarded limit approaches
\begin{equation}
\label{4.39}
U(z_{A})=-\frac{\hbar c\alpha(0)d}
 {160\pi^2\varepsilon_0z_{A}^5}\,
 \biggl[\frac{14\varepsilon^2(0)-9}{\varepsilon(0)}
 -\frac{6\mu^2(0)-1}{\mu(0)}\biggr].
\end{equation}
In the non-retarded limit one can again distinguish between a
purely dielectric plate and a purely magnetic plate.
Equation~(\ref{4.38}) approaches
\begin{equation}
\label{4.40}
U(z_{A})=-\frac{3\hbar d}{64\pi^2\varepsilon_0z_{A}^4}
 \int_0^\infty\dif\xi\,\alpha(\mi\xi)\,
 \frac{\varepsilon^2(\mi\xi)-1}{\varepsilon(\mi\xi)}
\end{equation}
in the former case and
\begin{equation}
\label{4.41}
U(z_{A})=\frac{\hbar d}{64\pi^2\varepsilon_0z_{A}^2}
 \int_0^\infty\dif\xi\,\biggl(\frac{\xi}{c}\biggr)^2\alpha(\mi\xi)\,
 \frac{[\mu(\mi\xi)-1][3\mu(\mi\xi)+1]}{\mu(\mi\xi)}
\end{equation}
in the latter case. Comparing Eqs.~(\ref{4.39})--(\ref{4.41}) with
Eqs.~(\ref{4.32})--(\ref{4.34}), we see that the power laws change
from $z_{A}^{-n}$ to $z_{A}^{-(n+1)}$ when the plate
thickness changes from being infinitely large to being infinitely
small.

If the permeability is sufficiently big, a magneto-dielectric
plate of finite thickness features a potential wall \cite{0012,0019},
as illustrated in Fig.~\ref{fig4} for a two-level atom, with the
permittivity and permeability of the plate being again given by
Eq.~(\ref{4.35}). It is seen that for a thin plate the barrier is very
low. It raises with increasing thickness of the plate, reaches a
maximal height for some intermediate thickness and then lowers slowly
towards the asymptotic half space value as the thickness is further
increased.

%%%%%%%%%%%%%%%%%%%%%%%%%%%%%%%%%%%%%%%%%%%%%%%%%%%%%%%%%%%%%%%%%%%%%%

\paragraph{Planar cavity}
\label{sec4.1.3.4}

When an atom is placed between two magneto-electric plates, then the
competing effects of attractive and repulsive interaction with the two
plates can result in the formation of a trapping potential
\cite{0012,0019}. Let us model a magneto-electric planar cavity by
two identical half spaces of permittivity $\varepsilon(\omega)$ and
permeability $\mu(\omega)$ which are separated by a distance
$d$.\footnote{Purely electric planar cavities have been
modeled with various degrees of detail, e.g., by two perfectly
conducting plates \cite{0070,0321,0034,0278,0301,0063,0315,0314}, two
electric half spaces \cite{0314}, or even two electric plates of
finite thickness \cite{0032,0033}.} Substitution of the Fresnel
reflection coefficients (\ref{A.10}) into Eq.~(\ref{4.21}) yields the
potential of an atom placed within a cavity bounded by the half
spaces:
\begin{multline}
\label{4.42}
U(z_{A})=\frac{\hbar\mu_0}{8\pi^2}\int_0^\infty\dif\xi\,\xi^2
 \alpha(\mi\xi)\int_0^\infty\dif q\,\frac{q}{b}
 \bigl[\me^{-2bz_{A}}+\me^{-2b(d-z_{A})}\bigr]
 \biggl[\frac{1}{D_s}\,\frac{\mu(\mi\xi)b-b_1}{\mu(\mi\xi)b+b_1}
 \\[.5ex]
-\,\biggl(1+2\frac{q^2c^2}{\xi^2}\biggr)\frac{1}{D_p}\,
 \frac{\varepsilon(\mi\xi)b-b_1}
 {\varepsilon(\mi\xi)b+b_1}\biggr].
\end{multline}
%
%%%%%%%%%%%%%%%  F I G U R E %%%%%%%%%%%%%%%%%%%%%%%%%%%%%%%%%%%%%%%%%
\begin{figure}[!t!]
\begin{center}
\includegraphics[width=\linewidth]{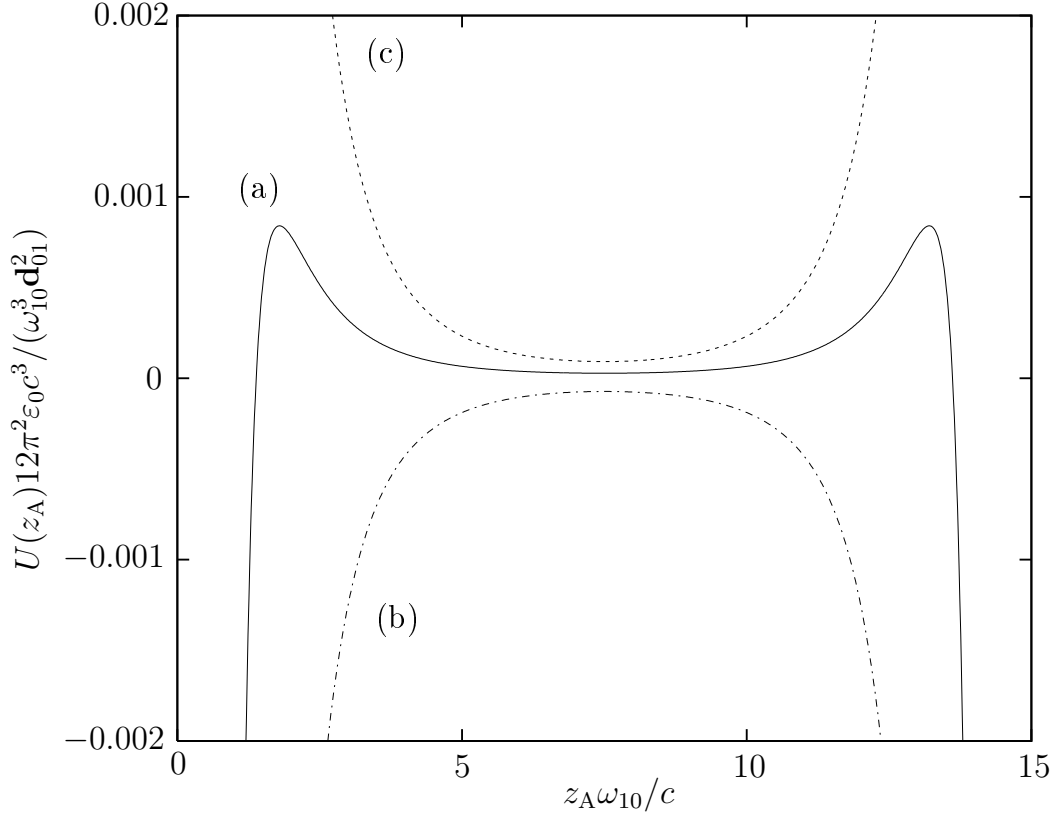}
\end{center}
\caption{
\label{fig5}
The potential of a ground-state two-level atom placed between two
(a) magneto-dielectric (all parameters as in Fig.~\ref{fig3}),
(b) purely dielectric [$\mu(\omega)$ $\!=$ $\!1$, other parameters as
in (a)], and (c) purely magnetic [$\varepsilon(\omega)$ $\!=$ $\!1$,
other parameters as in (a)] half spaces of separation $d$ $\!=$
$\!15c/\omega_{10}$, Eq.~(\ref{4.42}), is shown as a function of the
position of the atom.
}
\end{figure}%
%%%%%%%%%%%%%%%%%%%%%%%%%%%%%%%%%%%%%%%%%%%%%%%%%%%%%%%%%%%%%%%%%%%%%%
As expected, the potential is in general not the sum of the
potentials associated with the left and right plates separately,
as a comparison of Eq.~(\ref{4.42}) with Eq.~(\ref{4.31}) shows.
Clearly, the difference is due to the effect of multiple reflection
between the two plates which gives rise to the denominators
$D_\sigma$,
\begin{equation}
\label{4.43}
\frac{1}{D_\sigma}=\sum_{n=0}^\infty\big(r_{\sigma -}
 \me^{-bd}r_{\sigma +} \me^{-bd}\big)^n,
\end{equation}
recall Eq.~(\ref{3.33}).

The formation of a potential well is illustrated in Fig.~\ref{fig5}
where the potentials of an atom placed between purely dielectric and
purely magnetic plates, are also shown. It is seen that the attractive
(repulsive) potentials associated with each of two purely dielectric
(magnetic) plates combine to an infinite potential wall (well) at the
center of the cavity, while a potential well of finite depth can be
realized within the cavity in the case of two magneto-dielectric
plates of sufficiently large permeability.

%%%%%%%%%%%%%%%%%%%%%%%%%%%%%%%%%%%%%%%%%%%%%%%%%%%%%%%%%%%%%%%%%%%%%%

\subsubsection{Asymptotic power laws}
\label{sec4.1.4}

As we have seen, the dispersive interaction of
polarizable/magnetizable objects in their ground states can often be
given in terms of simple asymptotic power laws in the retarded and
non-retarded limits. Typical examples are given in Tab.~\ref{tab1}
where the asymptotic power laws for the dispersion force on an atom
interacting with another atom [Eqs.~(\ref{4.13}), (\ref{4.14}),
(\ref{4.19}) and (\ref{4.20})], a small sphere \cite{0113}, thin ring
\cite{0020}, a thin plate [Eqs.~(\ref{4.39})--(\ref{4.41})] and a
semi-infinite half space [Eqs.~(\ref{4.32})--(\ref{4.34})], and for
the force per unit area between two half spaces \cite{0134,0133}, are
shown.
%%%%%%%%%%%%%%%  T A B L E %%%%%%%%%%%%%%%%%%%%%%%%%%%%%%%%%%%%%%%%%%%
\begin{table}
\begin{center}
 \begin{tabular}{|c||c|c|c|c|}
\hline
 Distance $\rightarrow$
 &\multicolumn{2}{c|}{Retarded}&\multicolumn{2}{c|}{Nonretarded} \\
\hline
 Polarizability $\rightarrow$
 &$\mathrm{p}\leftrightarrow \mathrm{p}$
 &$\mathrm{p}\leftrightarrow \mathrm{m}$
 &$\mathrm{p}\leftrightarrow \mathrm{p}$
 &$\mathrm{p}\leftrightarrow \mathrm{m}$\\
\hline\hline
 \hspace*{1ex}(a)\hspace*{4.4cm}\begin{picture}(1,1)
 \put(-117,-17){\includegraphics[width=4cm]{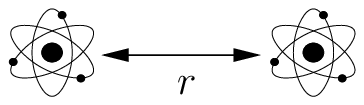}}
 \end{picture}
 &\parbox{5ex}{$$-\frac{1}{r^8}$$}
 &\parbox{5ex}{$$+\frac{1}{r^8}$$}
 &\parbox{5ex}{$$-\frac{1}{r^7}$$}
 &\parbox{5ex}{$$+\frac{1}{r^5}$$}\\
\hline
 \hspace*{1ex}(b)\hspace*{4.4cm}\begin{picture}(1,1)
 \put(-117,-17){\includegraphics[width=4cm]{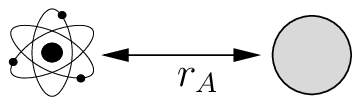}}
 \end{picture}
 &\parbox{5ex}{$$-\frac{1}{r_{A}^8}$$}
 &\parbox{5ex}{$$+\frac{1}{r_{A}^8}$$}
 &\parbox{5ex}{$$-\frac{1}{r_{A}^7}$$}
 &\parbox{5ex}{$$+\frac{1}{r_{A}^5}$$}\\
\hline
 \hspace*{1ex}(c)\hspace*{4.4cm}\begin{picture}(1,1)
 \put(-117,-17){\includegraphics[width=4cm]{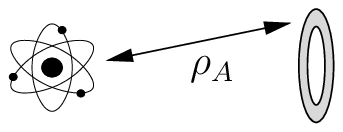}}
 \end{picture}
 &\parbox{5ex}{$$-\frac{1}{\rho_{A}^8}$$}
 &\parbox{5ex}{$$+\frac{1}{\rho_{A}^8}$$}
 &\parbox{5ex}{$$-\frac{1}{\rho_{A}^7}$$}
 &\parbox{5ex}{$$+\frac{1}{\rho_{A}^5}$$}\\
\hline
 \hspace*{1ex}(d)\hspace*{4.4cm}\begin{picture}(1,1)
 \put(-117,-17){\includegraphics[width=4cm]{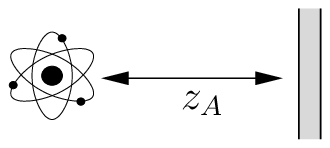}}
 \end{picture}
 &\parbox{5ex}{$$-\frac{1}{z_{A}^6}$$}
 &\parbox{5ex}{$$+\frac{1}{z_{A}^6}$$}
 &\parbox{5ex}{$$-\frac{1}{z_{A}^5}$$}
 &\parbox{5ex}{$$+\frac{1}{z_{A}^3}$$}\\
\hline
 \hspace*{1ex}(e)\hspace*{4.4cm}\begin{picture}(1,1)
 \put(-117,-17){\includegraphics[width=4cm]{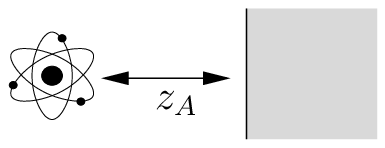}}
 \end{picture}
 &\parbox{5ex}{$$-\frac{1}{z_{A}^5}$$}
 &\parbox{5ex}{$$+\frac{1}{z_{A}^5}$$}
 &\parbox{5ex}{$$-\frac{1}{z_{A}^4}$$}
 &\parbox{5ex}{$$+\frac{1}{z_{A}^2}$$}\\
\hline
 \hspace*{1ex}(f)\hspace*{4.4cm}\begin{picture}(1,1)
 \put(-117,-17){\includegraphics[width=4cm]{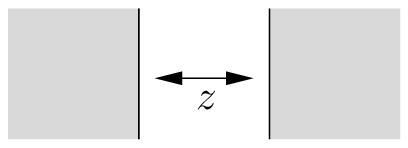}}
 \end{picture}
 &\parbox{5ex}{$$-\frac{1}{z^4}$$}
 &\parbox{5ex}{$$+\frac{1}{z^4}$$}
 &\parbox{5ex}{$$-\frac{1}{z^3}$$}
 &\parbox{5ex}{$$+\frac{1}{z}$$}\\
\hline
\end{tabular}
\end{center}
\vspace*{2ex}
\caption{%
Asymptotic power laws for the forces between (a) two atoms, (b) an
atom and a small sphere, (c) an atom and a thin ring, (d) an atom an a
thin plate, (e) an atom and a half space and (f) for the force per
unit area between two half spaces. In the table heading, $\mathrm{p}$
stands for a polarizable object and $\mathrm{m}$ for a magnetizable
one. The signs $+$ and $-$ denote repulsive and attractive forces,
respectively.}
\label{tab1}
\end{table}%
%%%%%%%%%%%%%%%%%%%%%%%%%%%%%%%%%%%%%%%%%%%%%%%%%%%%%%%%%%%%%%%%%%%%%%

Comparing the dispersion forces between objects of different shapes
and sizes, it is seen that the signs are always the same, while the
leading inverse powers are the same or changed by some global power
when moving from one row of the table to another. This can be
understood by assuming that the leading inverse power is determined by
the contribution to the force which results from the two-atom
interaction [row (a)] by pairwise summation. Obviously, integration
of two-atom forces over the (finite) volumes of a small sphere (b) or
a thin ring (c) does not change the respective power law, while
integration over an infinite volume lowers the leading inverse power
according to the number of infinite dimensions. So, the leading
inverse powers are lowered by two and three for the interaction of an
atom with a thin plate of infinite lateral extension (d) and a half
space (e), respectively. The power laws for the force between two half
spaces (f) can then be obtained from the atom--half-space force
(e) by integrating over the three infinite dimensions where
integration over $z$ lowers the leading inverse powers by one while
the trivial integrations over $x$ and $y$ yield an infinite force,
i.e., a finite force per unit area. It follows from the table that
many-atom interactions do not change the leading power laws resulting
from the summation of pairwise interactions, but only modify the
proportionality factors.

All dispersion forces in Tab.~\ref{tab1} are seen to be attractive
for two polarizable objects and repulsive for a polarizable object
interacting with a magnetizable one. It can further be noted that
in the retarded limit the forces decrease more rapidly with increasing
distance than might be expected from considering only the non-retarded
limit. Finally, the table shows that the retarded dispersion forces
between polarizable/polarizable and polarizable/magnetizable objects
follow the same power laws, while in the non-retarded limit the forces
between polarizable and magnetizable objects are weaker than those
between polarizable objects by two powers in the object separation.
This can be understood by regarding the forces as being due to the
electromagnetic field created by the first object interacting with the
second. While the electric and magnetic far fields created by an
oscillating electric dipole display the same distance dependence, the
electric near field (which can interact with a second polarizable
object) is stronger than the magnetic near field (which interacts with
a second magnetizable object) by one power in the object separation
(giving rise to a difference of two powers in second-order
perturbation theory).

%%%%%%%%%%%%%%%%%%%%%%%%%%%%%%%%%%%%%%%%%%%%%%%%%%%%%%%%%%%%%%%%%%%%%%

\subsection{Excited atoms}
\label{sec4.2}

In a first attempt, dispersion forces on atoms in excited energy
eigenstates can also be derived from perturbative energy shifts. A
straightforward generalization of the calculation outlined in
Sec.~\ref{sec4.1.1} to an atom prepared in an arbitrary energy
eigenstate $|m\rangle$ yields the potential \cite{0008,0012,0018}
\begin{equation}
\label{4.15x}
U_m(\vect{r}_{A})=U_m^\mathrm{or}(\vect{r}_{A})
 +U_m^\mathrm{r}(\vect{r}_{A})
\end{equation}
where
\begin{equation}
\label{4.16x}
U_m^\mathrm{or}(\vect{r}_{A})=\frac{\hbar\mu_0}{2\pi}
 \int_0^\infty\dif\xi\,\xi^2 \trace\bigl[\bm{\alpha}_m(\mi\xi)
 \sprod\ten{G}^{(1)}(\vect{r}_{A},\vect{r}_{A},\mi\xi)
 \bigr]
\end{equation}
and
\begin{equation}
\label{4.17x}
U_m^\mathrm{r}(\vect{r}_{A})
 =-\mu_0\sum_k \Theta(\omega_{mk})\omega_{mk}^2
 \vect{d}_{mk}\sprod\mathrm{Re}\,
 \ten{G}^{(1)}(\vect{r}_{A},\vect{r}_{A},\omega_{mk})
 \sprod\vect{d}_{km}
\end{equation}
[$\Theta(z)$, unit step function] are the off-resonant and resonant
contributions to the potential, respectively, and
\begin{align}
\label{4.18x}
\bm{\alpha}_m(\omega)
=\lim_{\epsilon\to 0}\frac{1}{\hbar}\sum_k\biggl[
 \frac{\vect{d}_{mk}\tprod\vect{d}_{km}}
 {\omega_{km}-\omega-\mi\epsilon}
 +\frac{\vect{d}_{km}\tprod\vect{d}_{mk}}
 {\omega_{km}+\omega+\mi\epsilon}
 \biggr]
\end{align}
is the atomic polarizability tensor. Equation~(\ref{4.15x}) [together
with Eqs.~(\ref{4.16x}) and (\ref{4.17x})] obviously reduces to the
ground-state result (\ref{4.6-1}) in the special case $m$ $\!=$ $\!0$.
Note that the resonant contribution vanishes in the ground state; it
is only present for an excited atom that can undergo real transitions.
Equations~(\ref{4.15x})--(\ref{4.17x}) which can also be obtained
by means of linear-response theory \cite{0235,0042}, have been
used to calculate the potential of an excited atom near a perfectly
conducting \cite{0235,0042}, dielectric \cite{0042}, birefringent
dielectric half space \cite{0045} and an electric cylinder
\cite{0442}.

The above mentioned approaches are time-independent and essentially
perturbative and inspection of Eqs.~(\ref{4.15x})--(\ref{4.18x})
reveals that the application of (static) perturbative methods to
excited atoms is problematic in several respects. Firstly, the
potential is determined by quantities that are attributed to the
unperturbed atomic transitions which do not take into account the
effect of line broadening, whereas in practice finite line widths are
observed which are known to strongly affect resonant transitions.
Secondly, the potential and hence also the force remains constant in
time; this is not very realistic for excited atoms which undergo
spontaneous decay with the allowed (dipole-) transitions being the
same as those entering the potential. And thirdly, perturbation theory
does not apply to the case of strong atom--field coupling. These
problems can be overcome by a dynamical approach to the calculation of
forces acting on excited atoms.

%%%%%%%%%%%%%%%%%%%%%%%%%%%%%%%%%%%%%%%%%%%%%%%%%%%%%%%%%%%%%%%%%%%%%%

\subsubsection{Dynamical approach}
\label{sec4.2.1}

Instead of deriving the dispersion force from an energy shift by some
means or other, we return to the origin of the force by starting from
the Lorentz force acting on an atom and calculating its expectation
value for a given initial state. In particular, when the
electromagnetic field is initially in its ground state, then this
expression yields the sought dispersion force which is genuinely
time-dependent for atoms initially prepared in excited states.

Summing the physical momenta $m_\alpha\dot{\hat{\vect{r}}}_\alpha$ of
the particles constituting the atom [as given by Eq.~(\ref{2.92})],
one obtains for the atom as a whole
\begin{equation}
\label{4.45}
m_{A}\dot{\hat{\vect{r}}}_{A}
=\hat{\vect{p}}_{A}
 +\int\dif^3 r\,\hat{\vect{P}}_\mathrm{at}(\vect{r})
 \vprod\hat{\vect{B}}(\vect{r}).
\end{equation}
Hence, the center-of-mass motion is governed by the Newton equation
\begin{equation}
\label{4.46}
m_{A}\ddot{\hat{\vect{r}}}_{A}
 =\hat{\vect{F}}_\mathrm{L}
\end{equation}
where, according to Eq.~(\ref{4.45}), the Lorentz force is given by
\begin{equation}
\label{4.47}
\hat{\vect{F}}_\mathrm{L}
 =\frac{\mi}{\hbar}\bigl[\hat{H},
 \hat{\vect{p}}_{A}\bigr]
 +\frac{\dif}{\dif t}
 \int\dif^3 r\,
 \hat{\vect{P}}_\mathrm{at}(\vect{r})
 \vprod\hat{\vect{B}}(\vect{r})
\end{equation}
with $\hat{\vect{P}}_\mathrm{at}(\vect{r})$ from Eq.~(\ref{2.64}).
The first term in Eq.~(\ref{4.47}) can be further evaluated by
recalling Eq.~(\ref{2.82}) and using the commutation relations
(\ref{2.60}). By making use of the identity
$\bm{\nabla}_{\!\!{A}}\tprod\hat{\vect{P}}_\mathrm{at}(\vect{r})$
$\!=$ $-\bm{\nabla}\tprod\hat{\vect{P}}_\mathrm{at}(\vect{r})$ [recall
Eq.~(\ref{2.64})], one can show that
\begin{equation}
\label{4.48}
 \frac{\mi}{\hbar}\biggl[
 \frac{1}{2\varepsilon_0}\int\dif^3r\,
 \hat{\vect{P}}^2_\mathrm{at}(\vect{r}),
 \hat{\vect{p}}_{A}\biggr]
 =\frac{1}{2\varepsilon_0}\int\dif^3r\,
 \bm{\nabla}\hat{\vect{P}}_\mathrm{at}^2(\vect{r})
 =\vect{0}
\end{equation}
and by recalling Eq.~(\ref{2.86}) and using the
definitions~(\ref{2.64}) and (\ref{2.65}), one derives
\begin{multline}
\label{4.49}
\frac{\mi}{\hbar}\biggl[
 \sum_\alpha\frac{1}{2 m_\alpha}
 \Bigl\{\hat{\vect{p}}_\alpha
 +\int\dif^3 r\,\hat{\bm{\Xi}}_\alpha(\vect{r})
 \vprod\hat{\vect{B}}(\vect{r})\Bigr\}^2,
 \hat{\vect{p}}_{A}\biggr]\\[.5ex]
=\bm{\nabla}_{\!\!{A}}\Bigl\{
 \int\dif^3r\,\bigl[\hat{\vect{M}}_\mathrm{at}(\vect{r})
 +\hat{\vect{P}}_\mathrm{at}(\vect{r})\vprod
 \dot{\hat{\vect{r}}}_{A}\bigr]
 \sprod\hat{\vect{B}}(\vect{r})\Bigr\}.
\end{multline}
Equations~(\ref{4.48}) and (\ref{4.49}) then imply that
Eq.~(\ref{4.47}) can be written as
\begin{align}
\label{4.50}
\hat{\vect{F}}_\mathrm{L}
=&\,\bm{\nabla}_{\!\!{A}}\Bigl\{\int\dif^3r\,
 \hat{\vect{P}}_\mathrm{at}(\vect{r})\sprod\hat{\vect{E}}(\vect{r})
 +\int\dif^3r\,\bigl[\hat{\vect{M}}_\mathrm{at}(\vect{r})
 +\hat{\vect{P}}_\mathrm{at}(\vect{r})\vprod
 \dot{\hat{\vect{r}}}_{A}\bigr]
 \sprod\hat{\vect{B}}(\vect{r})\Bigr\}
 \nonumber\\[.5ex]
&\,+\frac{\dif}{\dif t}\int\dif^3r\,
 \hat{\vect{P}}_\mathrm{at}(\vect{r})\vprod\hat{\vect{B}}(\vect{r}).
\end{align}
It should be mentioned that by using
Eqs.~(\ref{2.66c})--(\ref{2.66d2}) together with
the Maxwell equations (\ref{2.71}) and (\ref{2.73}), this equation
can be given in the equivalent form
\begin{equation}
\label{4.51}
\hat{\vect{F}}_\mathrm{L}
=\int\dif^3r\bigl[\hat{\rho}_\mathrm{at}(\vect{r})
 \hat{\vect{E}}(\vect{r})
 +\hat{\vect{j}}_\mathrm{at}(\vect{r})
 \vprod\hat{\vect{B}}(\vect{r})\bigr]
\end{equation}
which corresponds to the Eq.~(\ref{3.4}) used in Sec.~\ref{sec3} as a
starting point for calculating dispersion forces on
bodies.\footnote{Note that the field created by the atom only gives
rise to internal forces, so that one may equivalently write
Eq.~(\ref{4.51}) with the total fields
$\hat{\bm{\mathcal{E}}}(\vect{r})$ and
$\hat{\bm{\mathcal{B}}}(\vect{r})$ [Eq.~(\ref{2.69})] instead of
$\hat{\vect{E}}(\vect{r})$ and $\hat{\vect{B}}(\vect{r})$.}

{F}rom Eq.~(\ref{4.50}), the Lorentz force in long-wavelength
approximation can be obtained by performing a leading-order expansion
in the relative particle coordinates
$\hat{\overline{\vect{r}}}_\alpha$, resulting in
\begin{equation}
\label{4.52}
\hat{\vect{F}}_\mathrm{L}
=\bm{\nabla}_{\!\!{A}}\bigl[
 \hat{\vect{d}}\sprod\hat{\vect{E}}(\hat{\vect{r}}_{A})
 +\hat{\vect{m}}\sprod\hat{\vect{B}}(\hat{\vect{r}}_{A})
 +\hat{\vect{d}}\vprod\dot{\hat{\vect{r}}}_{A}
 \sprod\hat{\vect{B}}(\hat{\vect{r}}_{A})\bigr]
 +\frac{\dif}{\dif t}\bigl[\hat{\vect{d}}\vprod
 \hat{\vect{B}}(\hat{\vect{r}}_{A})\bigr],
\end{equation}
[recall Eqs.~(\ref{2.66}) and (\ref{2.66b})] where
\begin{align}
\label{4.53}
\frac{\dif}{\dif t}\bigl[\hat{\vect{d}}\vprod
 \hat{\vect{B}}(\hat{\vect{r}}_{A})\bigr]
=&\;\frac{\mi}{\hbar}\bigl[\hat{H},\hat{\vect{d}}\vprod
 \hat{\vect{B}}(\hat{\vect{r}}_{A})\bigr]
 =\dot{\hat{\vect{d}}}\vprod
 \hat{\vect{B}}(\hat{\vect{r}}_{A})
 +\hat{\vect{d}}\vprod\dot{\hat{\vect{B}}}(\vect{r})
 \big|_{\vect{r}=\hat{\vect{r}}_{A}}\nonumber\\[.5ex]
&+{\textstyle\frac{1}{2}}\hat{\vect{d}}\vprod
 \bigl[\bigl(\dot{\hat{\vect{r}}}_{A}
 \sprod\bm{\nabla}_{\!\!{A}}\bigr)
 \hat{\vect{B}}(\hat{\vect{r}}_{A})
 +\hat{\vect{B}}(\hat{\vect{r}}_{A})
 \bigl(\overleftarrow{\bm{\nabla}}_{\!\!{A}}\sprod
 \dot{\hat{\vect{r}}}_{A}\bigr)\bigr].
\end{align}
Discarding all terms proportional to $\dot{\hat{\vect{r}}}_{A}$ (which
are of the order $v/c$ and thus negligible for nonrelativistic
center-of-mass motion), as well as the contribution from the magnetic
interactions, Eq.~(\ref{4.52}) reduces to
\begin{equation}
\label{4.54}
\hat{\vect{F}}_\mathrm{L}
=\left\{\bm{\nabla}
 \left[\hat{\vect{d}}\sprod\hat{\vect{E}}(\vect{r})\right]
 +\frac{\dif}{\dif t}\left[\hat{\vect{d}}\vprod
 \hat{\vect{B}}(\vect{r})\right]
 \right\}_{\vect{r}=\hat{\vect{r}}_{A}},
\end{equation}
whose expectation value
\begin{equation}
\label{4.54-1}
\vect{F} =\langle\hat{\vect{F}}_\mathrm{L}\rangle
\end{equation}
quite generally provides a basis for calculating electromagnetic
forces on non-magnetic atoms, including dispersion forces. Needless to
say that Eq.~(\ref{4.54}) is valid regardless of the state the atom
and the body-assisted field are prepared in.

At this point we recall that, according to Eqs.~(\ref{2.24-1}) and
(\ref{2.31-1}), the electric and the magnetic induction fields are
expressed in terms of the dynamical variables
$\hat{\vect{f}}_\lambda(\vect{r},\omega)$ and
$\hat{\vect{f}}_\lambda^\dagger(\vect{r},\omega)$. It is not difficult
to prove that in electric-dipole approximation,
$\hat{\vect{f}}_\lambda(\vect{r},\omega)$ obeys the Heisenberg
equation of motion
\begin{equation}
\label{4.55}
\dot{\hat{\vect{f}}}_\lambda(\vect{r},\omega)
=\frac{\mi}{\hbar}\bigl[\hat{H},
 \hat{\vect{f}}_\lambda(\vect{r},\omega)\bigr]
 =-\mi\omega\hat{\vect{f}}_\lambda(\vect{r},\omega)
 +\frac{\mi}{\hbar}\,\hat{\vect{d}}\sprod
\ten{G}_\lambda^\ast(\hat{\vect{r}}_{A},\vect{r},\omega)
\end{equation}
[recall the Hamiltonian (\ref{2.82-1}) together with
Eqs.~(\ref{2.83}), (\ref{2.84}) and (\ref{2.90-1})],
whose formal solution reads
\begin{equation}
\label{4.56}
\hat{\vect{f}}_\lambda(\vect{r},\omega,t)
=\hat{\vect{f}}_{\lambda\mathrm{free}}(\vect{r},\omega,t)
 +\hat{\vect{f}}_{\lambda\mathrm{source}}(\vect{r},\omega,t)
\end{equation}
where
\begin{equation}
\label{4.57}
 \hat{\vect{f}}_{\lambda\mathrm{free}}(\vect{r},\omega,t)
 =\me^{-\mi\omega (t-t_0)}
 \hat{\vect{f}}_\lambda(\vect{r},\omega)
\end{equation}
and
\begin{equation}
\label{4.58}
\hat{\vect{f}}_{\lambda\mathrm{source}}(\vect{r},\omega,t)
=\frac{\mi}{\hbar}\int_{t_0}^t\dif\tau\,\me^{-\mi\omega(t-\tau)}
 \hat{\vect{d}}(\tau)\sprod\ten{G}_\lambda^\ast
 [\hat{\vect{r}}_{A}(\tau),\vect{r},\omega]
\end{equation}
($t_0$, initial time), respectively, determine the free-field parts
$\underline{\hat{\vect{E}}}_\mathrm{free}(\vect{r},\omega,t)$ and
$\underline{\hat{\vect{B}}}_\mathrm{free}(\vect{r},\omega,t)$
and the source-field parts
$\underline{\hat{\vect{E}}}_\mathrm{source}(\vect{r},\omega,t)$ and
$\underline{\hat{\vect{B}}}_\mathrm{source}(\vect{r},\omega,t)$
of the electric and the induction field in the $\omega$ domain.
Substitution of Eqs.~(\ref{4.56})--(\ref{4.58}) together with
Eqs.~(\ref{2.24-1}) and (\ref{2.31-1}) into Eq.~(\ref{4.54-1})
together with Eq.~(\ref{4.54}) and use of Eq.~(\ref{2.30b}) leads to
the following expression for the mean force \cite{0008,0018}:
\begin{equation}
\label{4.59}
\vect{F}(t)=\vect{F}_\mathrm{free}(t)+\vect{F}_\mathrm{source}(t)
\end{equation}
with
\begin{multline}
\label{4.60}
\vect{F}_\mathrm{free}(t)
= \int_0^\infty\dif\omega
 \biggl\{\bm{\nabla}
 \bigl\langle\hat{\vect{d}}(t)\sprod
 \underline{\hat{\vect{E}}}_\mathrm{free}(\vect{r},\omega,t)
 \bigr\rangle\\[.5ex]
+\frac{\dif}{\dif t}\bigl[\bigl\langle\hat{\vect{d}}(t)\vprod
 \underline{\hat{\vect{B}}}_\mathrm{free}(\vect{r},\omega,t)
 \bigr\rangle\bigr]
 \biggr\}_{\vect{r}=\hat{\vect{r}}_{A}(t)}+\mathrm{C.c.}
\end{multline}
and
\begin{equation}
\label{4.60b}
\vect{F}_\mathrm{source}(t)
 =\vect{F}_\mathrm{source}^\mathrm{el}(t)
+\vect{F}_\mathrm{source}^\mathrm{mag}(t)
\end{equation}
where the components
\begin{multline}
\label{4.61}
\vect{F}_\mathrm{source}^\mathrm{el}(t)
=\biggl\{\frac{\mi\mu_0}{\pi}\int_0^\infty\dif\omega\,\omega^2
 \int_{t_0}^t\dif\tau\,\me^{-\mi\omega(t-\tau)}\\[.5ex]
 \times\bm{\nabla}\bigl\langle\hat{\vect{d}}(t)\sprod
 \mathrm{Im}\,\ten{G}[\vect{r},\hat{\vect{r}}_{A}(\tau),\omega]\sprod
 \hat{\bf d}(\tau)\bigr\rangle
 \biggr\}_{\vect{r}=\hat{\vect{r}}_{A}(t)}+\mathrm{C.c.}
\end{multline}
and
\begin{multline}
\label{4.62}
\vect{F}_\mathrm{source}^\mathrm{mag}(t)
=\biggl\{\frac{\mu_0}{\pi}\int_{0}^\infty
 \dif\omega\,\omega\,
 \frac{\dif}{\dif t}\int_{t_0}^t\dif\tau\,\me^{-\mi\omega(t-\tau)}
 \\[.5ex]
\times\bigl\langle\hat{\vect{d}}(t)\vprod\bigl(\bm{\nabla}\vprod
 \mathrm{Im}\,\ten{G}[\vect{r},\hat{\vect{r}}_{A}(\tau),\omega]
 \bigr)\sprod\hat{\vect{d}}(\tau)\bigr\rangle
 \biggr\}_{\vect{r}=\hat{\vect{r}}_{A}(t)}+\mathrm{C.c.}
\end{multline}
are related to the source-field parts of the electric and the
induction field, respectively.

While Eq.~(\ref{4.59}) together with Eqs.~(\ref{4.60})--(\ref{4.62})
gives the force on a (non-magnetic) atom subject to an arbitrary
electromagnetic field, the pure dispersion force can be obtained by
considering the case where the (body-assisted) electromagnetic field
is initially prepared in the ground state $|\{0\}\rangle$ so that
\begin{equation}
\label{4.62-1}
\langle\{0\}|\cdots\underline{\hat{\vect{E}}}_\mathrm{free}
 (\vect{r},\omega,t)|\{0\}\rangle
= \langle\{0\}|\cdots\underline{\hat{\vect{B}}}_\mathrm{free}
 (\vect{r},\omega,t)|\{0\}\rangle=0
\end{equation}
[recall Eq.~(\ref{2.36})] which implies that
$\vect{F}_\mathrm{free}(t)$ $\!=$ $\!\vect{0}$. Hence,
Eq.~(\ref{4.59}) simply reduces to
\begin{equation}
\label{4.65}
\vect{F}(t)= \vect{F}_\mathrm{source}(t)
\end{equation}
in this case. In particular, for chosen atomic position,
$\hat{\vect{r}}_A$ may be regarded as a time-independent c-number
parameter [$\hat{\vect{r}}_A(t)\mapsto\vect{r}_A$], so that the
expectation values to be taken in Eqs.~(\ref{4.61}) and (\ref{4.62})
only refer to the internal state of the atom. It should be pointed out
that the concept is not restricted to the calculation of the mean
force but can be extended to higher-order force moments (for a
discussion of force fluctuations, see also Ref.~\cite{0072}).

%%%%%%%%%%%%%%%%%%%%%%%%%%%%%%%%%%%%%%%%%%%%%%%%%%%%%%%%%%%%%%%%%%%%%%

\subsubsection{Weak atom--field coupling}
\label{sec4.2.2}

The remaining task now consists in the determination of the
dipole--dipole correlation function
\begin{equation}
\label{4.66}
\left\langle\hat{\bf d}(t)\tprod\hat{\bf d}(\tau)\right\rangle
=\sum_{m,n}\sum_{m',n'}\vect{d}_{mn}\tprod\vect{d}_{m'n'}
\left\langle\hat{A}_{mn}(t)\hat{A}_{m'n'}(\tau)\right\rangle
\end{equation}
in Eqs.~(\ref{4.61}) and (\ref{4.62}) [$\hat{A}_{mn}$ $\!=$
$\!|m\rangle\langle n|$, recall Eq.~(\ref{2.84})]. To that end, the
problem of the internal atomic dynamics must be solved. Let first
consider the case of weak atom--field coupling where the Markov
approximation can by used to considerably simplify the problem.
Under the assumption that the relevant atomic transition frequencies
are well separated from one another so that diagonal and off-diagonal
density matrix elements evolve independently, application of the
quantum-regression theorem (see,~e.g., Ref.~\cite{0605}) yields the
familiar result
\begin{equation}
\label{4.67}
\left\langle\hat{A}_{mn}(t)\hat{A}_{m'n'}(\tau)\right\rangle
 =\delta_{nm'}\left\langle\hat{A}_{mn'}(\tau)\right\rangle
 \me^{\{\mi\tilde{\omega}_{mn}(\vect{r}_{A})
 -[\Gamma_m(\vect{r}_{A})
 +\Gamma_n(\vect{r}_{A})]/2\}(t-\tau)}
\end{equation}
($t$ $\!\ge$ $\!\tau$, $m$ $\!\neq$ $n$). Here,
\begin{equation}
\label{4.68}
\tilde{\omega}_{mn}(\vect{r}_{A})
=\omega_{mn}+\delta\omega_m(\vect{r}_{A})
 -\delta\omega_n(\vect{r}_{A})
\end{equation}
are the atomic transition frequencies including the position-dependent
energy-level shifts\footnote{The Lamb shifts observed in free space
are thought of as being already included in the frequencies
$\omega_{mn}$.}
\begin{gather}
\label{4.69}
\delta\omega_m(\vect{r}_{A})
 =\sum_k \delta\omega_m^k(\vect{r}_{A}),\\[.5ex]
\label{4.70}
\delta\omega_m^k(\vect{r}_{A})=\frac{\mu_0}{\pi\hbar}\,
 \mathcal{P}\int_0^\infty\dif\omega\,\omega^2\,
 \frac{\vect{d}_{km}\sprod\mathrm{Im}\,
 \ten{G}^{(1)}(\vect{r}_{A},\vect{r}_{A},\omega)\sprod
 \vect{d}_{mk}}{\tilde{\omega}_{mk}(\vect{r}_{A})-\omega}\,
\end{gather}
(cf.~also Refs.~\cite{0186,0439,0187}) which are due to the
interaction of the atom with the body-assisted electromagnetic field,
and similarly,
\begin{gather}
\label{4.71}
\Gamma_m(\vect{r}_{A})
 = \sum_k \Gamma_m^k(\vect{r}_{A}),\\[.5ex]
\label{4.72}
\Gamma_m^k(\vect{r}_{A})=\frac{2\mu_0}{\hbar}\,
 \Theta[\tilde{\omega}_{mk}(\vect{r}_{A})]
 \tilde{\omega}_{mk}^2(\vect{r}_{A})\vect{d}_{km}\sprod
 \mathrm{Im}\,\ten{G}[\vect{r}_{A},\vect{r}_{A},
 \tilde{\omega}_{mk}(\vect{r}_{A})]\sprod\vect{d}_{mk}
\end{gather}
are the position-dependent level widths. Note that the
position-dependent energy shifts $\hbar\delta\omega_m(\vect{r}_{A})$
as given by Eq.~(\ref{4.69}) together with Eq.~(\ref{4.70}) reduce to
those obtained by leading-order perturbation theory,
Eqs.~(\ref{4.15x})--(\ref{4.17x}), if the frequency shifts in the
denominator on the r.h.s. of Eq.~(\ref{4.70}) are ignored.

Substituting Eqs.~(\ref{4.66}) and (\ref{4.67}) into
Eqs.~(\ref{4.60b})--(\ref{4.62}), one can then show that the force on
an atom that is initially prepared in an arbitrary state can be
represented in the form \cite{0008,0012,0018}
\begin{equation}
\label{4.73}
\vect{F}(t)
 =\sum_{m,n}\sigma_{nm}(t)\vect{F}_{mn}(\vect{r}_{A})
\end{equation}
where the atomic density matrix elements $\sigma_{nm}(t)$ $\!=$
$\!\langle\hat{A}_{mn}(t)\rangle$ solve the intra-atomic master
equation together with the respective initial condition, and we have 
\begin{equation}
\label{4.74}
\vect{F}_{mn}(\vect{r}_{A})
=\vect{F}_{mn}^\mathrm{el,or}(\vect{r}_{A})
 +\vect{F}_{mn}^\mathrm{el,r}(\vect{r}_{A})
 +\vect{F}_{mn}^\mathrm{mag,or}(\vect{r}_{A})
 +\vect{F}_{mn}^\mathrm{mag,r}(\vect{r}_{A})
\end{equation}
with the various electric/magnetic, off-resonant/resonant force
components being given as follows:
\begin{multline}
\label{4.75}
\vect{F}_{mn}^\mathrm{el,or}(\vect{r}_{A})
 =-\frac{\hbar\mu_0}{2\pi}
 \int_0^\infty\dif\xi\,\xi^2\\[.5ex]
\times\bigl(\bm{\nabla}\,\trace\bigl\{
 [\bm{\alpha}_{mn}(\vect{r}_{A},\mi\xi)
 +\bm{\alpha}_{mn}(\vect{r}_{A},-\mi\xi)]
 \sprod\ten{G}^{(1)}(\vect{r}_{A},\vect{r},\mi\xi)
 \bigr\}\bigr)_{\vect{r}=\vect{r}_{A}},
\end{multline}
\begin{multline}
\label{4.76}
\vect{F}_{mn}^\mathrm{el,r}(\vect{r}_{A})
=\mu_0\sum_{k}\Theta({\tilde{\omega}}_{nk})
 \Omega^2_{mnk}(\vect{r}_{A})\\[.5ex]
\times\bigl\{\bm{\nabla}\vect{d}_{mk}\sprod
 \ten{G}^{(1)}[\vect{r},\vect{r}_{A},\Omega_{mnk}(\vect{r}_{A})]
 \sprod\vect{d}_{kn}\bigr\}_{\vect{r}=\vect{r}_{A}}
 +\mathrm{C.c.},
\end{multline}
\begin{multline}
\label{4.77}
\vect{F}_{mn}^\mathrm{mag,or}(\vect{r}_{A})
=\frac{\hbar\mu_0}{2\pi}
 \int_0^\infty\dif\xi\,\xi^2\,\trace\biggl\{\biggl[
 \frac{\tilde{\omega}_{mn}(\vect{r}_{A})}{\mi\xi}\,
 \bm{\alpha}_{mn}^\trans(\vect{r}_{A},\mi\xi)\\[.5ex]
-\,\frac{\tilde{\omega}_{mn}(\vect{r}_{A})}{\mi\xi}\,
 \bm{\alpha}_{mn}^\trans(\vect{r}_{A},-\mi\xi)\biggr]
 \vprod\bigl[\bm{\nabla}\vprod
 \ten{G}^{(1)}(\vect{r},\vect{r}_{A},\mi\xi)\bigr]
 \biggr\}_{\vect{r}=\vect{r}_{A}},
\end{multline}
\begin{multline}
\label{4.78}
\vect{F}_{mn}^\mathrm{mag,r}(\vect{r}_{A})
=\mu_0\sum_{k}\Theta({\tilde{\omega}_{nk}})
 \tilde{\omega}_{mn}(\vect{r}_{A})
 \Omega_{mnk}(\vect{r}_{A})\\[.5ex]
 \times\bigl(\vect{d}_{mk}\vprod\bigl\{
 \bm{\nabla}\vprod\ten{G}^{(1)}[\vect{r},\vect{r}_{A},
 \Omega_{mnk}(\vect{r}_{A})]\sprod\vect{d}_{kn}
 \bigr\}\bigr)_{\vect{r}=\vect{r}_{A}}+\mathrm{C.c.}
\end{multline}
Here, $\Omega_{mnk}(\vect{r}_{A})$ and
$\bm{\alpha}_{mn}(\vect{r}_{A},\omega)$, respectively, are the complex
atomic transition frequencies and the generalized polarizability
tensor:
\begin{equation}
\label{4.79}
\Omega_{mnk}(\vect{r}_{A})
=\tilde{\omega}_{nk}(\vect{r}_{A})
 +\mi[\Gamma_m(\vect{r}_{A})+\Gamma_k(\vect{r}_{A})]/2,
\end{equation}
\begin{equation}
\label{4.80}
\bm{\alpha}_{mn}(\vect{r}_{A},\omega)
=\frac{1}{\hbar}\sum_k\left[
 \frac{\vect{d}_{mk}\tprod\vect{d}_{kn}}
 {-\Omega_{mnk}(\vect{r}_{A})-\omega}
 +\frac{\vect{d}_{kn}\tprod\vect{d}_{mk}}
 {-\Omega_{nmk}^\ast(\vect{r}_{A})+\omega}
 \right].
\end{equation}

Equations~(\ref{4.73})--(\ref{4.78}) show that the force components
$\vect{F}_{mn}(\vect{r}_A)$ ($m$ $\!\neq$ $\!n$) associated with
(non-vanishing) off-diagonal elements of the atomic density matrix
contain contributions arising from the interaction of the atom with
both the electric and the magnetic field, where the magnetic force
components display a vector structure which is entirely different from
that of the electric ones. Since under the assumptions made, diagonal
and off-diagonal density matrix elements are not coupled to each other
so that
\begin{equation}
\label{4.83}
\sigma_{nm}(t)
=\me^{\{\mi\tilde{\omega}_{mn}(\vect{r}_{A})
 -[\Gamma_m(\vect{r}_{A})+\Gamma_n(\vect{r}_{A})]/2\}(t-t_0)}
 \sigma_{nm}(t_0)
 \end{equation}
($m$ $\!\neq$ $\!n$), force components associated with off-diagonal
density-matrix elements can only be observed if the atom is initially
prepared in an at least partially coherent superposition of energy
eigenstates. Accordingly, if the atom is initially prepared in an
incoherent superposition of energy eigenstates, then only force
components $\vect{F}_{mm}(\vect{r}_A)$ which are associated with
diagonal density-matrix elements and which are electrical by their
nature,
\begin{equation}
\label{4.80-1}
\vect{F}_{mm}(\vect{r}_A)
 =\vect{F}_{mm}^\mathrm{el,or}(\vect{r}_A)
 +\vect{F}_{mm}^\mathrm{el,r}(\vect{r}_A)
 \equiv\vect{F}_{mm}^\mathrm{or}(\vect{r}_A)
 +\vect{F}_{mm}^\mathrm{r}(\vect{r}_A),
\end{equation}
are observed, with the density matrix elements obeying the balance
equations
\begin{equation}
\label{4.84}
\dot{\sigma}_{mm}(t)
=-\Gamma_m(\vect{r}_{A})\sigma_{mm}(t)
 +\sum_k\Gamma_k^m(\vect{r}_{A})\sigma_{kk}(t).
\end{equation}
%
%%%%%%%%%%%%%%%  F I G U R E %%%%%%%%%%%%%%%%%%%%%%%%%%%%%%%%%%%%%%%%%
\begin{figure}[!t!]
\begin{center}
\includegraphics[width=\linewidth]{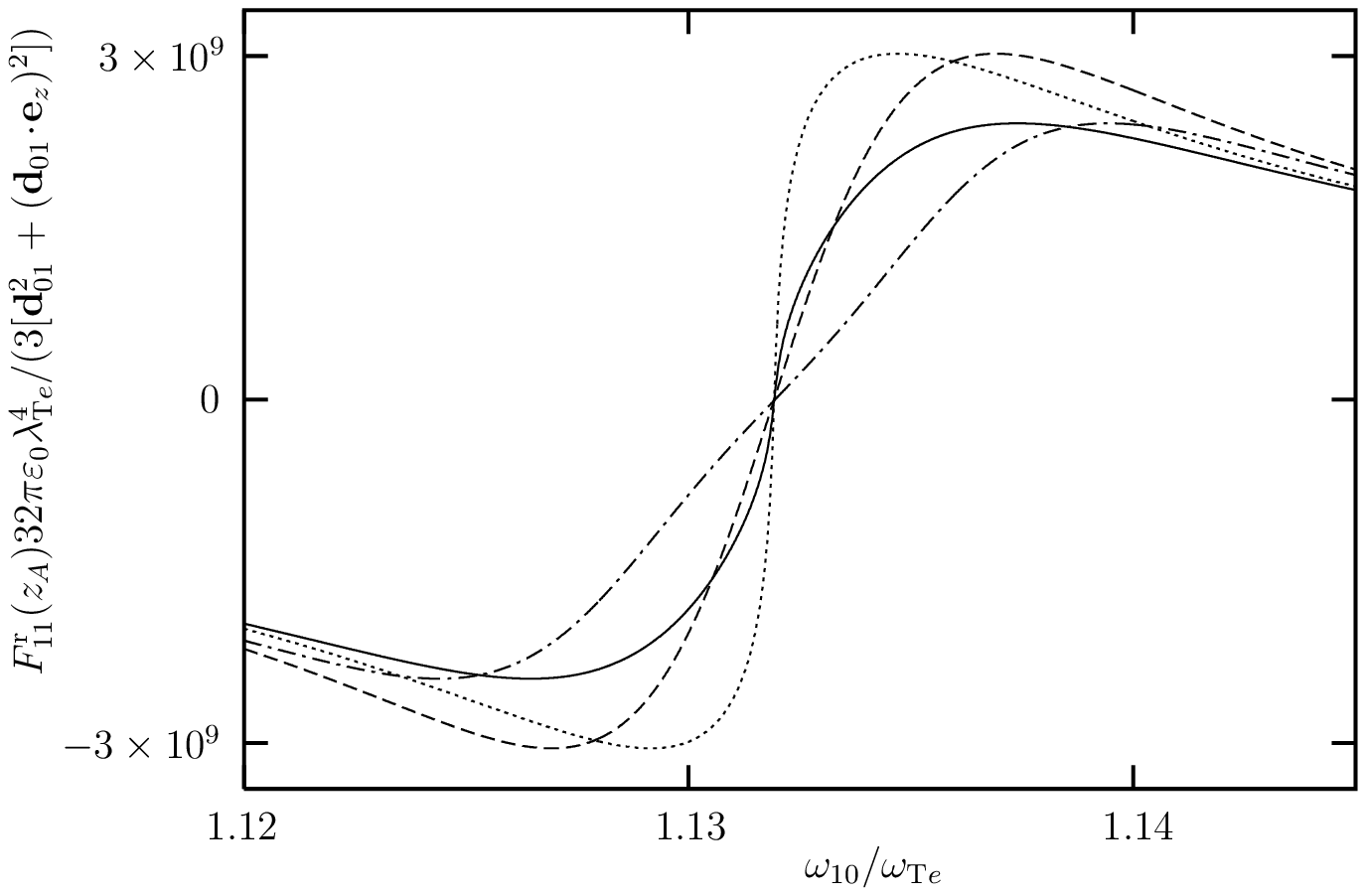}
\end{center}
\caption{
\label{fig6}
The resonant component of the force on a two-level atom in the upper
state placed in front of a dielectric half space, Eq.~(\ref{4.87}), is
shown as a function of the unperturbed transition frequency
$\omega_{10}$ (solid line)
($\omega_{\mathrm{P}e}/\omega_{\mathrm{T}e}$ $\!=$ $\!0.75$,
$\gamma_{e}/\omega_{\mathrm{T}e}$ $\!=$ $\!0.01$,
$\omega_{\mathrm{T}e}^2[\vect{d}_{01}^2$
$\!+$ $\!(\vect{d}_{01}\sprod\vect{e}_z)^2]/3
\pi\hbar\varepsilon_0 c^3$ $\!=$
$\!10^{-7}$,
$z_{A}/\lambda_{\mathrm{T}e}$ $\!=$ $\!0.0075$,
$\lambda_{\mathrm{T}e}$ $\!=$ $\!2\pi c/\omega_{\mathrm{T}e}$).
For comparison, both the perturbative result as obtained from
Eq.~(\ref{4.17x}) (dashed lines) and the separate effects of level
shifting (dotted lines) and level broadening (dash-dotted lines), are
shown.
}
\end{figure}%
%%%%%%%%%%%%%%%%%%%%%%%%%%%%%%%%%%%%%%%%%%%%%%%%%%%%%%%%%%%%%%%%%%%%%%
In particular, when the level shifts and broadenings are neglected,
then the force components $\vect{F}_{mm}^\mathrm{or}(\vect{r}_A)$ and
$\vect{F}_{mm}^\mathrm{r}(\vect{r}_A)$ as follow from
Eqs.~(\ref{4.75}) and (\ref{4.76}) obviously reduce to those that are
obtained from the perturbative potential (\ref{4.15x})--(\ref{4.17x})
by means of Eq.~(\ref{4.3}). Note that the gradient in
Eqs.~(\ref{4.75}) and (\ref{4.76}) acts only on the Green tensor and
not on the additional position-dependent quantities, so that this
result cannot be derived from a potential in the usual way. Since the
force components associated with excited-state density matrix elements
are transient, they are only observable on time scales of the order of
magnitude of the respective decay times $\Gamma_m^{-1}(\vect{r}_A)$
which are known to sensitively depend on the atomic position
\cite{0605}. Needless to say that the force $\vect{F}(t)$ that acts on
an initially excited atom approaches the ground-state force
$\vect{F}_{00}(\vect{r}_A)$ after sufficiently long times,
$\lim_{t\to\infty}\langle\vect{F}(t)\rangle$ $\!=$
$\vect{F}_{00}(\vect{r}_{A})$.

In order to illustrate the effect of the body-induced level shifting
and broadening on the force, let us consider a two-level atom which is
situated at distance $z_{A}$ very close to a dielectric half space. By
means of the respective Green tensor (App.~\ref{appA}), it turns out
that in the non-retarded limit the shift and width of the transition
frequency are determined by
\begin{equation}
\label{4.85}
\delta\omega(z_{A})=\delta\omega_1(z_{A})-\delta\omega_0(z_{A})
=-\frac{\vect{d}_{01}^2+(\vect{d}_{01}\sprod\vect{e}_z)^2}
 {32\pi\hbar\varepsilon_0 z_{A}^3}\,
\frac{|\varepsilon[\omega_{10}+\delta\omega(z_{A})]|^2-1}
{|\varepsilon[\omega_{10}+\delta\omega(z_{A})]+1|^2}
\end{equation}
and
\begin{equation}
\label{4.86}
\Gamma(z_{A})=\Gamma_1(z_{A})
=\frac{\vect{d}_{01}^2+(\vect{d}_{01}\sprod\vect{e}_z)^2}
 {8\pi\hbar\varepsilon_0 z_{A}^3}\,
\frac{\mathrm{Im}\,\varepsilon[\omega_{10}+\delta\omega(z_{A})]}
{|\varepsilon[\omega_{10}+\delta\omega(z_{A})]+1|^2}\,,
\end{equation}
respectively, where the transition-dipole matrix element has
been assumed to be real and the (small) off-resonant contribution to
the frequency shift has been omitted. Note that due to the appearance
of the frequency shift on the r.h.s. of Eq.~(\ref{4.85}), this
equation determines the shift only implicitly. The dominant
contribution to the force on the atom in the upper state is the
resonant one, i.e., $\vect{F}_{11}(\vect{r}_A)$ $\!\simeq$
$\!\vect{F}_{11}^\mathrm{r}(\vect{r}_A)$. Substituting the half-space
Green tensor (App.~\ref{appA}) into Eqs.~(\ref{4.76}) and
(\ref{4.80-1}), one can show that
[$\vect{F}_{11}^\mathrm{r}(\vect{r}_A)$ $\!=$
$\!F_{11}^\mathrm{r}(z_A)\vect{e}_z$]
\cite{0008,0012,0018}
\begin{equation}
\label{4.87}
F_{11}^\mathrm{r}(z_{A})
=-\frac{3[\vect{d}_{01}^2+(\vect{d}_{01}\sprod\vect{e}_z)^2]}
 {32\pi\hbar\varepsilon_0 z_{A}^4}\,
\frac{|\varepsilon[\Omega_{110}(z_{A})]|^2-1}
{|\varepsilon[\Omega_{110}(z_{A})]+1|^2}
\end{equation}
where, according to Eq.~(\ref{4.79}),
\begin{equation}
\label{4.88}
\Omega_{110}(z_{A})=\tilde{\omega}_{10}(z_{A})
 +\mi\Gamma(z_{A})/2
 =\omega_{10}+\delta\omega(z_{A})
 +\mi\Gamma(z_{A})/2.
\end{equation}
In particular, for a medium whose permittivity is of Drude--Lorentz
type, Eq.~(\ref{4.35}) leads to ($\gamma_e,\Gamma$ $\!\ll$
$\!\omega_{\mathrm{T}e}$)
\begin{equation}
\label{4.89}
\varepsilon[\Omega_{110}(z_{A})]
=1+\frac{\omega_{\mathrm{P}e}^2}{\omega_{\mathrm{T}e}^2
-\tilde{\omega}_{10}^2(z_{A})
-\mi [\Gamma(z_{A})+\gamma_e]\tilde{\omega}_{10}(z_{A})}\,,
\end{equation}
showing that the absorption parameter of the half-space medium,
$\gamma_e$, is replaced with the total absorption parameter, i.e., the
sum of $\gamma_e$ and the spon\-tan\-eous-decay constant
$\Gamma(z_{A})$ of the atom. Figure~\ref{fig6} displays the resonant
component of the force on a two-level atom in the upper state placed
near a (single-resonance) dielectric half space as a function of the
unperturbed transition frequency $\omega_{10}$.
%%%%%%%%%%%%%%%  F I G U R E %%%%%%%%%%%%%%%%%%%%%%%%%%%%%%%%%%%%%%%%%
\begin{figure}[!t!]
\begin{center}
\includegraphics[width=\linewidth]{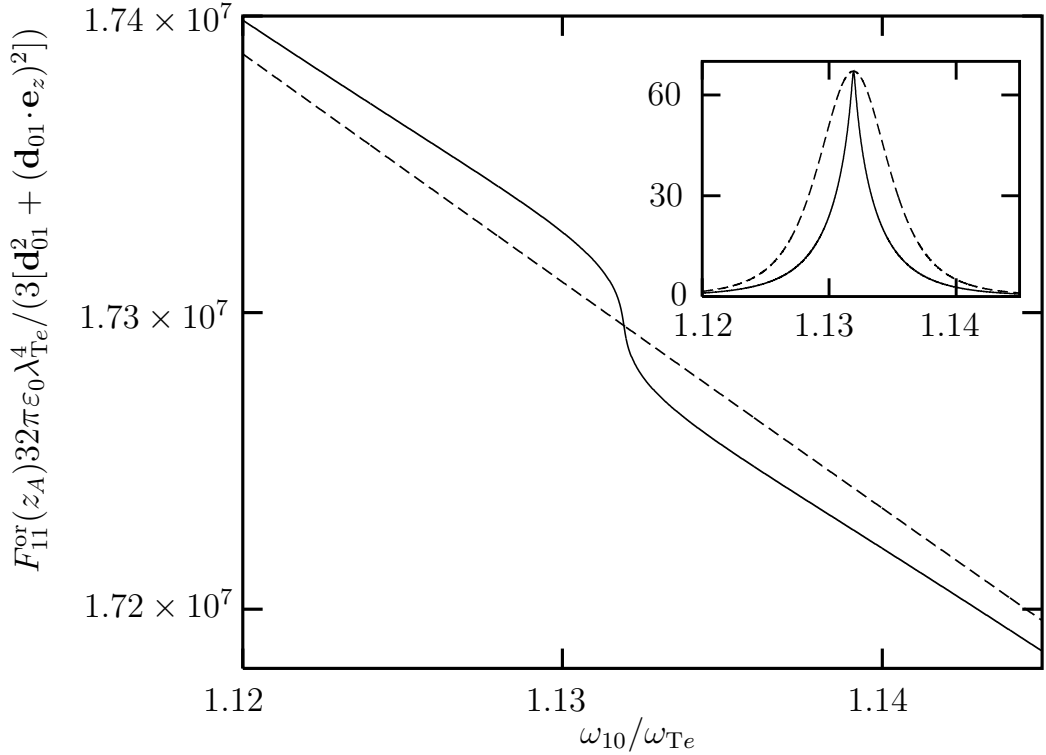}
\end{center}
\caption{
\label{fig7}
The off-resonant component of the force on a two-level atom in the
upper state placed in front of a dielectric half space,
Eq.~(\ref{4.90}), is shown as a function of the unperturbed transition
frequency (solid line), the parameters being the same as in
Fig.~\ref{fig6}. For comparison, the perturbative result is also shown
(dashed lines). The inset displays the difference between the force
with and without consideration of level broadening (solid lines). In
addition, the same difference is displayed when the level shifts are
ignored (dashed lines).
}
\end{figure}%
%%%%%%%%%%%%%%%%%%%%%%%%%%%%%%%%%%%%%%%%%%%%%%%%%%%%%%%%%%%%%%%%%%%%%%
It is seen that in the vicinity of the (surface-plasmon) frequency
\mbox{$\omega_\mathrm{S}$ $\!=$
$\!\sqrt{\omega_{\mathrm{T}e}^2+\omega_{\mathrm{P}e}^2/2}$},
an enhanced force is observed which is attractive (repulsive) for red
(blue) detuned atomic transition frequencies $\omega_{10}$ $\!<$
$\!\omega_\mathrm{S}$ ($\omega_{10}$ $\!>$ $\!\omega_\mathrm{S}$)---a
result already known from perturbation theory \cite{0042}. However, it
is also seen that due to body-induced level shifting and broadening
the absolute value of the force can be noticeably reduced.
Interestingly, the positions of the extrema of the force remain nearly
unchanged, because level shifting and broadening give rise to
competing effects that almost cancel.

The calculation of the off-resonant component of the force,
Eq.~(\ref{4.80-1}) together with Eq.~(\ref{4.75}), leads to
[$\vect{F}_{11}^\mathrm{or}(\vect{r}_A)$ $\!=$
$F_{11}^\mathrm{or}(z_A)\vect{e}_z$]
\begin{multline}
\label{4.90}
F_{11}^\mathrm{or}(z_{A})
=\frac{3[\vect{d}_{01}^2+(\vect{d}_{01}\sprod\vect{e}_z)^2]}
 {32\pi^2\hbar\varepsilon_0 z_{A}^4}
 \int_0^\infty\dif\xi\,
 \frac{\varepsilon(\mi\xi)-1}{\varepsilon(\mi\xi)+1}\\[.5ex]
\times\frac{\tilde{\omega}_{10}(z_{A})}
 {\tilde{\omega}_{10}^2(z_{A})+[\xi+\Gamma(z_{A})/2]^2}\,
 \frac{\tilde{\omega}_{10}^2(z_{A})+\xi^2+\Gamma^2(z_{A})/4}
 {\tilde{\omega}_{10}^2(z_{A})+[\xi-\Gamma(z_{A})/2]^2}\,.
\end{multline}
Equation~(\ref{4.90}) reveals that the off-resonant component of the
force is only weakly influenced by the level broadening [the
leading-order dependence being $O(\Gamma^2)$] which is in agreement
with the physical requirement that the virtual emission and absorption
processes governing the off-resonant component should be only weakly
affected by decay-induced broadening. Formally, the absence of a
linear-order term $O(\Gamma)$ is due to the fact that the atomic
polarizability (\ref{4.80}) enters the off-resonant force
components (\ref{4.75}) only in the combination
$[\bm{\alpha}_{mm}(\vect{r}_{A},\mi\xi)
+\bm{\alpha}_{mm}(\vect{r}_{A},-\mi\xi)]$, which could not have been
anticipated from the perturbative result (\ref{4.16x}) [where in
fact, $\bm{\alpha}_{m}(\vect{r}_{A},\mi\xi)$ and
$\bm{\alpha}_{m}(\vect{r}_{A},-\mi\xi)$ coincide, recall
Eq.~(\ref{4.18x})]. The effects of level shifting and broadening are
illustrated in Fig.~\ref{fig7}. Compared to the perturbative result,
the frequency shift has the effect of raising or lowering the force
for $\omega_{10}$ $\!<$ $\!\omega_\mathrm{S}$ or $\omega_{10}$ $\!>$
$\!\omega_\mathrm{S}$, respectively, whereas the effect of broadening
is not visible in the curves. Only by plotting the difference between
the results with and without broadening, a slight reduction of the
force becomes visible in the vicinity of $\omega_\mathrm{S}$ where
$\Gamma$ is largest. Since this behavior is generally typical of
off-resonant components, the perturbative result may be regarded as a
good approximation for the forces on ground-state atoms where no
resonant components are present.

%%%%%%%%%%%%%%%%%%%%%%%%%%%%%%%%%%%%%%%%%%%%%%%%%%%%%%%%%%%%%%%%%%%%%%

\subsubsection{Strong atom--field coupling}
\label{sec4.2.3}

Strong atom--field coupling may occur if an initially excited atom
interacts resonantly with a sharply peaked (quasi-)mode of a
body-assisted field, as it is observed in cavity-like systems.
In this case, the atom--field dynamics can no longer be described
within the Markov approximation. Typically, a single atomic transition
is in resonance with such a mode, so the (resonant part of the)
dispersion force can be studied, to a good approximation, by employing
the two-level model in rotating-wave approximation with respect to the
interaction Hamiltonian (\ref{2.90-1}) (see, e.g., Ref.~\cite{0605}).
To be more specific, let us consider the interaction of a two-level
atom initially prepared in the upper state $|1\rangle$ with the
body-assisted electromagnetic field in the ground state
$|\{0\}\rangle$ and calculate the electric part of the resonant
component of the force acting on the atom, i.e., in the
Schr\"{o}dinger picture,
\begin{equation}
\label{4.91}
\vect{F}(t)
\simeq\langle\psi(t)|\bigl\{\bm{\nabla}
 \bigl[\hat{\vect{d}}\sprod\hat{\vect{E}}(\vect{r})
 \bigr]\bigr\}_{\vect{r}=\vect{r}_{A}}|\psi(t)\rangle,
\end{equation}
with the state vector $|\psi(t)\rangle$ [$|\psi(t\!=\!t_0)\rangle$
$\!=$ $\!|\{0\}\rangle|1\rangle$] being represented in the form
\begin{multline}
\label{4.92}
|\psi(t)\rangle=\psi_1(t)|\{0\}\rangle|1\rangle\\[.5ex]
 +\sum_{\lambda={e},{m}}
 \int\dif^3r\int_0^\infty\dif\omega\,\frac{\psi_0(\omega,t)}
 {\hbar g(\vect{r}_{A},\omega)}\,\vect{d}_{01}\sprod
 \ten{G}^\ast_\lambda(\vect{r}_{A},\vect{r},\omega)\sprod
 |\vect{1}(\vect{r}_{A},\omega)\rangle|0\rangle.
\end{multline}
It is normalized to unity provided that
\begin{equation}
\label{4.93}
|\psi_1(t)|^2+\int_0^\infty\dif\omega\,|\psi_0(\omega,t)|^2=1
\end{equation}
and
\begin{equation}
\label{4.94}
g^2(\vect{r}_{A},\omega)
 =\frac{\mu_0}{\pi\hbar}\,\omega^2
 \vect{d}_{10}\sprod
 \mathrm{Im}\ten{G}(\vect{r}_{A},\vect{r}_{A},\omega)
 \sprod\vect{d}_{01}.
\end{equation}
Substituting Eq.~(\ref{4.92}) into the Schr\"{o}dinger equation
\begin{equation}
\label{4.95}
\mi\hbar\frac{\partial}{\partial t}|\psi(t)\rangle
=\hat{H}|\psi(t)\rangle,
\end{equation}
with $\hat{H}$ being given according to Eq.~(\ref{2.82-1}) together
with Eqs.~(\ref{2.83}), (\ref{2.84}) and (\ref{2.90-1}), one obtains
the following coupled differential equations for $\psi_1(t)$ and
$\psi_0(\omega,t)$:
\begin{align}
\label{4.96}
&\dot{\psi}_1(t)=-\frac{\mi}{\hbar}\,E_1\psi_1(t)
 +\mi\int_0^\infty\dif\omega\,g(\vect{r}_{A},\omega)
 \psi_0(\omega,t),\\[.5ex]
\label{4.97}
&\dot{\psi}_0(\omega,t)=-\frac{\mi}{\hbar}\,(E_0+\hbar\omega)
 \psi_0(\omega,t)+\mi g(\vect{r}_{A},\omega)\psi_1(t).
\end{align}
Equation~(\ref{4.97}) together with the initial condition
$\psi_0(\omega,t\!=\!t_0)$ $\!=$ $\!0$ can be formally integrated in a
straightforward way. Inserting the result into Eqs.~(\ref{4.92}) and
(\ref{4.96}) then yields
\begin{multline}
\label{4.98}
|\psi(t)\rangle
 =\psi_1(t)|\{0\}\rangle|1\rangle
 +\frac{\mi}{\hbar}\sum_{\lambda={e},{m}}\int\dif^3r
 \int_0^\infty\dif\omega
 \int_{t_0}^t\dif\tau\,
 \me^{-\mi(E_0/\hbar+\omega)(t-\tau)}\psi_1(\tau)\\[1ex]
 \times\vect{d}_{01}\sprod
 \ten{G}^\ast_\lambda(\vect{r}_{A},\vect{r},\omega)\sprod
 |\vect{1}(\vect{r}_{A},\omega)\rangle|0\rangle
\end{multline}
and
\begin{equation}
\label{4.99}
\dot{\psi}_1(t)=-\mi\,\frac{E_1}{\hbar}\,\psi_1(t)
 -\int_0^\infty\dif\omega\,g^2(\vect{r}_{A},\omega)
 \int_{t_0}^t\dif\tau\,\me^{-\mi(E_0/\hbar+\omega)(t-\tau)}
 \psi_1(\tau),
\end{equation}
respectively. Combining Eqs.~(\ref{4.91}) and (\ref{4.98}) and making
use of the integral relation (\ref{2.30b}), one finds that
\begin{multline}
\label{4.100}
\vect{F}(t)
=\frac{\mi\mu_0}{\pi}\int_0^\infty\dif\omega\,\omega^2
 \left\{\bm{\nabla}\left[\vect{d}_{10}\sprod
 \mathrm{Im}\,\ten{G}^{(1)}(\vect{r},\vect{r}_{A},\omega)
 \sprod\vect{d}_{01}\right]\right\}_{\vect{r}=\vect{r}_{A}}
 \\[.5ex]
\times\int_{t_0}^t\dif\tau\,\psi_1^\ast(t)\psi_1(\tau)
 \me^{-\mi(E_0/\hbar+\omega)(t-\tau)}
 +\mathrm{C.c.}
\end{multline}

Since so far nothing has been said about the strength of the
atom--field coupling, Eq.~(\ref{4.100}) gives the electric part
of the resonant component of the dispersion force on a two-level atom
which is initially prepared in the upper state for arbitrary coupling
strengths. Let us now approximate that part of the excitation spectrum
of the body-assisted electromagnetic field which may give rise to
strong atom--field coupling in the resonant transition by a quasi-mode
(labeled by $\nu$) of Lorentzian shape,
\begin{equation}
\label{4.101}
g^2(\vect{r}_{A},\omega)=g^2(\vect{r}_{A},\omega_\nu)\,
 \frac{\gamma_\nu^2/4}{(\omega-\omega_\nu)^2+\gamma_\nu^2/4} +
 g'{^2}(\vect{r}_{A},\omega)
\end{equation}
and assume that the effect of the residual part of the field which is
described by the term $g'^{2}(\vect{r}_{A},\omega)$ is weakly coupled
to the atom, so that it can be treated in the Markov approximation.
{F}rom Eq.~(\ref{4.99}) it then follows that \cite{0719}
\begin{equation}
\label{4.103}
\psi_1(t)=\me^{[-\mi E_1/\hbar-\mi\delta\omega'_1(\vect{r}_{A})
 -\Gamma_1'(\vect{r}_{A})/2](t-t_0)}\phi_1(t)
\end{equation}
where $\phi_1(t)$ is the solution to the differential equation
\begin{equation}
\label{4.107}
\ddot{\phi}_1(t)
 +\left\{\mi\Delta\omega(\vect{r}_{A})+\left[\gamma_\nu
 -\Gamma_1'(\vect{r}_{A})\right]/2\right\}
 \dot{\phi}_1(t)
 +{\textstyle\frac{1}{4}}\Omega_\mathrm{R}^2(\vect{r}_{A})
 \phi_1(t)=0
\end{equation}
together with the initial conditions $\phi_1(t\!=\!t_0)$ $\!=$ $\!1$,
$\dot{\phi}_1(t\!=\!t_0)$ $\!=$ $\!0$. Here,
\mbox{$\Delta\omega(\vect{r}_{A})$ $\!=$
$\!\omega_\nu-\tilde{\omega}'_{10}(\vect{r}_{A})$} and
$\Omega_\mathrm{R}(\vect{r}_{A})$ $\!=$ $\!\sqrt{2\pi\gamma_\nu
g^2(\vect{r}_{A},\omega_\nu)}$, respectively, are the detuning and the
vacuum Rabi frequency and
\begin{equation}
\label{4.104}
\delta\omega_1'(\vect{r}_{A})
=\delta\omega_1(\vect{r}_{A})
 +\frac{\Omega_\mathrm{R}^2(\vect{r}_{A})}{4}\,
 \frac{\Delta\omega(\vect{r}_{A})}
 {[\Delta\omega(\vect{r}_{A})]^2+\gamma_\nu^2/4}
\end{equation}
and
\begin{equation}
\label{4.105}
\Gamma_1'(\vect{r}_{A})=\Gamma_1(\vect{r}_{A})
 -\frac{\Omega_\mathrm{R}^2(\vect{r}_{A})}{4}\,
 \frac{\gamma_\nu}{[\Delta\omega(\vect{r}_{A})]^2
 +\gamma_\nu^2/4}\,,
\end{equation}
respectively, are the shift and width of the upper level associated
with the residual part of the field where
$\delta\omega_1(\vect{r}_{A})$ and $\Gamma_1(\vect{r}_{A})$,
are defined according to Eqs.~(\ref{4.69})--(\ref{4.72}) with the
shifted transition frequency being given by\footnote{Note that
contrary to Eq.~(\ref{4.68}), the ground-state shift is absent here as
a consequence of the rotating-wave approximation.}
\begin{equation}
\label{4.105-1}
\tilde{\omega}'_{10}(\vect{r}_{A})
 =\omega_{10}+\delta\omega_1'(\vect{r}_{A})
\end{equation}
in place of Eq.~(\ref{4.68}). Equation (\ref{4.107}) can easily be
solved to obtain
\begin{equation}
\label{4.108}
\phi_1(t)=c_+(\vect{r}_{A})\me^{\Omega_+(\vect{r}_{A})(t-t_0)}
 +c_-(\vect{r}_{A})\me^{\Omega_-(\vect{r}_{A})(t-t_0)}
\end{equation}
where
\begin{equation}
\label{4.109}
c_{\pm}(\vect{r}_{A})=\frac{\Omega_{\mp}(\vect{r}_{A})}
 {\Omega_{\mp}(\vect{r}_{A})-\Omega_{\pm}(\vect{r}_{A})}
\end{equation}
and
\begin{align}
\label{4.110}
\Omega_\pm(\vect{r}_{A})
 =&-{\textstyle\frac{1}{2}}\bigl\{
 \mi\Delta\omega(\vect{r}_{A})\!+\![\gamma_\nu\!
 -\!\Gamma_1'(\vect{r}_{A})]/2\bigr\}
 \nonumber\\[.5ex]
&\mp{\textstyle\frac{1}{2}}
 \sqrt{\bigl\{\mi\Delta\omega(\vect{r}_{A})\!+\![\gamma_\nu\!
 -\!\Gamma_1'(\vect{r}_{A})]/2\bigr\}^2
 -\Omega_\mathrm{R}^2(\vect{r}_{A})}\,.
\end{align}
Combination of Eqs.~(\ref{4.100}), (\ref{4.103}) and (\ref{4.108})
then leads to the sought force \cite{0719}:
\begin{equation}
\label{4.111}
\vect{F}(t)
=\frac{\mu_0}{\pi}\int_0^\infty\dif\omega\,\omega^2
 s(\vect{r}_{A},\omega,t-t_0)\Bigl\{\bm{\nabla}\bigl[\vect{d}_{10}
 \sprod\mathrm{Im}\,\ten{G}^{(1)}(\vect{r},\vect{r}_{A},\omega)
 \sprod\vect{d}_{01}\bigr]\Bigr\}_{\vect{r}=\vect{r}_{A}}\!\!\!
 +\mathrm{C.c.}
\end{equation}
with
\begin{multline}
\label{4.112}
s(\vect{r}_{A},\omega,t)\\[.5ex]
=|c_+(\vect{r}_{A})|^2\,
 \frac{\me^{[-\Gamma_1'(\vect{r}_{A})+\Omega_+^\ast(\vect{r}_{A})
 +\Omega_+(\vect{r}_{A})]t}
 -\me^{\{\mi[\tilde{\omega}'_{10}(\vect{r}_{A})-\omega]
 -\Gamma_1'(\vect{r}_{A})/2+\Omega_+^\ast(\vect{r}_{A})\}t}}
 {\omega-\tilde{\omega}'_{10}(\vect{r}_{A})
 +\mi\Gamma_1'(\vect{r}_{A})/2-\mi\Omega_+(\vect{r}_{A})}
 \hspace{6ex}\\[.5ex]
+c_+^\ast(\vect{r}_{A}) c_-(\vect{r}_{A})\,
 \frac{\me^{[-\Gamma_1'(\vect{r}_{A})+\Omega_+^\ast(\vect{r}_{A})
 +\Omega_-(\vect{r}_{A})]t}
 -\me^{\{\mi[\tilde{\omega}'_{10}(\vect{r}_{A})-\omega]
 -\Gamma_1'(\vect{r}_{A})/2+\Omega_+^\ast(\vect{r}_{A})\}t}}
 {\omega-\tilde{\omega}'_{10}(\vect{r}_{A})
 +\mi\Gamma_1'(\vect{r}_{A})/2-\mi\Omega_-(\vect{r}_{A})}\\[.5ex]
+c_-^\ast(\vect{r}_{A}) c_+(\vect{r}_{A})\,
 \frac{\me^{[-\Gamma_1'(\vect{r}_{A})+\Omega_-^\ast(\vect{r}_{A})
 +\Omega_+(\vect{r}_{A})]t}
 -\me^{\{\mi[\tilde{\omega}'_{10}(\vect{r}_{A})-\omega]
 -\Gamma_1'(\vect{r}_{A})/2+\Omega_-^\ast(\vect{r}_{A})\}t}}
 {\omega-\tilde{\omega}'_{10}(\vect{r}_{A})
 +\mi\Gamma_1'(\vect{r}_{A})/2 -\mi\Omega_+(\vect{r}_{A})}\\[.5ex]
+|c_-(\vect{r}_{A})|^2\,
 \frac{\me^{[-\Gamma_1'(\vect{r}_{A})+\Omega_-^\ast(\vect{r}_{A})
 +\Omega_-(\vect{r}_{A})]t}
 -\me^{\{\mi[\tilde{\omega}'_{10}(\vect{r}_{A})-\omega]
 -\Gamma_1'(\vect{r}_{A})/2+\Omega_-^\ast(\vect{r}_{A})\}t}}
 {\omega-\tilde{\omega}'_{10}(\vect{r}_{A})
 +\mi\Gamma_1'(\vect{r}_{A})/2-\mi\Omega_-(\vect{r}_{A})}\,.
\end{multline}

Let us first make contact with result obtained in the limit of weak
atom--field coupling where the first term under the square root in
Eq.~(\ref{4.110}) is much larger than the second one. This is the
case when for given transition dipole moment, the spectrum of the
field in the resonance region is sufficiently flat,
\begin{equation}
\label{4.113}
\gamma_\nu \gg 2\Omega_\mathrm{R}(\vect{r}_{A})
\end{equation}
or when the atomic transition is sufficiently far detuned from the
field resonance,
\begin{equation}
\label{4.113-0}
|\Delta\omega(\mathbf{r}_{A})|\gg
 2\Omega^2_\mathrm{R}(\mathbf{r}_{A})/\gamma_\nu.
\end{equation}
By means of Taylor expansion it can then be shown that
\begin{align}
\label{4.113-1}
\Omega_\pm(\vect{r}_{A})\simeq\begin{cases}
-\mi\Delta\omega(\vect{r}_{A})
 -[\gamma_\nu-\Gamma_1'(\vect{r}_{A})]/2,\\ \\[-1.5ex]
 {\displaystyle\frac{\mi\Omega_\mathrm{R}^2(\vect{r}_{A})}{4}\,
 \frac{\Delta\omega(\vect{r}_{A})}
 {[\Delta\omega(\vect{r}_{A})]^2+\gamma_\nu^2/4}
 -\frac{\Omega_\mathrm{R}^2(\vect{r}_{A})}{8}\,
 \frac{\gamma_\nu}{[\Delta\omega(\vect{r}_{A})]^2
 +\gamma_\nu^2/4}}\,,
\end{cases}
\end{align}
so $c_+(\vect{r}_{A})$ $\!\simeq$ $\!1$, $c_-(\vect{r}_{A})$
$\!\simeq$ $\!0$ and Eq.~(\ref{4.100}) [together with
Eqs.~(\ref{4.103}) and (\ref{4.108})] approximates to \cite{0719}
\begin{align}
\label{4.115}
\vect{F}(t)=&\;\me^{-\Gamma_1(\vect{r}_{A})(t-t_0)}\,
 \frac{\mu_0}{\pi}\int_0^\infty\dif\omega\,\omega^2
 \frac{\left[\bm{\nabla}\vect{d}_{10}\sprod
 \mathrm{Im}\,\ten{G}^{(1)}(\vect{r},\vect{r}_{A},\omega)
 \sprod\vect{d}_{01}\right]_{\vect{r}=\vect{r}_{A}}}
 {\omega-\tilde{\omega}_{10}(\vect{r}_{A})
 -\mi\Gamma_1(\vect{r}_{A})/2}
 +\mathrm{C.c.}\nonumber\\[.5ex]
\simeq&\;\me^{-\Gamma_1(\vect{r}_{A})(t-t_0)}\vect{F}_1(\vect{r}_A)
\end{align}
with
\begin{equation}
\label{4.115-1}
\vect{F}_1(\vect{r}_A)
=\mu_0\Omega^2_{10}(\vect{r}_{A})
 \bigl\{\bm{\nabla}\vect{d}_{10}\sprod
 \ten{G}^{(1)}[\vect{r},\vect{r}_{A},\Omega_{10}(\vect{r}_{A})]
 \sprod\vect{d}_{01}\bigr\}_{\vect{r}=\vect{r}_{A}}
 +\mathrm{C.c.}
\end{equation}
and
\begin{equation}
\label{4.115-2}
\Omega_{10}(\vect{r}_{A})
=\tilde{\omega}_{10}(\vect{r}_{A})+\mi\Gamma_1(\vect{r}_{A})/2
\end{equation}
which corresponds to the term
$\sigma_{11}(t)\vect{F}_{11}^\mathrm{el,r}(\vect{r}_A)$
in Eqs.~(\ref{4.73}) and (\ref{4.74}) with
$\vect{F}_{11}^\mathrm{el,r}(\vect{r}_A)$ being given
according to Eq.~(\ref{4.76}).

The strong-coupling limit is realized if the spectrum of the field
features a sufficiently sharp peak and the atomic transition is near
resonant with this peak, such that
\begin{equation}
\label{4.116}
\gamma_\nu\le 2\Omega_\mathrm{R}(\vect{r}_{A})
 \quad\mathrm{and}\quad
 |\Delta\omega(\mathbf{r}_{A})|\ll
 2\Omega^2_\mathrm{R}(\mathbf{r}_{A})/\gamma_\nu.
\end{equation}
In this case, the square root in Eq.~(\ref{4.110}) becomes
approximately real,
\begin{equation}
\label{4.117}
\Omega_{\pm}(\vect{r}_{A})
 \simeq -{\textstyle\frac{1}{2}}\left\{\mi
 \Delta\omega(\vect{r}_{A})
 +{\textstyle\frac{1}{2}}[\gamma_\nu-\Gamma_1'(\vect{r}_{A})]\right\}
 \mp{\textstyle\frac{1}{2}}\mi\Omega(\vect{r}_{A})
\end{equation}
where
\begin{equation}
\label{4.118}
\Omega(\vect{r}_{A})=\sqrt{\Omega_\mathrm{R}^2(\vect{r}_{A})
 +[\Delta\omega(\vect{r}_{A})]^2
 -[\gamma_\nu-\Gamma_1'(\vect{r}_{A})]^2/4}
\end{equation}
so that the coefficients $c_\pm(\vect{r}_{A})$ [Eq.~(\ref{4.109})]
can be given in the form
\begin{equation}
\label{4.119}
c_\pm(\vect{r}_{A})
 =\frac{\Omega(\vect{r}_{A})\mp
 \Delta\omega(\vect{r}_{A})\pm\mi
 [\gamma_\nu\!-\!\Gamma'_1(\vect{r}_{A})]/2}
 {2\Omega(\vect{r}_{A})}\,.
\end{equation}
Substituting Eqs.~(\ref{4.117}) and (\ref{4.119}) into
Eq.~(\ref{4.111}) [together with Eq.~(\ref{4.112})], one finds
that for real dipole matrix elements $\vect{F}(t)$ approximates to
\cite{0719}
\begin{multline}
\label{4.121}
\vect{F}(t)
=2\me^{-[\gamma_\nu + \Gamma_1'(\vect{r}_{A})](t-t_0)/2}
 \sin^2[\Omega(\vect{r}_{A})(t-t_0)/2]\\
 \times\frac{[\Delta\omega(\vect{r}_{A})]^2
 -[\gamma_\nu\!-\!\Gamma'_1(\vect{r}_{A})]^2/4}
 {\Omega_\mathrm{R}^2(\vect{r}_{A})+
 [\Delta\omega(\vect{r}_{A})]^2
 -[\gamma_\nu\!-\!\Gamma'_1(\vect{r}_{A})]^2/4}\,
 \mathbf{F}_1(\vect{r}_{A})
\end{multline}
where $\mathbf{F}_1(\vect{r}_{A})$ is given according to
Eq.~(\ref{4.115-1}) with
\begin{equation}
\label{4.122}
\Omega'_{10}(\vect{r}_{A})
 =\tilde{\omega}'_{10}(\vect{r}_{A})+\mi\Gamma'_1(\vect{r}_{A})/2
\end{equation}
in place of Eq.~(\ref{4.115-2}). Note that
$[\gamma_\nu+\Gamma_1'(\vect{r}_A)]/2$ $\!=$ $\!\gamma_\nu/2$ if
$|\Delta\omega(\vect{r}_A)|$ $\!\ll$ $\gamma_\nu/2$ and
$[\gamma_\nu+\Gamma_1'(\vect{r}_A)]/2$ $\!=$
$\!\Gamma_1(\vect{r}_A)/2$ if $\gamma_\nu/2$ $\!\ll$
$\!|\Delta\omega(\vect{r}_A)|$ $\!\ll$
$\!2\Omega^2_\mathrm{R}(\mathbf{r}_{A})/\gamma_\nu$.

Comparing Eq.~(\ref{4.121}) with Eq.~(\ref{4.115}), we see that while
the resonant component of the force in the limit of weak atom--field
coupling simply exponentially decreases as a function of time, Rabi
oscillations of the force are typically observed in the
strong-coupling limit---in agreement with the well-known features of
spontaneous emission in the two coupling regimes. As a consequence of
the appearance of Rabi oscillations, the magnitude of the force
changes periodically; for appropriate spatial structure of the
resonantly interacting quasi-mode, the atom may be trapped with the
trap being set by the atom itself, cf.~also Refs.~\cite{0409,0410}.
Rabi oscillations do not occur if the system is initially prepared in
a dressed state; in this case the (exponentially decaying) force is
simply given by the gradient of the position-dependent part
$\pm\hbar\Omega(\vect{r}_{A})/2$ [recall Eq.~(\ref{4.118})] of the
respective dressed-state energy \cite{0179,0407}.

%%%%%%%%%%%%%%%%%%%%%%%%%%%%%%%%%%%%%%%%%%%%%%%%%%%%%%%%%%%%%%%%%%%%%%

\section{Concluding remarks}\vspace*{-1ex}
\label{sec5}

Dispersion forces are a particular signature of the quantum nature of
the interaction of matter with the electromagnetic field. As soon as
the interacting matter consists of a large number of elementary atomic
particles, exact microscopic calculations become very involved.
Therefore most theoretical approaches to dispersion forces make use of
assumptions typical of macroscopic electrodynamics by
introducing---sooner oder later---familiar macroscopic concepts such
as boundary conditions at surfaces of discontinuity and/or
constitutive relations averaged over a sufficiently large number of
the elementary constituents of the respective material objects.
Macroscopic electrodynamics whose applicability surprisingly ranges
even to nano-structures, has the benefit of being universally valid,
because it uses only very general physical properties, without the
need of involved ab initio calculations. Moreover, all the relevant
quantities used for characterizing the material objects can easily be
inferred from measurements. This concept does not only apply to
classical electrodynamics but also to QED. Macroscopic QED has been
well elaborated for the case of locally responding media described in
terms of complex-valued, position- and frequency-dependent
permittivities and permeabilities. It can be extended to arbitrary
linear media, including spatially dispersing media, since the
description of the quantized field in terms of current densities and
the Green tensor associated with the macroscopic Maxwell equations is
independent of the particular medium description. When supplemented
with standard atom--field coupling terms, a powerful tool for studying
medium-assisted quantum effects in QED is obtained. Clearly, the
applicability of the theory is restricted to body--body and body--atom
separations that are sufficiently large compared with the length scale
on which the atomistic structure of the bodies begins to play a
role.

In particular, the so established macroscopic QED provides a unified
approach to the various types of dispersion forces---an approach which
incorporates the benefits of normal-mode and linear-response
approaches, while exactly taking into account real material
properties. In particular, dispersion forces between electrically
neutral, unpolarized and unmagnetized ground-state bodies simply
reflect the forces which are due to the action of the fluctuating
body-assisted electromagnetic vacuum on the fluctuating charge and
current densities of the bodies. Since all the relevant
characteristics of the bodies enter the so obtained force formulas via
the Green tensor of the associated macroscopic Maxwell equations, they
are valid for arbitrary (linear) bodies. Both the Casimir stress and
the Casimir force density can thus be introduced in a natural way.
Moreover, by appropriate Born-series expansions of the Green tensor,
relations between dispersion forces on bodies and dispersion forces on
atoms can be established which clearly demonstrate the common origin
of all these forces. In particular, the force on an atom in the
presence of arbitrary bodies as well as the force between two atoms
can be obtained as limiting cases of the body--body force.

In this article we have restricted our attention to ground-state
bodies, i.e, to bodies that at zero temperature interact with the
electromagnetic field where the effect of dispersion forces is purely
quantum by nature. As outlined, an extension of the central results to
include equilibrium systems at finite temperatures can be obtained in
a straightforward way by simply replacing the vacuum averages by
thermal averages. In this context it should be pointed out that the
central assumption of linear-response theories according to which the
thermal average of the field fluctuations is related to the imaginary
part of the field response function, is explicitly fulfilled within
the framework of macroscopic QED.\vspace*{-1ex}\pagebreak

As we have seen, dispersion forces on ground-state atoms turn out to
be limiting cases of dispersion forces on macroscopic bodies, so the
corresponding formulas can be obtained without explicitly addressing
the underlying atom--field interaction. Of course, they can also be
derived by explicitly solving the quantum-mechanical problem of
individual atoms interacting with the electromagnetic field, with the
presence of macroscopic bodies being again described within the
framework of macroscopic QED. For ground-state atoms where only
virtual transitions occur, this leads to results that agree with the
ones obtained from the purely macroscopic approach, as expected. In
fact, explicitly addressing the atom--field interaction is more
flexible, because it can also be applied to non-equilibrium systems,
such as initially excited atoms where also real transitions are
involved in the atom--field interaction. In this case, a dynamical
description is in general preferred to be employed, leading to
time-dependent expressions for the forces, according to the
temporal evolution of the atomic quantum state. In particular
for weak atom--field coupling, the force on an initially excited atom
is a sum of components whose temporal evolution follows that of the
associated atomic density matrix elements which is in turn governed
by the familiar master equation of an atomic system undergoing
radiative damping. For strong atom--field coupling, damped Rabi
oscillations may occur which periodically change the magnitude of the
force. The dynamical approach could serve as a starting point for
studying dispersion forces on bodies that are not in thermal
equilibrium, by appropriately modeling such bodies as collections of
excited atoms \cite{0333,0522}.\vspace*{-1ex}

Including linearly responding bodies in macroscopic QED has the
advantage that from the very beginning of all calculations the effect
of the bodies is taken into account in a consistent manner, without
the need to specify the properties of the bodies at an early stage. In
this way very general results of broad applicability can be obtained.
This naturally applies not only to dispersion forces, but also to
other quantum phenomena of radiation--matter interaction which are
strongly influenced by the presence of macroscopic bodies---phenomena
that may be subsumed under the term Casimir effect in the broadest
sense of the word. Typical examples are the enhancement and inhibition
of spontaneous emission, resonant energy transfer between atoms or
molecules and the wide field of cavity-QED effects.

\ack
We would like to acknowledge fruitful collaboration with Ho Trung
Dung, T. Kampf, C. Raabe and H. Safari. Furthermore, we are grateful
to L. Arntzen, G. Barton, I. Bondarev, M. DeKieviet, A. Guzm\'{a}n, A.
Lambrecht, S. Linden, E. Shamonina, Y. Sherkunov, L. Rizzuto, M. S.
Toma\v{s} and C. V\'{i}llarreal for stimulating discussions.

%%%%%%%%%%%%%%%%%%%%%%%%%%%%%%%%%%%%%%%%%%%%%%%%%%%%%%%%%%%%%%%%%%%%%%
%%%%%%%%%%%%%%%%%%%%%%%%%%%%%%%%%%%%%%%%%%%%%%%%%%%%%%%%%%%%%%%%%%%%%%

\appendix

%%%%%%%%%%%%%%%%%%%%%%%%%%%%%%%%%%%%%%%%%%%%%%%%%%%%%%%%%%%%%%%%%%%%%%

\section{Overview over scenarios}
\label{app1}

In order to provide for an overview over the various theoretical
works on dispersion forces, references containing work on body--body,
atom--body and atom--atom forces are given in separate tables. In the
tables, the references are structured according to the scenarios
considered, regardless of the methods used to address these
scenarios.\vspace{4ex}

%%%%%%%%%%%%%%%  T A B L E %%%%%%%%%%%%%%%%%%%%%%%%%%%%%%%%%%%%%%%%%%%
\begin{table}[!h!]
\begin{center}
 \begin{tabular}{|c||c|c|c|}
\hline
Material $\rightarrow$ & & &Magneto- \\ \cline{1-1}
Geometry $\downarrow$ & \raisebox{2.5ex}[0pt]{Perfect cond.}
&\raisebox{2.5ex}[0pt]{Electric}
&electric \\ \hline\hline
&\cite{0373,0068,0131,0746}
&\cite{0120,0650,0649,0341,0630,0631,%
0651,0333,0522}
&\cite{0122,0123}
\\
&\cite{0602,0677,0674}
&\cite{0611,0652,0642,0641,0656,0640,0668,0667,0626}
&\cite{0659,0134}
\\
Half space
&\cite{0637,0744}
&\cite{0644,0132,0676,0690,0607,0657,0680,0628,0621,0604}
&\cite{0124,0133}
\\
+ half space
&\cite{0616}$^\mathrm{T}$
&\cite{0675,0672,0687,0048,0638,0606,0603},
\cite{0666}$^\mathrm{T}$
&\cite{0125}$^\mathrm{T}$
\\
&\cite{0747,0748,0749}$^\mathrm{L}$
&\cite{0688,0625,0645,0682,0681,%
0057,0264,0646,0691,0689,0683}$^\mathrm{T}$
&\cite{0126}$^\mathrm{T}$
\\
&
&\cite{0673}$^\mathrm{L}$,
\cite{0684,0743}$^\mathrm{Tq}$
&\cite{0127}$^\mathrm{T}$
\\ \hline
Half space
&
&
&\cite{0670,0669}
\\
+ plate
&
&
&\cite{0661,0198}
\\ \hline
Half space
&\cite{0695,0637,0744}
&\cite{0641,0638},
\cite{0625}$^\mathrm{T}$
&
\\
+ sphere
&
&
&
\\ \hline
Half space
&\cite{0750,0695}
&
&
\\
+ cylinder
&
&
&
\\ \hline
Plate
&
&\cite{0197,0612,0678,0665,0203,0647,0660}
&
\\
+ plate
&
&\cite{0639}$^\mathrm{Tq}$
&
\\ \hline
Plate
&
&\cite{0071}
&
\\
+ sphere
&
&
&
\\ \hline
Sphere
&\cite{0124,0637}
&\cite{0343,0654,0377}
&
\\
+ sphere
&
&
&
\\ \hline
\end{tabular}
\end{center}
\vspace*{2ex}
\caption{%
Schedular summary of references associated with theoretical work on
body--body forces. The superscripts indicate that finite temperature
(T), lateral forces (L) and/or torques (Tq) are included.
}
\label{tab4}
\end{table}%
%%%%%%%%%%%%%%%%%%%%%%%%%%%%%%%%%%%%%%%%%%%%%%%%%%%%%%%%%%%%%%%%%%%%%%

\clearpage

%%%%%%%%%%%%%%%  T A B L E %%%%%%%%%%%%%%%%%%%%%%%%%%%%%%%%%%%%%%%%%%%
\begin{table}[!h!]
\begin{center}
 \begin{tabular}{|c||c|c|c|}
\hline
Material $\rightarrow$ & & &Magneto- \\ \cline{1-1}
Geometry $\downarrow$ & \raisebox{2.5ex}[0pt]{Perfect conductor}
&\raisebox{2.5ex}[0pt]{Electric}
&electric \\ \hline\hline
&\cite{0030,0022,0288,0282,0289}
&\cite{0119,0120,0288,0282,0277,0025,0023,0026,0027}
&\cite{0043,0392}
\\
&\cite{0325,0412,0047,0055,0054,0056,0320}
&\cite{0269,0281,0273,0286,0276,0341,0348,0272,0283,0290,0036}
&\cite{0012,0018}
\\
&\cite{0321,0072,0035,0039}
&\cite{0285,0287,0295,0051,0053,0031,0039}
&\cite{0019}, \cite{0330}$^\mathrm{M}$
\\
&\cite{0041}, \cite{0367,0061}$^\mathrm{E}$
&\cite{0041,0077,0653,0400,0280,0048,0388}
&
\\
\raisebox{2.5ex}[0pt]{Half space}
&\cite{0062,0296,0235}$^\mathrm{E}$
&\cite{0020},\cite{0347,0028,0333,0522,0296}$^\mathrm{E}$
&
\\
&\cite{0042}$^\mathrm{E}$,
\cite{0095}$^\mathrm{M}$
&\cite{0308,0331,0042,0044,0045,0443}$^\mathrm{E}$
&
\\
&\cite{0376,0037}$^\mathrm{T}$
&\cite{0008,0012,0018}$^\mathrm{E}$,
\cite{0275}$^\mathrm{M}$,
\cite{0057}$^\mathrm{T}$
&
\\
&
&\cite{0264,0046,0394,0399}$^\mathrm{T}$
&
\\ \hline
Plate
&
&\cite{0391}$^\mathrm{T}$
&\cite{0012,0019}
\\ \hline
&\cite{0110,0346}
&\cite{0110,0270,0060,0069,0077,0349,0372,0305}
&\cite{0113}
\\
\raisebox{2.5ex}[0pt]{Sphere}
&
&\cite{0017}
&
\\ \hline
&
&\cite{0040,0397,0284,0077,0395,0316,0305}
&
\\
\raisebox{2.5ex}[0pt]{Cylinder}
&
&\cite{0186,0439,0187},
\cite{0442}$^\mathrm{E}$,
\cite{0391}$^\mathrm{T}$
&
\\ \hline
&\cite{0070,0321,0032,0314}
&\cite{0032,0314},
\cite{0033}$^\mathrm{T}$
&\cite{0012,0019}
\\
\raisebox{2.5ex}[0pt]{Planar}
&\cite{0327,0292,0278,0301,0063,0315}$^\mathrm{E}$
&
&
\\
\raisebox{2.5ex}[0pt]{Cavity}
&\cite{0293}$^\mathrm{E,M}$,
\cite{0034}$^\mathrm{T}$
&
&
\\ \hline
Spher. cav.
&
&\cite{0393,0396,0313,0305},
\cite{0441}$^\mathrm{E}$
&
\\ \hline
Cyl. cav.
&
&\cite{0316,0305}
&
\\ \hline
Parab. cav.
&\cite{0390,0389}
&
&
\\ \hline
\end{tabular}
\end{center}
\vspace*{2ex}
\caption{%
Schedular summary of references associated with theoretical work on
atom--body forces. Unless otherwise stated, nonmagnetic ground-state
atoms are considered. The superscripts indicate that excited atoms
(E), magnetic atoms (M) and/or finite temperature (T) are included.}
\label{tab3}
\end{table}%
%%%%%%%%%%%%%%%%%%%%%%%%%%%%%%%%%%%%%%%%%%%%%%%%%%%%%%%%%%%%%%%%%%%%%%

\clearpage

%%%%%%%%%%%%%%%  T A B L E %%%%%%%%%%%%%%%%%%%%%%%%%%%%%%%%%%%%%%%%%%%
\begin{table}[!h!]
\begin{center}
 \begin{tabular}{|c||c|c|c|}
\hline
Material $\rightarrow$ & & &Magneto- \\ \cline{1-1}
Geometry $\downarrow$ & \raisebox{2.5ex}[0pt]{Perfect conductor}
&\raisebox{2.5ex}[0pt]{Electric}
&electric \\ \hline\hline
&\multicolumn{3}{c|}{
\cite{0030,0374,0515,0510,0507,0119,0120,%
0514,0513,0237,0011,0521,0325,0498,%
0047,0055,0054,0056,0320,0051,0323,%
0007,0067,0201,0200,0490,0035,0048,0491,0020}}\\
Free space
&\multicolumn{3}{c|}{
\cite{0527,0526,0493,0099,0098,0333,0522}$^\mathrm{E}$,%
\cite{0496,0497}$^\mathrm{E,M}$,
\cite{0499,0121,0189,0492,0089,0095,%
0094,0097,0537,0096,0491}$^\mathrm{M}$}\\
&\multicolumn{3}{c|}{
\cite{0104}$^\mathrm{M,T}$,
\cite{0103,0376,0105,0101,0057,0264,0037,0102}$^\mathrm{T}$
}
\\ \hline
&
&\cite{0351}
&\cite{0009,0113,0739}\\
\raisebox{2.5ex}[0pt]{Bulk medium}
&
&
&\cite{0669}$^\mathrm{M}$
\\ \hline
Half Space
&\cite{0361,0367,0309,0679},
\cite{0093}$^\mathrm{T}$
&\cite{0518,0036,0107,0653,0523}
&\cite{0009,0491}
\\ \hline
Planar cavity
&\cite{0092},
\cite{0093}$^\mathrm{T}$
&\cite{0107},
\cite{0108}$^\mathrm{T}$
&
\\ \hline
\end{tabular}
\end{center}
\vspace*{2ex}
\caption{%
Schedular summary of references associated with theoretical work on
atom--atom forces, possibly in the presence of bodies. Unless
otherwise stated, nonmagnetic ground-state atoms are considered. The
superscripts indicate that excited atoms (E), magnetic atoms (M)
and/or finite temperature (T) are included.}
\label{tab2}
\end{table}%
%%%%%%%%%%%%%%%%%%%%%%%%%%%%%%%%%%%%%%%%%%%%%%%%%%%%%%%%%%%%%%%%%%%%%%

%%%%%%%%%%%%%%%%%%%%%%%%%%%%%%%%%%%%%%%%%%%%%%%%%%%%%%%%%%%%%%%%%%%%%%

\section{Green tensors}
\label{appA}

The Green tensor in free space is given by \cite{0003}
\begin{equation}
\label{A.1}
\ten{G}_\mathrm{free}(\vect{r},\vect{r}',\mi\xi)
 =\frac{1}{3}\Bigl(\frac{c}{\xi}\Bigr)^2\delta(\bm{\rho})\ten{I}
 +\frac{c^2\me^{-\xi\rho/c}}{4\pi\xi^2\rho^3}
 \bigl[a(\xi\rho/c)\ten{I}
 -b(\xi\rho/c)
 \vect{e}_\rho\tprod\vect{e}_\rho\bigr]
\end{equation}
($\bm{\rho}$ $\!=$ $\!\vect{r}-\vect{r}'$; $\rho$ $\!=$
$\!|\bm{\rho}|$; $\vect{e}_\rho$ $\!=$ $\!\bm{\rho}/\rho$) where
\begin{equation}
\label{A.2}
a(x)=1+x+x^2,\qquad b(x)=3+3x+x^2.
\end{equation}

The scattering Green tensor for the planar magneto-electric
structure characterized by Eqs.~(\ref{3.27}) and (\ref{3.28}) is given
by \cite{0217,0215}
\begin{equation}
\label{A.3}
\ten{G}^{(1)}(\vect{r},\vect{r}',\mi\xi)
=\int\dif^2q\,\me^{\mi\vect{q}\,\sprod\,(\vect{r}-\vect{r}')}
\ten{G}^{(1)}(\vect{q},z,z',\mi\xi)\quad\mbox{for }0<z,z'<d
\end{equation}
($\vect{q}\perp\vect{e}_z$) with
\begin{multline}
\label{A.4}
\ten{G}^{(1)}(\vect{q},z,z',\mi\xi)
 =\frac{\mu(\mi\xi)}{8\pi^2b}
 \sum_{\sigma=s,p}\biggl\{\frac{r_{\sigma -}r_{\sigma +}
 \me^{-2bd}}{D_\sigma}
 \Bigl[\vect{e}_\sigma^+\tprod\vect{e}_\sigma^+\me^{-b(z-z')}
 +\vect{e}_\sigma^-\tprod\vect{e}_\sigma^-\me^{b(z-z')}\Bigr]\\[.5ex]
+\,\frac{1}{D_\sigma}
 \Bigl[\vect{e}_\sigma^+\tprod\vect{e}_\sigma^-r_{\sigma -}
 \me^{-b(z+z')}
 +\vect{e}_\sigma^-\tprod\vect{e}_\sigma^+r_{\sigma +}
 \me^{-2bd}e^{b(z+z')}\Bigr]\biggr\}.
\end{multline}
Here, $b$ and $D_\sigma$ are defined by Eqs.~(\ref{3.32}) and
(\ref{3.33}), respectively,
\begin{equation}
\label{A.5}
\vect{e}_s^\pm=\vect{e}_q\vprod\vect{e}_z,
 \quad\vect{e}_p^\pm=-\frac{1}{k}(\mi q\vect{e}_z
 \pm b\vect{e}_q)
\end{equation}
($\vect{e}_q$ $\!=$ $\!\vect{q}/q$, $q$ $\!=$ $\!|\vect{q}|$) with
\begin{equation}
\label{A.6}
k=\frac{\xi}{c}\sqrt{\varepsilon(\mi\xi)\mu(\mi\xi)}
\end{equation}
are the polarization vectors for $s$- and $p$-polarized waves
propagating in the positive ($+$) and negative ($-$) $z$-directions,
and $r_\sigma^\pm$ $\!=$ $\!r_\sigma^\pm(\xi,q)$ with $\sigma$ $\!=$
$\!s,p$ describe the reflection of these waves at the right ($+$) and
left ($-$) walls, respectively.

In particular, assume that both walls are multi-slab magneto-electrics
consisting of $N_\pm$ homogeneous layers of thicknesses $d^j_\pm$ ($j$
$\!=$ $1,\ldots,N_\pm$) with $d^{N_\pm}_\pm$ $\!=$ $\!\infty$,
permittivity $\varepsilon^j_\pm(\omega)$ and permeability
$\mu^j_\pm(\omega)$. In this case the reflection coefficients can be
obtained from the recurrence relations [$r_{\sigma\pm}$ $\!\equiv$
$\!r_{\sigma\pm}^0$; $d$ $\!\equiv$ $\!d^0_\pm$; $\varepsilon(\omega)$
$\!\equiv$ $\varepsilon^0_\pm(\omega)$; $\mu(\omega)$ $\!\equiv$
$\mu^0_\pm(\omega)$]
\begin{align}
\label{A.7}
&r_{s\pm}^j=
\frac{(\mu^{j+1}_\pm b^j_\pm-\mu^j_\pm b^{j+1}_\pm )
+(\mu^{j+1}_\pm b^j_\pm+\mu^j_\pm b^{j+1}_\pm )
\,\me^{-2b^{j+1}_\pm d^{j+1}_\pm }r_{s\pm}^{j+1}}
{(\mu^{j+1}_\pm b^j_\pm+\mu^j_\pm b^{j+1}_\pm )
+(\mu^{j+1}_\pm b^j_\pm-\mu^j_\pm b^{j+1}_\pm )
\,\me^{-2b^{j+1}_\pm d^{j+1}_\pm }r_{s\pm}^{j+1}}\,,\\[.5ex]
\label{A.8}
&r_{p\pm}^j=
\frac{(\varepsilon^{j+1}_\pm b^j_\pm-\varepsilon^j_\pm b^{j+1}_\pm )
+(\varepsilon^{j+1}_\pm b^j_\pm+\varepsilon^j_\pm b^{j+1}_\pm )
\,\me^{-2b^{j+1}_\pm d^{j+1}_\pm }r_{p\pm}^{j+1}}
{(\varepsilon^{j+1}_\pm b^j_\pm+\varepsilon^j_\pm b^{j+1}_\pm )
+(\varepsilon^{j+1}_\pm b^j_\pm-\varepsilon^j_\pm b^{j+1}_\pm )
\,\me^{-2b^{j+1}_\pm d^{j+1}_\pm }r_{p\pm}^{j+1}}
\end{align}
($j$ $\!=$ $\!0,\ldots,N_\pm$ $\!-$ $\!1$) with
$r_{\sigma\pm}^{N_\pm}$ $\!=$ $\!0$ where
\begin{equation}
\label{A.9}
b^j_\pm = \sqrt{\frac{\xi^2}{c^2}\
\varepsilon^j_\pm(\mi\xi)\mu^j_\pm(\mi\xi)+q^2}\,.
\end{equation}
For single, semi-infinite slabs, Eqs.~(\ref{A.7}) and (\ref{A.8})
reduce to the Fresnel coefficients
\begin{equation}
\label{A.10}
r_{s\pm}=
\frac{\mu^1_\pm b - \mu b^1_\pm}
{\mu^1_\pm b + \mu b^1_\pm}\,,\qquad
r_{p\pm}=
\frac{\varepsilon^1_\pm b - \varepsilon b^1_\pm}
{\varepsilon^1_\pm b +\varepsilon b^1_\pm}\,.
\end{equation}

%%%%%%%%%%%%%%%%%%%%%%%%%%%%%%%%%%%%%%%%%%%%%%%%%%%%%%%%%%%%%%%%%%%%%%
%%%%%%%%%%%%%%%%%%%%%%%%%%%%%%%%%%%%%%%%%%%%%%%%%%%%%%%%%%%%%%%%%%%%%%

\bibliographystyle{elsart-num}
\bibliography{review}

%%%%%%%%%%%%%%%%%%%%%%%%%%%%%%%%%%%%%%%%%%%%%%%%%%%%%%%%%%%%%%%%%%%%%%

\end{document}